\documentclass[12pt,preprint]{aastex}
\input psfig.sty

\slugcomment{ApJ, In Press}
\shorttitle{Challenges to TTV}
\shortauthors{}

\begin{document}
\title{Quantifying the challenges of detecting unseen planetary companions with transit timing variations}
\author{Dimitri Veras\altaffilmark{1,2}, Eric B. Ford\altaffilmark{1}, Matthew J. Payne\altaffilmark{1}}
\altaffiltext{1}{Astronomy Department, University of Florida, 211 Bryant Space Sciences Center, Gainesville, FL 32111, USA}
\altaffiltext{2}{Currently at:  Institute of Astronomy, University of Cambridge, Madingley Road, Cambridge CB3 0HA, UK}
\email{veras@astro.ufl.edu}
\begin{abstract}
Both ground and space-based transit observatories are poised to significantly increase the number of known transiting planets and the number of precisely measured transit times.  The variation in a planet's transit times may be used to infer the presence of additional planets.  Deducing the masses and orbital parameters of such planets from transit time variations (TTVs) alone is a rich and increasingly relevant dynamical problem.  In this work, we evaluate the extent of the degeneracies in this process, systematically explore the dependence of TTV signals on several parameters and provide phase space plots that could aid observers in planning future observations.  Our explorations are focused on a likely-to-be prevalent situation: a known transiting short-period Neptune or Jupiter-sized planet and a suspected external low-mass perturber on a nearly-coplanar orbit.  Through $\sim 10^7$ N-body simulations, we demonstrate how TTV signal amplitudes may vary by orders of magnitude due to slight variations in any one orbital parameter ($10^{-3}$ AU in semimajor axis, 0.005 in eccentricity, or a few degrees in orbital angles), and quantify the number of consecutive transit observations necessary in order to obtain a reasonable opportunity to characterize the unseen planet ($\gtrsim 50$ observations).  Planets in or near period commensurabilities of the form $p$:$q$, where $p \le 20$ and $q \le 3$, produce distinct TTV signatures, regardless of whether the planets are actually locked in a mean motion resonance.  We distinguish these systems from the secular systems in our explorations.  Additionally, we find that computing the autocorrelation function of a TTV signal can provide a useful diagnostic for identifying possible orbits for additional planets and suggest that this method could aid integration of TTV signals in future studies of particular exosystems.


%
%
\end{abstract}


\section{Introduction}

\subsection{Motivation}

Of the first $\approx 60$ exoplanets detected by transit photometry, none were accompanied by additional planets in the same system.  The discovery of HAT-P-13c \citep{baketal2009} and CoRoT-7 \citep{queetal2009} broke new ground as the first systems to contain both a planet observed to transit and a second planet detected by other means.  Then Kepler-9 became the first (and so far only) confirmed system with multiple transiting planets \citep{holmanetal2010}, although five other Kepler target stars now show evidence of multiple transiting candidates \citep{steetal2010}.  With nearly a third of all known transiting planets formally announced in the year 2009\footnote{http://exoplanets.org/}, the relentless pace of transit detections suggests that soon investigators will discover an abundance of multi-planet systems containing at least one known transiting planet.  

This trend in exoplanet astrophysics highlights the importance of performing follow-up observations for 1) single-planet transiting systems, as a way to determine if additional planets exist, and 2) multi-planet systems with at least one transiting planet, in order to better constrain the parameters of all planets in those systems.  The gravitational tug of planetary perturbers on a known transiting Hot Jupiter will cause variations in the mid-transit times of that planet.  Encoded in these transit timing variations (TTVs) is the influence of the hidden planet(s), and the process of extracting the mass and orbital parameters of these planets will become increasingly important as the number of transit detections increases.  Up to now, the only planets definitiely confirmed by TTVs is for a system where {\it both} planets transit their parent star \citep{holmanetal2010}.

\subsection{Observational Studies}

Previous studies have attempted to solve this ``inverse'' problem of deducing the mass and orbital parameters of an unseen perturber from a limited number of TTVs for a few specific systems.  \cite{steago2005} performed one of the first TTV analysis on a particular exosystem when they studied 12 transit observations for TrES-1, and demonstrated that the data could have identified a hypothetical perturber in that system that is at the order of an Earth-mass, or lower. \cite{agoste2007} then combined 13 transit observations with 68 radial velocity measurements for HD 209458 in order to constrain the presence of additional planets in that system. \cite{miletal2008a} obtained up to 12 consecutive transits from MOST data sets of HD 209458b in 2004 and 2005, and obtained no TTV signatures above $80$ s.  \cite{miletal2008b} obtained up to 10 consecutive transits from MOST data sets of HD 189733b in 2006, and obtained no TTVs above $45$ s.  The authors of both papers then used these parameters in order to restrict the possible existence of additional planets in these systems.  \cite{diazetal2008} analyzed just 5 transit data points to conclude that OGLE-TR-111b cannot produce the variations in signal; they rule out the presence of a satellite causing the TTV, and instead suggest an exterior Earth-mass planet could be the source of the variation. \cite{adaetal2010}, using data from 6 additional tranits, has subsequently challenged these claims.  \cite{couetal2008} analyzed 28 transit observations for the approximately Neptune-mass planet, Gl 436b, and were able to rule out the existence of any planets which cause a TTV of over 60 s. \cite{csietal2010} considered the TTV on 36 transits of CoRoT-1b and did not find any periodic signals, so as to rule out additional planets in the form of Super Earths, Saturn-like planets and Jupiter-like planets each in particular regions of parameter space.  

All these investigations report fewer than 50 transits.  We will demonstrate that in most cases, {\it at least} 50 transits are needed in order to appreciably narrow the phase space of possible solutions to the inverse problem.  However, the ongoing space-based missions CoRoT and Kepler provide cause for optimism.  CoRoT has already discovered at least 7 planets \citep{baretal2008,aloetal2008,aigetal2008,deletal2008,rauetal2009,legetal2009,frietal2010} and Kepler at least 5 \citep{boretal2010}.  The nearly continuous observational coverage for $0.5 - 3.5$ yr by both missions should allow for hundreds of transits of a single planet to be observed. 

\subsection{Theoretical Studies}

Purely theoretical studies have appealed to both analytical considerations and N-body simulations.  The nature of the inverse problem coupled with the observed constraints on the architecture of observable planetary systems with a transiting planet dictate which relevant branches of perturbation theory are useful for TTV studies.  When N-body simulations are used, they need to be run for half of one year to several years.  Such constraints help provide the context for the several important theoretical contributions which have established a foundation on which future TTV studies may be based and which we now proceed to outline.

\cite{holmur2005} and \cite{agoetal2005}, in two early studies, considered systems with two coplanar planets. \cite{holmur2005} reported a) that disorderly and aperiodic TTV signals may arise from the presence of multiple planets (their Fig. 1), b) on the correlation between planetary period ratio and timing signal for fixed values of several other orbital parameters and masses (their Figs. 3-4), and c) an estimate for the amplitude of this signal using Laplace-Lagrange secular theory (their Eq. 1). These results helped demonstrate that secular theory can produce the correct order of magnitude signal in many cases, but fail to describe the spikes in TTV signal which appear when the planetary period ratio is commensurate.

\cite{agoetal2005} explored a greater region of phase space, and focused their investigations on four regions in particular:  a) a non-transiting internal perturber whose interaction with the transiting planet is negligible, b) a non-transiting external perturber that is both eccentric and very well-separated (with a semimajor axis ratio $> 5$) from the transiting planet, c) the secular case with both planets on nearly circular orbits, and d) the resonant case with both planets on circular orbits.  In all cases, the authors analytically estimate the timing variations caused by the non-transiting planet.  For the middle two cases, the agreement between the theory and N-body simulations is particularly promising (their Figs. 2-3).  The authors thereby help establish a collection of quantitative estimates for TTV signal amplitudes and provide details of their derivations.

\cite{nesmor2008}, \cite{nesvorny2009} and \cite{nesbea2010} developed semianalytic expressions to approximate the magnitude of TTV signals from planets with given masses and orbital parameters.  They also attempt to solve the inverse problem in a manner that is $\sim 10^4$ faster to compute than direct N-body simulations in the planar case with a circular transiting planet and non-transiting external planet.  \cite{nesmor2008} provide a formulation for solving the inverse problem in low and mid-eccentricity ($\lesssim 0.4$) secular regimes (their Fig. 4) given a ``cutoff parameter'' which determine the number of Fourier terms to use in their analysis.  \cite{nesvorny2009} extended these results to the case of eccentric transiting planets and non-coplanar planets, and for these systems illustrated the phase space regions in which he can solve the inverse problem to various levels of accuracy (his Figs. 1 and 3).  Such expressions provide a valuable guide for determining which configurations should be explained in more detail although the models have trouble reproducing the correct behavior at strong MMRs.

Several studies have addressed other important aspects of TTVs.  \cite{forhol2007} considered how TTVs can be used to determine the existence of and characterize Trojan planets -- planets that approximately share the same orbit as an observed transiting planet.  They analytically estimated the resulting TTV signal and demonstrate the goodness of their relation with N-body simulations.  A related system architecture, which, in principle, could be identified through TTVs, is that of a transiting planet containing a large (terrestrial-mass or larger) satellite.  During their orbits, the Earth leads or trails the Earth-Moon barycenter by up to 2.5 minutes, and Saturn leads or trails the Saturn-Titan barycenter by up to 30 seconds \citep{deeg2002}.  Hence, \cite{deeg2002} claimed that a (now realistic; see \citealt*{knuetal2007}) detectability threshold of $10$ s would be needed to reliably detect a moon in the Saturn-Titan system through TTVs.  \cite{simetal2007} furthered the theory of detecting exomoons with TTVs by showing how a planet-satellite system can be represented by a single theoretical body on the planet-satellite line at the ``photocenter''.  Therefore, the phase space explored in this work may be extended to planet-moon systems through such a relation.  \cite{kipping2009a} and \cite{kipping2009b} have described how exomoons can be characterized through the use of transit duration variations (TDVs) as a way of helping to break degeneracies of the inverse problem from TTVs alone. \cite{marlau2010} recently demonstrated how Doppler velocity measurements may be combined with TTVs in order to help remove the degeneracies inherent in identifying the architecture of the system.  They considered systems similar in configuration to HAT-P-13 and HD 40307, as well as one with a transiting giant planet and a terrestrial-mass companion trapped in low-order mean motion resonance.

Both the high precision (a few seconds; \citealt*{knuetal2007}) with which transits can now be measured and the close proximity (sometimes within $0.02$ AU of their parent stars\footnote{http://exoplanet.eu/},\footnote{http://exoplanets.org/}) of Hot Jupiters to their parent stars necessitate the consideration of physical effects (e.g. general relativistic precession, sunspots, internal gravity waves) that have previously been neglected in exoplanet studies.  For TTVs, most of the effects cause changes which are either well below the current detectability threshold or act on timescales much greater than a few years.  For example, \cite{miraldaescude2002} finds that over the course of a typical observing campaign (several years), both general relativity and the quadrupole moment of a star will induce a pericenter change of a Hot Jupiter of much less than one degree.  However, \cite{heygla2007} propose a method for using TTVs in order to determine the $J_2$ moment of a parent star, and \cite{palkoc2008} focus on detecting the effect of general relativity from both TTV and TDV observations.

\subsection{Our Methodology}

This paper primarily explores TTV phase space regions which are likely to be observationally relevant and are {\it not} well described by analytical theories.  This choice dictates that we rely on N-body simulations. Running $\sim 10^7$ simulations is not prohibitively time-consuming since each simulation runs for just $\le 10$ yr.  This regime includes a circular transiting Hot Jupiter and an exterior coplanar lower-mass planet, with no additional restrictions.  We emphasize the sensitivity of TTV profiles to the initial orbital angle configuration, an important relation that has often been glossed over in the literature.  We also showcase the difficulty in providing any constraints on systems with under 50 transit observations, despite several investigators' best efforts. Finally, we caution investigators on the dangers of confusing two planets which are locked ``in resonance'' with two planets whose periods are roughly commensurate. These results expand upon the preliminary investigations from \cite{verfor2009}.

In Section 2, we a) describe the simulations which serve as a template from which other plots in the paper may be compared, b) state our definitions, and c) provide qualitative estimates for the magnitudes involved in solving the inverse TTV problem.  In Sections 3-6, we demonstrate the dependence of TTV signal amplitude on number of transit observations, orbital angle configuration, planetary masses, and time evolution of orbital elements, respectively.  We introduce a method for modeling the shape of a TTV curve in Section 7, discuss radial velocity follow-up, light travel time, and extensions to this study in Section 8, and conclude in Section 9.  The Appendix details the derivation of libration widths for some of the relevant resonances considered here.

\section{Fiducial Simulations}

As a first step towards understanding the large parameter space for 2-planet systems, we establish a fiducial set of simulations which broadly characterize the TTV signals produced from a transiting hot Jupiter and an external terrestrial-mass planet.  We consider a $M_i = 1 M_J$ inner planet which transits its parent $M_{\star} = 1 M_{\odot}$ star on an initially circular orbit ($e_i = 0$) at $a_i = 0.05$ AU, and an external $M_o = 1 M_{\oplus}$ planet with semimajor axis, $a_o$, and eccentricity, $e_o$.  The subscripts ``i'' and ``o'' are abbreviations for ``inner'' and ``outer''.  The mean longitude, mean anomaly and longitude of pericenter will be denoted by $\lambda$, $\Pi$ and $\varpi$, respectively.  For any system, only one of $\lambda$ or $\Pi$ needs to be specified; either determines a planet's location along a Keplerian orbit, and the former is often used in planar resonance studies \citep{murder2000}.  Henceforth, all stated orbital parameters are assumed to be {\it initial values} unless an explicit time dependence is included.

\begin{figure*}
  \centering
  \begin{tabular}{c}
    \multicolumn{1}{c}{\includegraphics[width=1.00\textwidth,height=0.85\textheight]{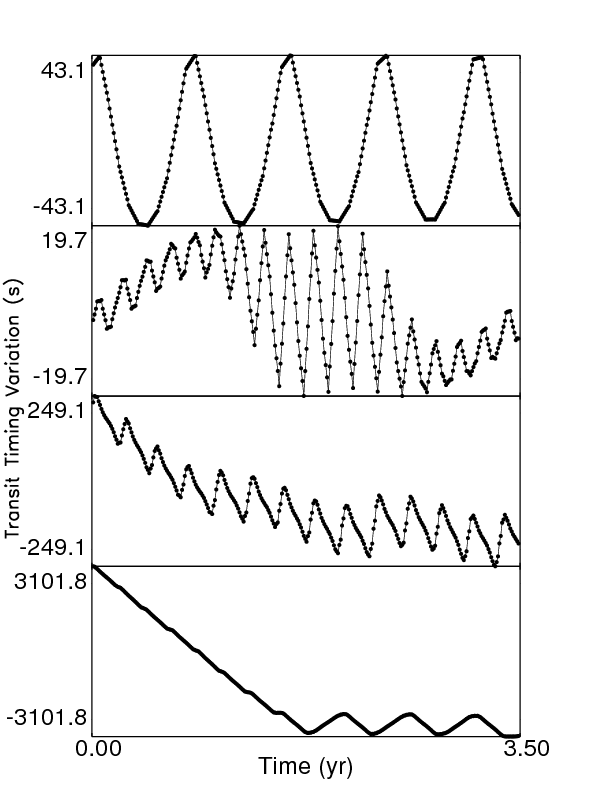}}
\\
  \end{tabular}
  \caption{A sample of transit timing variation (TTV) curves for four different systems over the course of the nominal Kepler mission lifetime ($3.5$ yr).  Each dot represents a transit, and the vertical axis labels are slightly offset for clarity, but do reflect the range of TTVs.  The curves from top to bottom correspond to 
$\lbrace{a_o/a_i = 3.6593, e_o = 0.596, \Pi_o = 70^{\circ}, \varpi_o = 180^{\circ} \rbrace}$,
$\lbrace{a_o/a_i = 2.3313, e_o = 0.395, \Pi_o = 0^{\circ}, \varpi_o = 0^{\circ} \rbrace}$,
$\lbrace{a_o/a_i = 1.5812, e_o = 0.204, \Pi_o = 0^{\circ}, \varpi_o = 0^{\circ} \rbrace}$, and
$\lbrace{a_o/a_i = 1.5812, e_o = 0.219, \Pi_o = 0^{\circ}, \varpi_o = 0^{\circ} \rbrace}$, with $\Pi_i = 0^{\circ}$ for all curves.  Note the wide variety of patterns exhibited by the TTV curves.}
  \label{sample}
\end{figure*}

After establishing values for $a_o/a_i, e_o, \varpi_o$, $\Pi_i$ or $\lambda_i$, and $\Pi_o$ or $\lambda_o$, we integrate the system over 10 years and tabulate the times of transits, and subtract the best fit Keplerian model to generate a TTV curve.  The deviations of the transit times from strict periodicity tend to form periodic patterns which can have timescales extending to the duration of the simulations.  We display a sample of these patterns and periodicities in Fig. \ref{sample} over the course of the nominal Kepler mission lifetime ($3.5$ yrs) in order to motivate the other figures in this paper.  The dots represent transit times, and the amplitude of their variations ranges over three orders of magnitude across the panels.  The variety of the patterns seen hint at the difficulties in deriving masses and orbital parameters of the external planet from TTV curves alone.  The top panel curve ($a_o/a_i = 3.6593, e_o = 0.596, \Pi_o = 70^{\circ}, \Pi_i = 0^{\circ}, \varpi_o = 180^{\circ}$) may be fitted well with a single sinusoid.  The curve in the next panel below ($a_o/a_i = 2.3313, e_o = 0.395, \Pi_i = \Pi_o = 0^{\circ}, \varpi_o = 0^{\circ}$) exhibits several modulations in addition to a simple sinusoid.  Up until $\approx 1$ yr, and between $\approx 2.5-3.5$ yr, this curve appears to exhibit a long period ($> 3.5$ yr) trend with a modulated sinusoid of amplitude less than $10$ s.  In between, at $\approx 1-2.5$ yr, the curve appears to oscillate with an amplitude of $20$ s without modulation.  The curve in the second panel from bottom ($a_o/a_i = 1.5812, e_o = 0.204, \Pi_i = \Pi_o = 0^{\circ}, \varpi_o = 0^{\circ}$) appears to exhibit a very long periodic trend ($\ge 3.5$ yr) with a high amplitude (hundreds of seconds) and a sawtooth-like modulation of amplitude $< 100$ s .  Note the distribution of spacings between the transit times on the modulated sinusoids.  These spacings may provide hints as to the architecture of the system (see Section 7).  The bottom panel ($a_o/a_i = 1.5812, e_o = 0.219, \Pi_i = \Pi_o = 0^{\circ}, \varpi_o = 0^{\circ}$) displays TTVs which vary by thousands of seconds over the first $\approx 2$ yrs before abruptly taking on the form of a simple sinusoid with amplitude $< 10^3$ s.  This panel especially highlights the importance of considering a sufficient number of observed transits with any type of analysis in order to avoid spurious conclusions.  We emphasize that the four curves in Fig. \ref{sample} are not representative of all curves henceforth studied, but rather illustrate different types of trends and patterns that one might uncover.

\begin{figure*}
  \centering
  \begin{tabular}{c}
    \multicolumn{1}{c}{\includegraphics[width=1.00\textwidth,height=0.90\textheight]{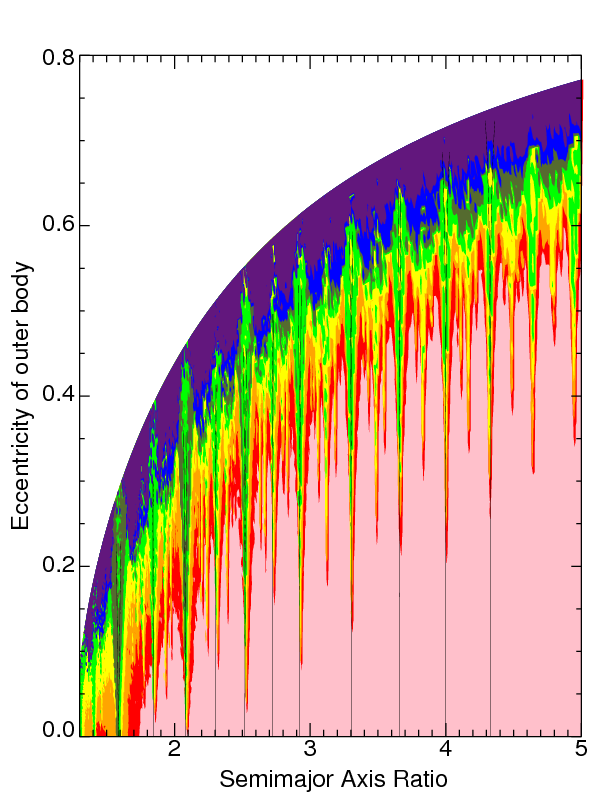}}
\\
  \end{tabular}
  \put(-414,98){\includegraphics[trim = 0mm 0mm 0mm 0mm, clip, width=0.28\textwidth]{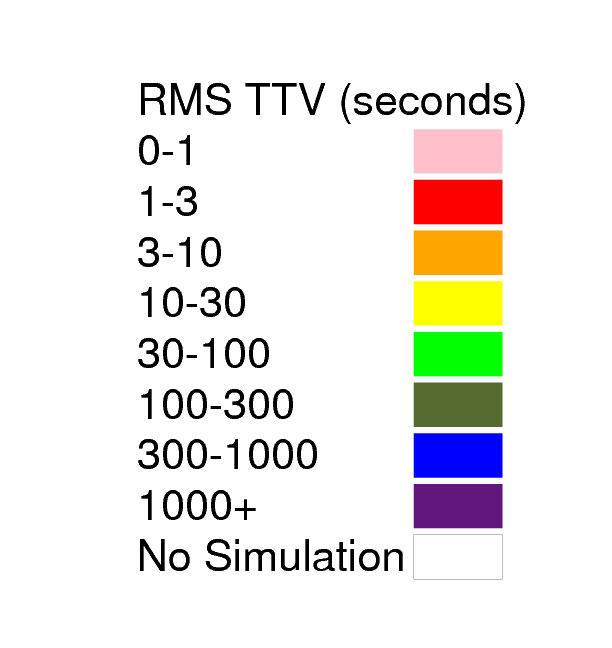}}
%
%
%
  \caption{``The Flames of Resonance''; the median RMS TTV amplitude ($S({\bf Q})$) for 5 different initial orbital configurations and $\approx 10$ yrs-worth ($N = 874$) of consecutive transits for a transiting $1 M_J$ hot Jupiter on a circular orbit at $0.05$ AU and a $1 M_{\oplus}$ external perturber with the orbital parameters indicated on the plot.  The contour levels in seconds are: pink ($0-1$), red ($1-3$), orange ($3-10$), yellow ($10-30$), light green ($30-100$), olive ($100-300$), blue ($300-1000$), and purple ($> 1000$).  Overplotted are resonant libration widths for selected PCs (Period Commensurabilities).}
  \label{fiducial}
\end{figure*}

In principle, we may select any combination of transits to mimic observed data.  For our fiducial case, we select the TTV curves based on the first 10 years of transits ($N \approx 874$, where $N$ is defined as the number of transits). Figure \ref{fiducial} displays the results of integrations of 120,000 systems.  We sampled 400 logarithmically-spaced values of the semimajor axis ratio and 60 uniformly-spaced values of $e_o$ for each semimajor axis ratio.  For every given pair of $a_o/a_i$ and $e_o$ values, we simulated 5 systems initialized with random mean anomalies of both planets and a random longitude of pericenter of the outer planet.  The reported values represent the median root-mean-square (RMS) TTV deviation amplitude (in seconds) of each of these sets of 5 simulations.  We only sampled systems guaranteed to be stable according to the Hill Stability Limit \citep{gladman1993}, i.e. where $e_o < e_H$.  The value of $e_H$ is a function of both planets' masses, semimajor axes and eccentricities, and for $e_o < e_H$, the orbits are guaranteed to never cross.  As we vary these parameters, the bounding curve (where $e_o = e_H$) on the figure will change slightly. We caution that i) some (unsampled) systems which lie above the curve in Fig. 1 may be stable, and that ii) some systems close to the Hill stability boundary which we do sample might be Lagrange unstable (where the outer planet drifts outward, generally causing an increase in the RMS TTV signal).  We explore the extent of the systems that are likely to be Lagrange unstable in Section 8.1, but note briefly here that such systems are very unlikely to be observed.   The inner semimajor axis ratio bound of 1.3 was chosen to roughly correspond to the point where the Hill Stability curve intersects the x-axis.  The outer semimajor axis ratio bound of 5.0 is arbitrarily chosen to allow one to consider highly hierarchical (widely-separated) systems.  Additionally, 5.0 is the ratio beyond which analytic formulae achieve success at reproducing the RMS TTV amplitude given the orbital parameters of the exterior planet (Fig. 2 of \citealt*{agoetal2005}).

Figure \ref{fiducial} is contoured on a logarithmic scale.  If one takes $10$s as the current detectability threshold for TTV signals, then any region on the plot that is not white, pink or red should contain a detectable signal.  Hence, widely separated ($a_o/a_i > 3$) planets containing an external planet on a low-eccentricity ($e_1 \lesssim 0.2$) orbit produce currently undetectable signals.  Additionally, for almost any $a_o/a_i \le 5$ with an external planet whose orbital eccentricity is close to the Hill Stability limit, the resulting TTV signal will be easily ($ \ge 10^3$ s) detectable.  The figure demonstrates that the signal amplitude is highly (on the $10^{-3} - 10^{-2}$ AU scale) sensitive to the semimajor axis ratio.

The flame-like features on the plot, similar to those found in Fig. 5 of \cite{agoetal2005}, indicate regions where the two planets are near a period commensurability (PC) that can be expressed as a ratio between two small integers ($p$:$q$, where $p, q \le 20$).  The longest of these ``flames of resonance" correspond to period commensurabilities of the form $p$:$1$, where $p \le 11$.  These PC locations approximate well regions where both planets might be locked in or reside just outside of a mean motion resonance (MMR).  Whether or not the planets with a given $p:q$ ratio are {\it} in a MMR is subject to individual system study, and is often highly dependent on its orbital angle architecture.  In the exoplanet dynamics community, a widely used criterion for determining whether a system is ``in" MMR is to determine if at least one ``resonant angle" is librating.   Consideration of {\it multiple} resonant angles for a given $p$:$q$ can be found in Fig. 3 of \cite{laucha2001}, Section 4.2 of \cite{kleetal2005}, Section 4.1.1 of \cite{rayetal2008}, Fig. 9 of \cite{crietal2008}, and Section 5 of \cite{fabmur2010}.  However, \cite{micetal2008a,micetal2008b} claim that for a given $p$:$q$, a MMR is characterized just by {\it one} resonant angle, whereas in the planar case, the other independent angular variable should be a secular angle.  \cite{Mardling2008,Mardling2010} asserts that for a given $p$:$q$, the resonant angle can be expressed in terms of a single index that traces the contribution from a particular semimajor axis ratio order in the gravitational potential.

For our purposes, we can neglect the contribution of the inner planet's longitude of pericenter ($\varpi_i$) because $e_i = 0$ and the mass ratio of the outer planet to inner planet ($M_o/M_i$) is negligible ($\sim 0.001$).  Hence, $e_i(t)$ will remain low ($\le 0.01$) as $t$ increases, and the only resonant angle we consider is:

\begin{equation}
\phi_{p,q} \equiv p \lambda_o - q \lambda_i - (p-q) \varpi_o
.
\label{resangle}
\end{equation}

The libration width of this resonance, defined here to be a range of $a_o$ at a given $e_o$, bounds a region of phase space where this angle {\it might} librate.  We use the formalism in \cite{murder2000} to compute libration widths for a sampling of 1st-4th order resonances, where order $\equiv p - q$.  Stan Dermott (2010, private communication) also provided us with the coefficients needed to compute libration widths for 5th-8th order resonant libration widths, with $q = 1$.  The horizontal spacing between the thin vertical lines extending from the x-axis on the plots in Fig. \ref{fiducial} represent these libration widths.  Notice that the spacing is negligible until $e_o \gtrsim e_H/2$.  Further, the lines do not extend to the Hill Stability boundary, because of the Sundman convergence criterion \citep{ferrazmello1994,sidnes1994}.  See the Appendix for additional details and a derivation of this libration width.

The location and extent of the libration widths help confirm that pronounced RMS TTV amplitudes ($\equiv S({\bf Q}) \equiv S(t,a_o/a_i,e_o,\Pi_o,\Pi_i,\varpi_o)$) are likely caused by PCs and that MMR configurations are achieved only for a high enough eccentricity of the outer planet.  Additionally, the libration widths for the $9$:$1$ MMR demonstrate that two planets may be locked in that resonance despite the weakness of an 8th-order resonance (resonant strength scales as $\sim e^{p-q}$).  Such planets would likely produce a detectable TTV signal.  Note also that the length of each ``flame" corresponds to the value of $q$.  Planets locked in a high-order ($\ge 10$) MMR with $q = 1$ or $2$ can produce a detectable signal ($S({\bf Q}) > 10$s) for sufficiently large $e_o$.  Although the highest-order MMR which would produce a detectable signal is highly dependent on the characteristics of the systems studied, planets in MMR of order $20-30$ can in principle produce detectable signals if $e_o$ is high enough.  Thus, a transiting Hot Jupiter at 0.05 AU could have detectable TTVs due to an Earth-mass planet in the habitable zone of a K or M-type star.  Because $S({\bf Q})$ is highly (on the $10^{-3} - 10^{-2}$ AU scale) sensitive to the semimajor axis ratio, detectable MMRs may help identify planets with particular orbital parameters.  However, when considering high $q$ MMRs, one should also take into account the time dependence of resonant locations; as a system evolves, libration widths would likely oscillate in $a_o(t) - e_o(t)$ space.

\begin{figure*}
  \centering
  \begin{tabular}{cc}
    \multicolumn{2}{c}{\includegraphics[width=0.50\textwidth,height=0.40\textheight]{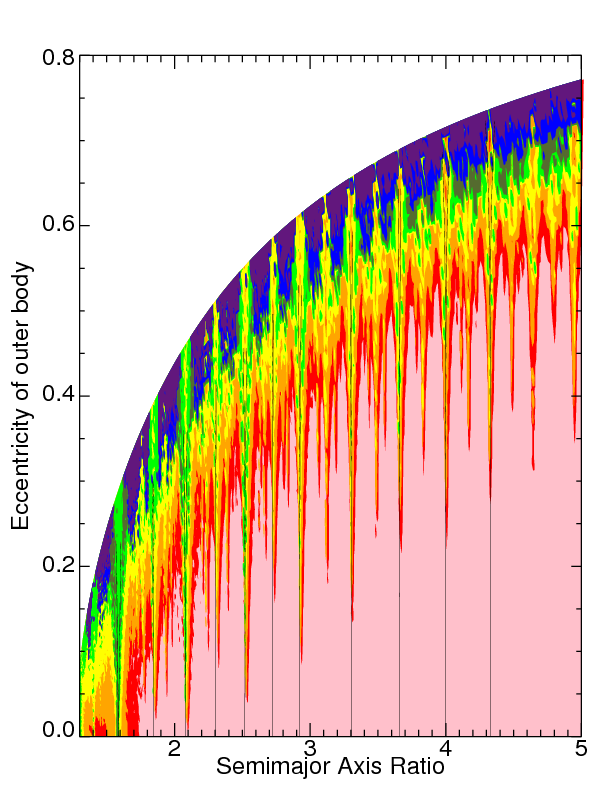}}
\\
    \multicolumn{2}{c}{\includegraphics[width=0.50\textwidth,height=0.40\textheight]{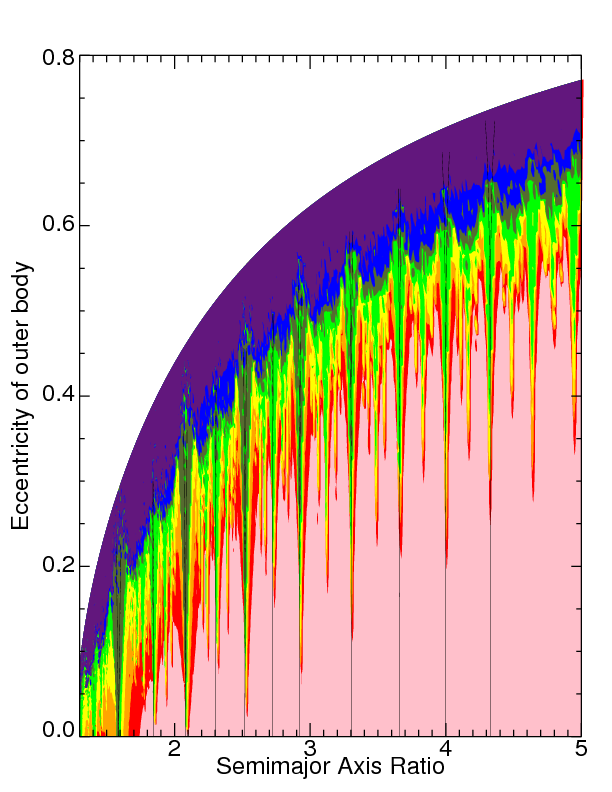}}
\\
  \end{tabular}
  \put(-365,198){\includegraphics[trim = 0mm 0mm 0mm 0mm, clip, width=0.30\textwidth]{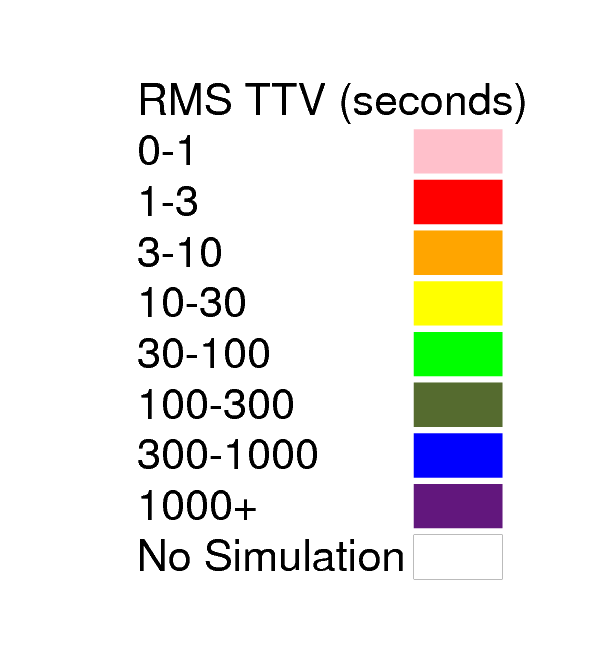}}
  \put(-350,150){\textbf{\LARGE Minimum} }
  \put(-350,-145){\textbf{\LARGE Maximum} }
%
%
  \caption{The minimum (upper panel) and maximum (lower panel) RMS TTV amplitude ($S({\bf Q})$) from the 5 different initial orbital configurations sampled in Fig. \ref{fiducial}.  Note that ($S({\bf Q})$) can vary by orders of magnitude depending on whether the minimum or maximum value is sampled.
}
  \label{minmax}
\end{figure*}

Figure \ref{fiducial} displays the median root-mean-square (RMS) TTV deviation amplitude (in seconds) of five different orbital angle configurations whereas Fig. \ref{minmax} reports the minimum and maximum deviations.  We emphasize that the following conclusions are for our fiducial case, with a given perturbing mass of $1 M_{\circ}$.  The differences in the two plots in Fig. \ref{minmax} are striking.  For a given $a_o$ and $e_o$, $S({\bf Q})$ may vary by orders of magnitude based on the initial values of the mean longitudes/mean anomalies and the outer planet longitude of pericenter ($\varpi_i$ is meaningless when $e_i = 0$).  The range of TTV signal amplitudes achieved from the different combinations of initial angles is comparable to the lower panel of Fig. \ref{minmax}, the maximum signal achieved.  Further, the median signals are on average greater than the mean signals (not shown).  The secular regime, away from PC, demonstrates little variation (at most a few seconds) in signal with initial orbital configuration.  However, detections in these regimes are unlikely without the advent of better-precision observational techniques.  The upper panel in Fig. \ref{minmax}, which displays the minimum signal achieved, best shows that for many strong (order $\le 7$) MMRs with $q=1$ or $2$, the RMS signal is {\it lower} than that from the high-eccentricity (at $e_H$) secular regime.  Therefore, for observations of a Hot Jupiter, 1) moderate (tens of seconds) values of $S({\bf Q})$ could indicate that $e_o$ may be as high as $e_H$, 2) high amplitude ($\gtrsim 10^3$) signals are not necessarily indicative of close proximity (within a few libration widths) to a strong PC, and 3) high amplitude ($\gtrsim 10^3$) $S({\bf Q})$ values are {\it always} indicative of a high value ($\gtrsim 0.8 e_H$) of $e_o$.

The choice of contour levels is arbitrary, but meant to show phase space structure and establish the detectability threshold of $10$s.  Increasing each contour level by one order of magnitude indicates that little structure can be discerned in the highest signal (purple) regions of Figs. \ref{fiducial} and \ref{minmax} and that $S({\bf Q})$ can reach well over $10^4$s.  Contrastingly, lowering each contour level by one order of magnitude demonstrates clear structure in the lowest signal regime (pink) of Figs. \ref{fiducial} and \ref{minmax}.  TTV curves in this regime have been successfully correlated with the mass and orbital elements of a hypothetical planet \citep{agoetal2005,nesmor2008}.  However, characteristic exoplanet detection thresholds ($\approx 10$ s) exceed typical signal amplitudes in this regime.  

\section{Correlations with the Number of Observations}

\begin{figure*}
  \centering
  \begin{tabular}{cccccc}
    \multicolumn{2}{c}{\includegraphics[width=0.33\textwidth]{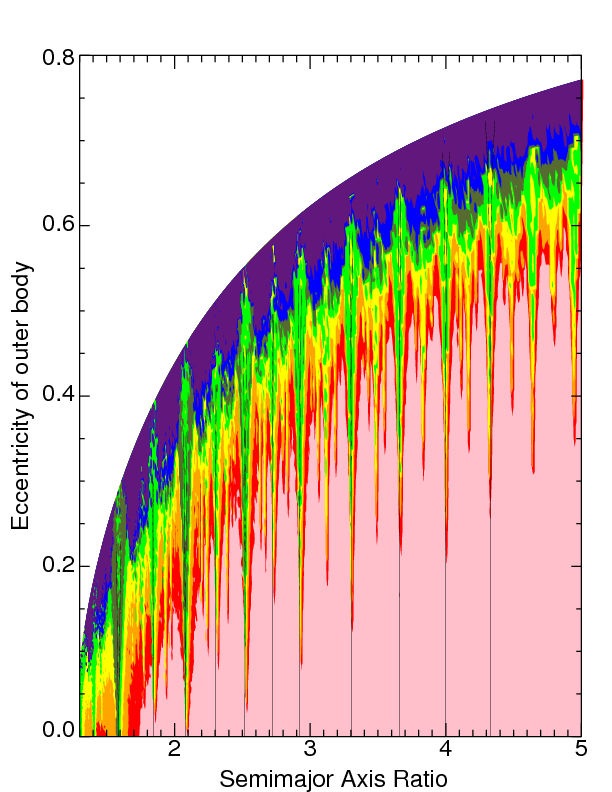}}&
    \multicolumn{2}{c}{\includegraphics[width=0.33\textwidth]{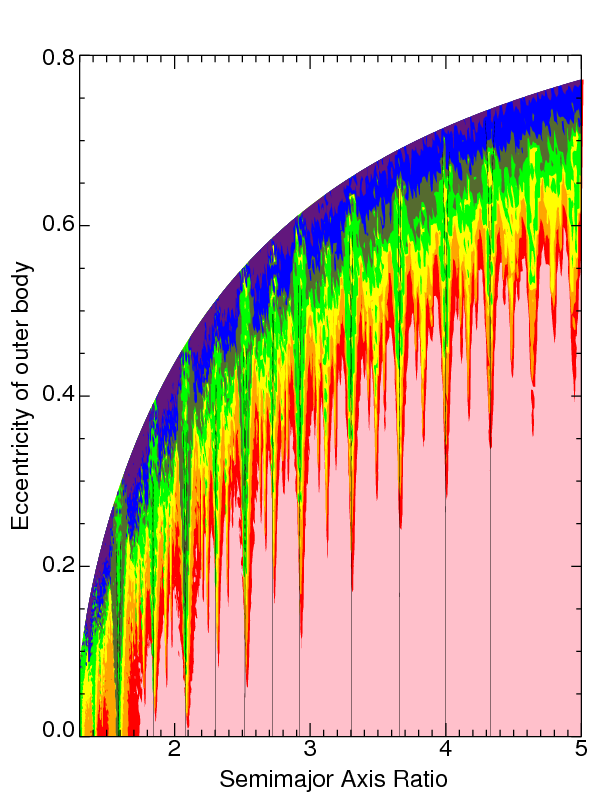}}&
    \multicolumn{2}{c}{\includegraphics[width=0.33\textwidth]{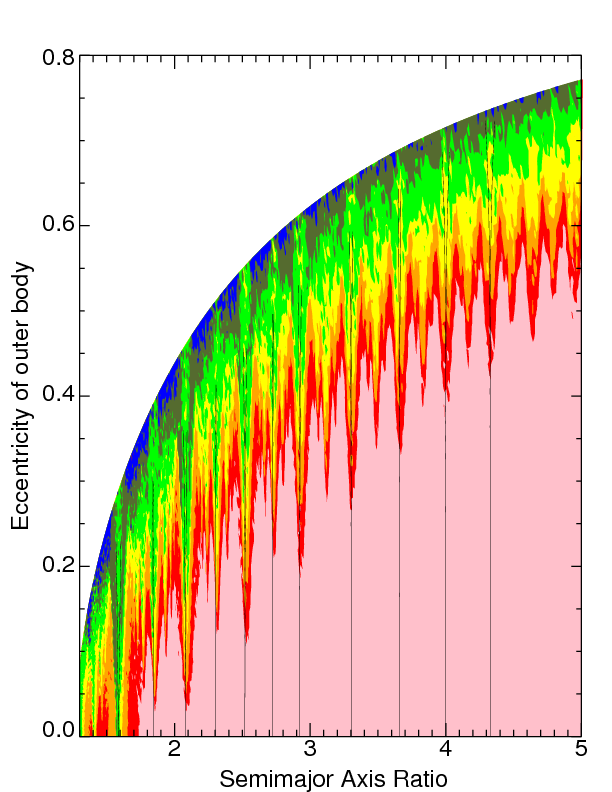}}\\
    \multicolumn{2}{c}{\includegraphics[width=0.33\textwidth]{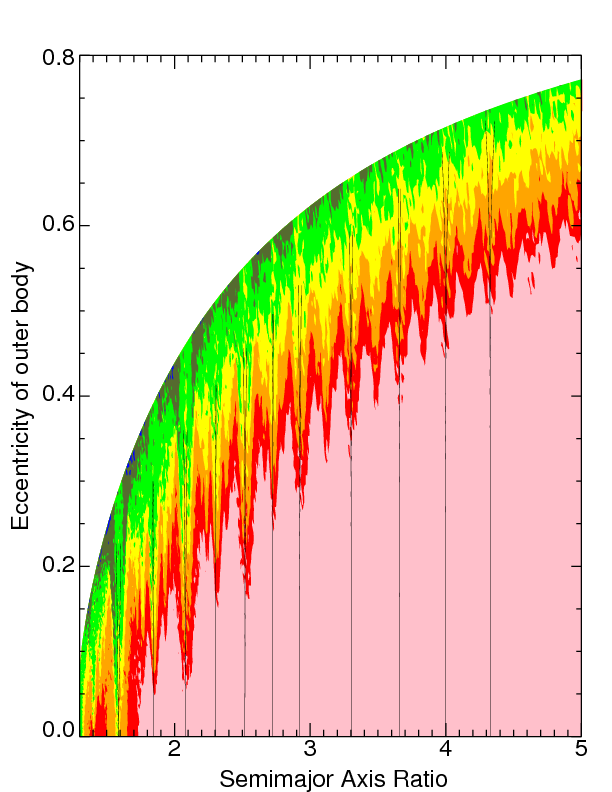}}&
    \multicolumn{2}{c}{\includegraphics[width=0.33\textwidth]{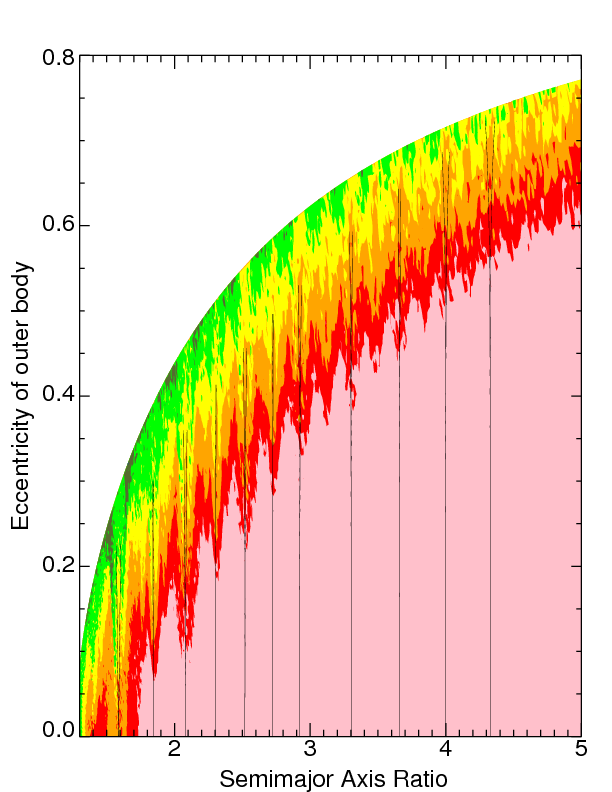}}&
    \multicolumn{2}{c}{\includegraphics[width=0.33\textwidth]{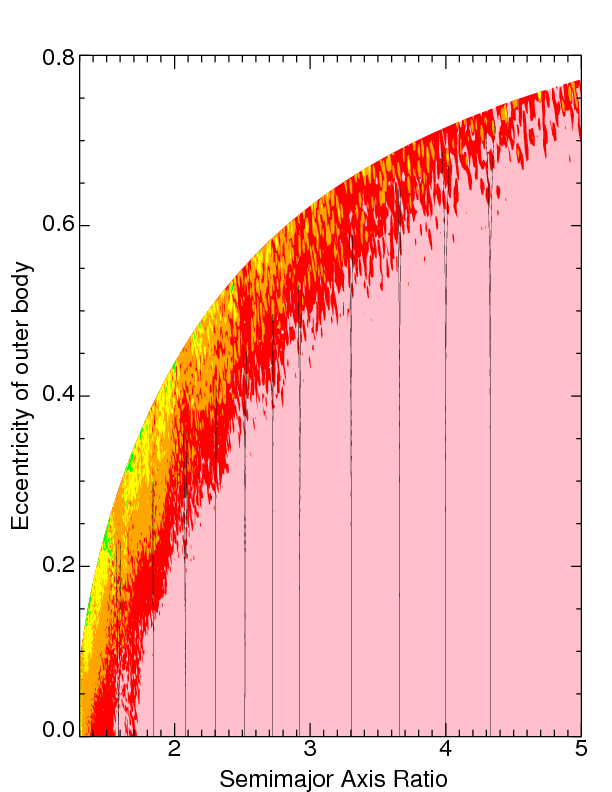}}
\\
  \end{tabular}
%
  \put(-471,175){\textbf{\large N=874} }
  \put(-305,175){\textbf{\large N=313} }
  \put(-133,175){\textbf{\large N=100} }
  \put(-469,-30){\textbf{\large N=50} }
  \put(-305,-30){\textbf{\large N=30} }
  \put(-135,-30){\textbf{\large N=10} }
%
%
  \caption{The Flames of Resonance (Fig. \ref{fiducial}) for different numbers of consecutive transit observations; $N = 874$ (corresponding to $\approx 10$ yr), $N = 313$ (corresponding to $\approx 3.5$ yr; the nominal Kepler lifetime), and $N = 100, 50, 30$ and $10$.  The contour levels are the same as in Figs. \ref{fiducial} and \ref{minmax}.  Overplotted are resonant libration widths for selected PCs (Period Commensurabilities).  Note that for detectable signals ($> 10$ s), often several years-worth of observations are needed to pinpoint the true TTV amplitude.}
  \label{Ncont}       
\end{figure*}

\begin{figure*}
  \centering
  \begin{tabular}{cccccc}
    \multicolumn{2}{c}{\includegraphics[width=0.33\textwidth]{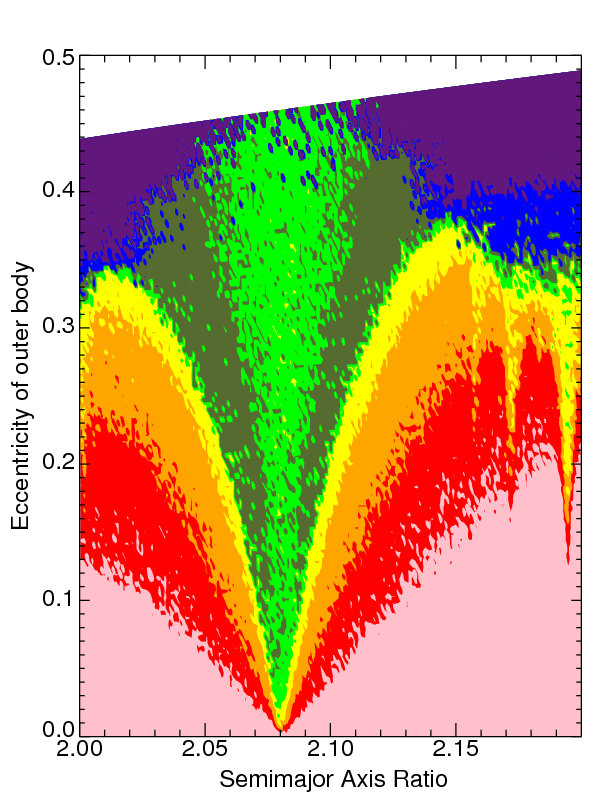}}&
    \multicolumn{2}{c}{\includegraphics[width=0.33\textwidth]{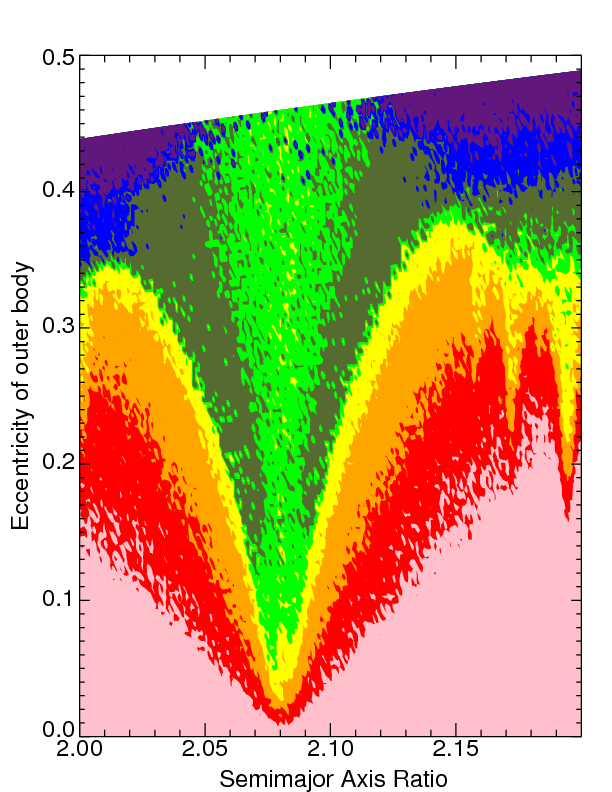}}&
    \multicolumn{2}{c}{\includegraphics[width=0.33\textwidth]{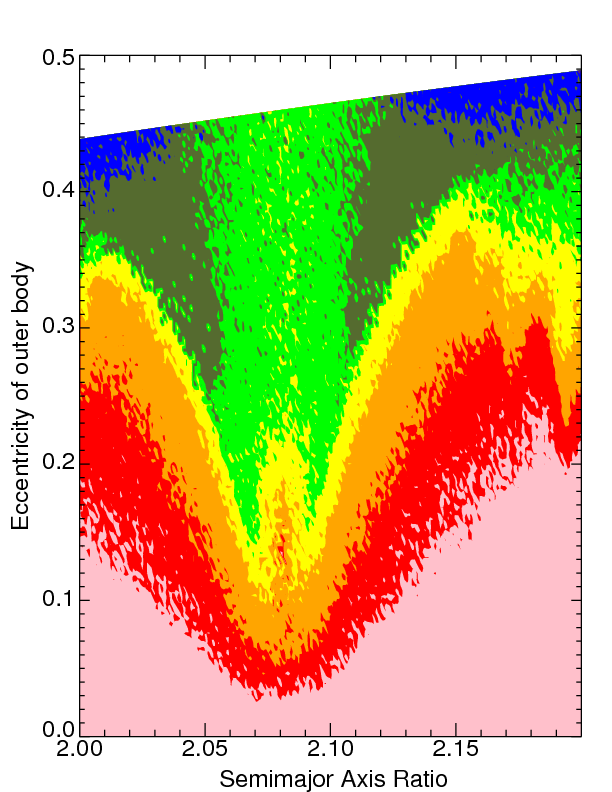}}\\
    \multicolumn{2}{c}{\includegraphics[width=0.33\textwidth]{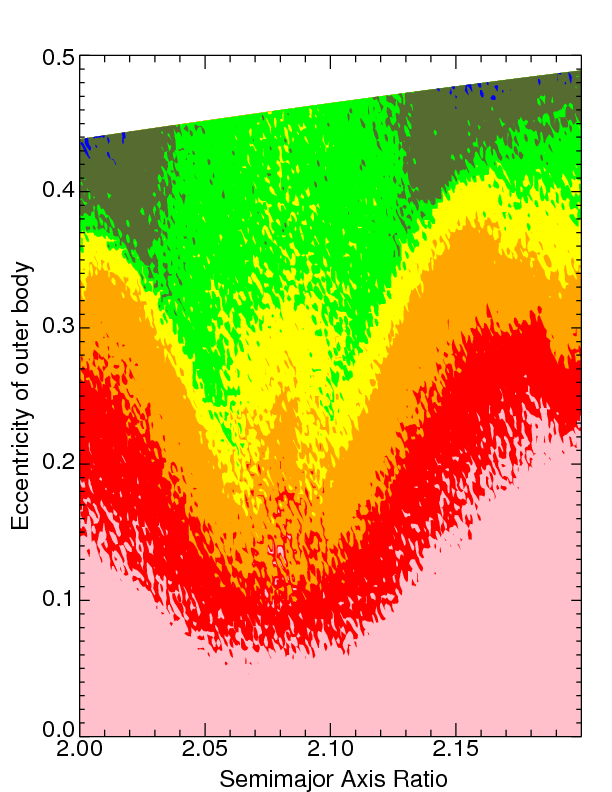}}&
    \multicolumn{2}{c}{\includegraphics[width=0.33\textwidth]{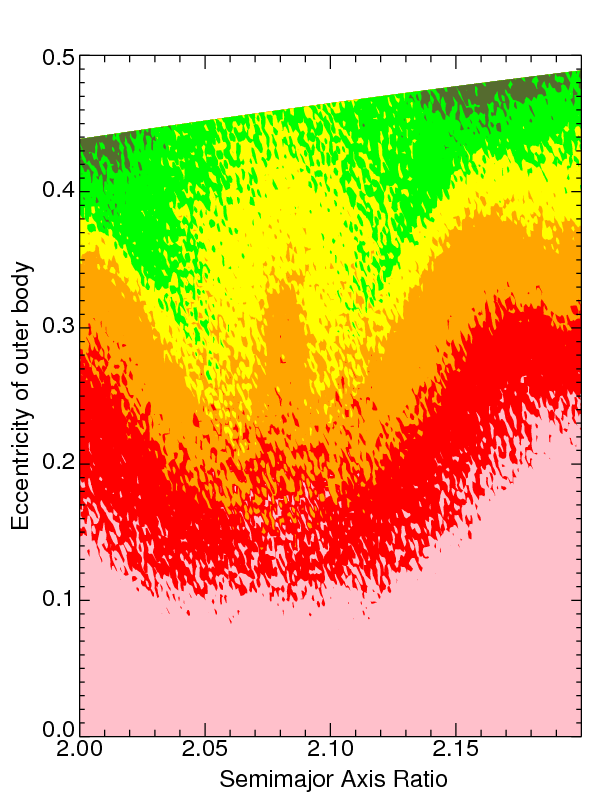}}&
    \multicolumn{2}{c}{\includegraphics[width=0.33\textwidth]{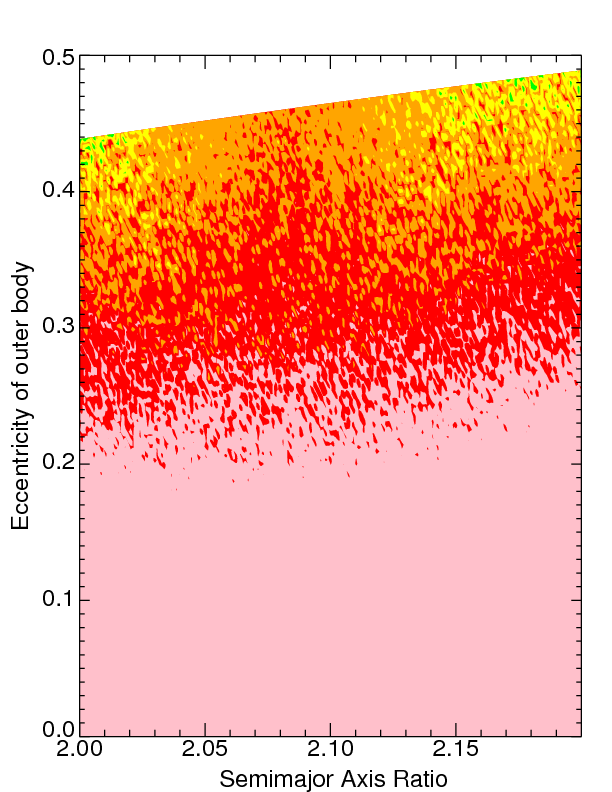}}
\\
  \end{tabular}
%
  \put(-400,30){\textbf{\large N=874} }
  \put(-235,30){\textbf{\large N=313} }
  \put(-70,30){\textbf{\large N=100} }
  \put(-400,-175){\textbf{\large N=50} }
  \put(-235,-175){\textbf{\large N=30} }
  \put(-70,-175){\textbf{\large N=10} }
%
%
%
  \caption{High-resolution (200 points in x and 120 points in y) RMS TTV signal amplitudes for the region around the $3$:$1$ PC for $N = 874$ (upper left), $N = 313$ (upper middle), $N = 100$ (upper right), $N = 50$ (lower left), $N = 30$ (lower middle), and $N = 10$ (lower right).  The contour levels are the same as in Figs. \ref{fiducial} and \ref{minmax}.  Note that after several years of observations, near-PC signal amplitudes are higher than in-PC amplitudes.}
  \label{N3to1}       
\end{figure*}

\begin{figure*}
  \centering
  \begin{tabular}{ccc}
    \multicolumn{3}{c}{\includegraphics[width=0.55\textwidth,height=0.30\textheight]{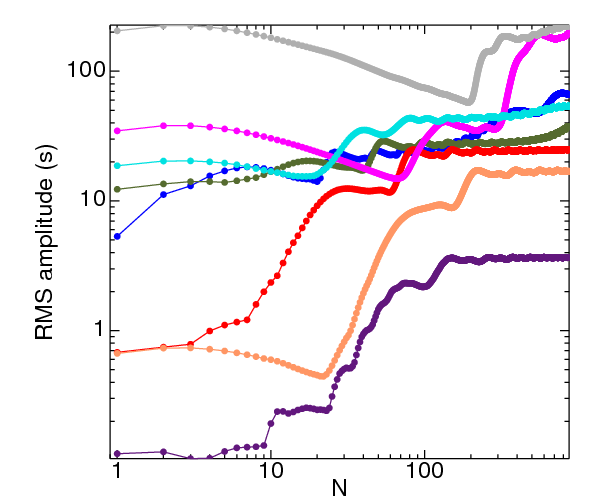}}\\
    \multicolumn{3}{c}{\includegraphics[width=0.55\textwidth,height=0.30\textheight]{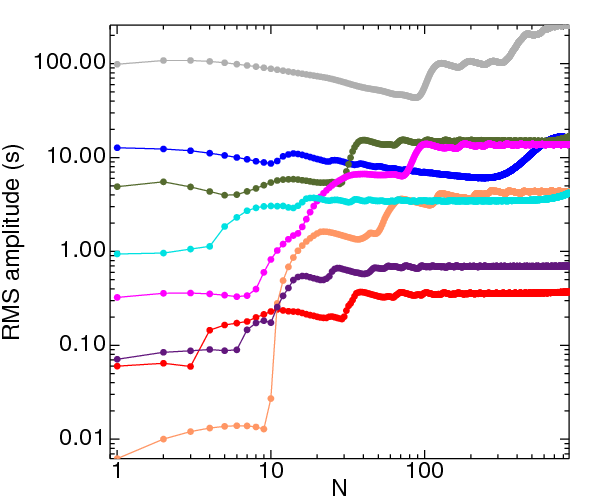}}\\
    \multicolumn{3}{c}{\includegraphics[width=0.55\textwidth,height=0.30\textheight]{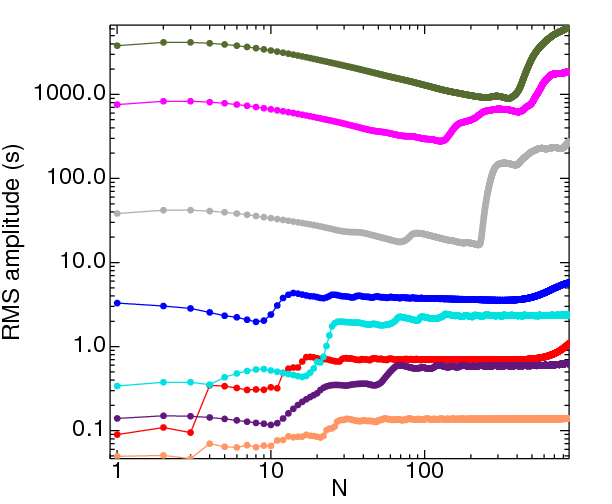}}\\
  \end{tabular}
%
  \put(-10,200){\textbf{\LARGE In-PC} }
  \put(-10,0){\textbf{\LARGE Near-PC} }
  \put(-10,-200){\textbf{\LARGE Secular} }
%
%
  \caption{The RMS TTV signal amplitude as a function of number of consecutive transit observations ($N$) for 8 systems in each of the following regimes: {\it upper plot}: in-PC (Period Commensurability), {\it middle plot}: near-PC, and {\it lower plot}: secular regimes. The initial orbital angles of these systems are fixed; their other initial conditions are provided in Table \ref{regimes}.  Note that generally TTV signal amplitude is a nonmonotonic function of $N$. }
  \label{Nline}
\end{figure*}

TTV signal amplitudes, $S({\bf Q})$, crucially depend on the number of observed transits, $N$ and the sampling rate.  Although one can define TTVs solely as the difference between consecutive transit times, \cite{agoetal2005} define TTVs as the deviation from an overall linear fit to the transit times of a system.  This definition does not rely on having successive transits.  Before missions such as CoRoT and Kepler, ground-based observations have struggled to yield $N \ge 100$.  These already-operating space-based missions, however, hope to observe {\it consecutive} transits for a half year (CoRoT) or multiple years (Kepler).  Therefore, we consider our fiducial system (Fig. \ref{fiducial}; we henceforth use the median value when averaging over multiple random angle initial configurations) for various durations corresponding to $N = 874, 313, 100, 50, 30$, and $10$ (Figs. \ref{Ncont}-\ref{Nline}) and by considering consecutive transits.  The first two values of $N$ listed correspond to observing campaigns of approximately 10 yrs and 3.5 yrs (roughly the maximum and nominal lifetime of Kepler). 

Figure \ref{Ncont} illustrates the TTV amplitude variation as a function of $N$.  For the $N = 10$ plot, almost no systems are detectable.  At $N = 30$, only a few systems featuring high eccentricity may be detectable.  For $N \ge 100$, systems close to a $q=1$ PC generate signals which change little with $N$ for $e_o < 0.75 e_H$.  However, $S({\bf Q})$ continues to increase for increasing $N$ for $e_o$ values closer to $e_H$.  


Because $S({\bf Q})$ is sensitive to the semimajor axis ratio on scales potentially smaller than $10^{-3}$ AU, higher resolution sampling of phase space will reveal additional features.  Similarly, as $N$ increases, larger sections of $a_o$-$e_o$ phase space will exhibit TTV curves with clearly discernible periods.  Hence, we have performed high resolution (200 points in x and 120 points in y) explorations of the phase space around the $3$:$1$ PC in Fig. \ref{N3to1}.  The figure demonstrates that as $N$ increases, the contour outlines become sharper and other weaker commensurabilities which neighbor the $3$:$1$ PC become more apparent.  Additionally, the double-lobed feature on the $3$:$1$ PC location at $e_o = 0.1$ disappears with increasing $N$. 

Importantly, the figure demonstrates that planets in a PC, and possibly in a MMR, do produce a distinct TTV signature, but not necessarily a {\it high amplitude} signature compared to its near-PC surroundings.  At $N = 30$, $S({\bf Q})$ at PC (yellow) is {\it lower} than the signal near PC (light green).  At $N=313$, $S({\bf Q})$ at PC (light green) is {\it lower} than the signal near PC (olive), and this result is {\it insensitive} to further increases of $N$.  The near-PC regime is sharply divided from the secular regime (note the thin light green contour between the regimes).  Additionally, after 10 yr, there is a sharp boundary (marked by sparse blue dots) between the near-PC regime and the high-eccentricity (at $\approx e_H$) secular regime.  Ten years might not be long enough for the some TTVs to produce periodic signals in the blue or purple areas.

The variation of TTV signal with $N$ is not necessarily monotonic, and is a function of the masses and orbital parameters of the planets.  We can look more closely at this $S({\bf Q})$ vs. $N$ dependence by fixing all other orbital parameters in specific cases.  Although identifying ``representative" systems is difficult because of the large size of the phase space, we sample 8 different particular systems in each of three regimes: ``in-PC'', ``near-PC'' and ``secular".  Table \ref{regimes} contains the $a_o/a_i$ and $e_o$ values for all 24 systems, and Fig. \ref{Nline} plots $S({\bf Q})$ vs $N$ for these systems.

\begin{deluxetable}{ c  c  c  c  c  c  c  c  c  c}
\label{TabLib}
\tabletypesize{\scriptsize}
\tablecaption{Sample Systems in 3 Regimes}
\tablewidth{0pt}
\tablehead{ 
   \colhead{designation} &
   \colhead{regime} &
   \colhead{close PC} &
   \colhead{$a_o/a_i$} &
   \colhead{$e_o$} &
   \colhead{$e_o/e_H$} &
   \colhead{Fig. \ref{Nline} color}
}
\renewcommand{\arraystretch}{1.5}
\startdata  
I & in-PC & $2$:$1$ & $(2/1)^{2/3}$ & $0.20$ & $0.69$ & blue 
\\
II & in-PC & $3$:$1$ & $(3/1)^{2/3}$ & $0.30$ & $0.66$ & olive
\\
III & in-PC & $4$:$1$ & $(4/1)^{2/3}$ & $0.365$ & $0.66$ & red
\\
IV & in-PC & $8$:$1$ & $(8/1)^{2/3}$ & $0.57$ & $0.80$ & magenta
\\
V & in-PC & $11$:$1$ & $(11/1)^{2/3}$ & $0.60$ & $0.78$ & salmon
\\
VI & in-PC & $5$:$2$ & $(5/2)^{2/3}$ & $0.20$ & $0.51$ & aqua
\\
VII & in-PC & $7$:$2$ & $(7/2)^{2/3}$ & $0.36$ & $0.71$ & gray
\\
VIII & in-PC & $11$:$3$ & $(11/3)^{2/3}$ & $0.243$ & $0.46$ & purple
\\
IX & near-PC & $2$:$1$ & $1.58$ & $0.11$ & $0.44$ & blue 
\\
X & near-PC & $3$:$1$ & $2.04$ & $0.30$ &  $0.66$ & olive
\\
XI & near-PC & $4$:$1$ & $2.57$ & $0.25$ & $0.44$ & red
\\
XII & near-PC & $8$:$1$ & $3.97$ & $0.57$ & $0.80$ &  magenta
\\
XIII & near-PC & $11$:$1$ & $4.98$ & $0.65$ & $0.85$ & salmon
\\
XIV & near-PC & $5$:$2$ & $1.87$ & $0.20$ & $0.49$ & aqua
\\
XV & near-PC & $7$:$2$ & $2.285$ & $0.36$ & $0.71$ &  gray
\\
XVI & near-PC & $11$:$3$ & $2.385$ & $0.243$ & $0.46$ &  purple
\\
XVII & secular & --- & $1.5$ & $0.035$ & $0.14$ & blue
\\
XVIII & secular & --- & $1.7$ & $0.28$ & $0.83$ &  olive
\\
XIX & secular & --- & $1.8$ & $0.10$ & $0.25$ &  red
\\
XX & secular & --- & $2.2$ & $0.45$ & $0.92$ &  magenta
\\
XXI & secular & --- & $2.45$ & $0.10$ & $0.19$ &  salmon
\\
XXII & secular & --- & $3.0$ & $0.455$ & $0.73$ &  aqua
\\
XXIII & secular & --- & $4.405$ & $0.665$ & $0.90$ &  gray
\\
XXIV & secular & --- & $4.88$ & $0.55$ & $0.71$ & purple
\\ \hline
\enddata
\tablecomments{\label{regimes} \small{
24 systems selected for individual analysis throughout the paper, with designations in Column 1.  Column 2 lists the regime the system resides in (where PC is Period Commensurability), and Column 3 lists the corresponding PC, if applicable.  The other columns list the initial semimajor axis ratio of the planets, the outer planet's eccentricity, and the color of the corresponding curve in Fig. \ref{Nline}.}}
\end{deluxetable}

We first consider the ``in-PC'' systems plotted in the upper plot of Fig. \ref{Nline}. 
In order to model the signal amplitude of any in-PC system with continuous observational coverage, one must observe at least 50 transits.  Only for $N > 50$ do any of the curves level off.  Generally, the higher the amplitude of $S({\bf Q})$, the greater the number of observations needed to level the curves (for a given perturber mass).  Only the red ($4$:$1$), salmon ($11$:$1$) and purple ($11$:$3$) curves are level at $N=874$, and only for the blue ($2$:$1$) curve is $S(\bf{Q})$ decreasing.  The signal curves are generally non-monotonic with $N$; additional observations may cause the signal amplitude to increase or decrease.  Note additionally that although eccentricities are high ($0.47 e_H \le e_o \le 0.80 e_H$), the signal amplitudes are relatively low ($S({\bf Q}) < 200$s) for all resonant curves for all $t \le 10$ yr.  Systems in a PC do not necessarily achieve signal amplitudes as high as those in other regimes, and preferentially have lower amplitudes, even after 10 yrs of observational sampling.

The prospects for detecting systems near but not on PC is more promising.  The middle panel of Fig. \ref{Nline} indicates that most of the curves level off after $N \approx 300$ (a little over 3 yrs).  However, at $N=800$, one can see the aqua ($5$:$2$) and olive ($3$:$1$) curves beginning to trend upwards.  The system near the $2$:$1$ PC (blue) exhibits similar oscillatory behavior from the $2$:$1$ in-PC system in the upper panel.  The two curves corresponding to the highest $e_o$ values (salmon with $e_o = 0.85 e_H$ and magenta with $e_o = 0.80 e_H$) maintain $S({\bf Q}) < 20$s for all $N$.  The salmon curves in the upper two panels of Fig. \ref{Nline} contain the same value of $e_o$ and both level at $20$s.  However, the in-PC system levels off $N \approx 230$ whereas the near-PC system levels off at $N \approx 110$, reflecting the general trend of near-PC systems having robust values of $S({\bf Q})$ with fewer observations.

The two secular systems with the greatest signals in the lower panel of Fig. \ref{Nline} (olive and magenta) undergo variations on the order of their initial values for at least 5 years.  Both of these systems feature $e_o > 0.80 e_H$ and $a_o/a_i \lesssim 2.2$. The strongly hierarchical ($a_o/a_i = 4.405$) system indicated by the gray curve raises its TTV signal by nearly one order of magnitude between $N=250-300$, and then doubles this value over the next 6 years.  The variation with $N$ of all three of these systems dwarfs the variation by the other 5 systems on the plot, including 2 with high eccentricities (aqua and purple, with $e_o = 0.73 e_H$ and $0.71 e_H$ respectively).  These two high eccentricity systems show little (a few percent) variation for $N > 200$.  Additionally, the signal for two of the lower eccentricity systems, indicated by the blue and red curves ($e_o = 0.14 e_H$ and $0.25 e_H$) appear to show negligible variation with $N$ for $50 \lesssim N \lesssim 400$.  However, both signals begin to increase for $N \gtrsim 400$ such that $S(N=874)/S(N=400) > 1.5$.  The qualitative difference in all these secular curves help illustrate that $S({\bf Q})$ is non-trivially dependent on $N$, even for secular systems far from a PC.

\section{Correlations with Orbital Angles}

Although we have already shown in Fig. \ref{fiducial} and Section 2 how randomly chosen sets of initial orbital angles (mean anomalies and longitudes of pericenter) for a given $a_o$ and $e_o$ can produce order of magnitude variations in TTV signals, here we study this dependency systematically.  In the following analyses, we keep all orbital parameters but one fixed in order to gain insight into the physical origin of the TTV signals. 

\begin{figure*}
  \centering
\resizebox{6in}{!} {
  \begin{tabular}{ccc ccc ccc}
    \multicolumn{3}{c}{\includegraphics[width=0.33\textwidth]{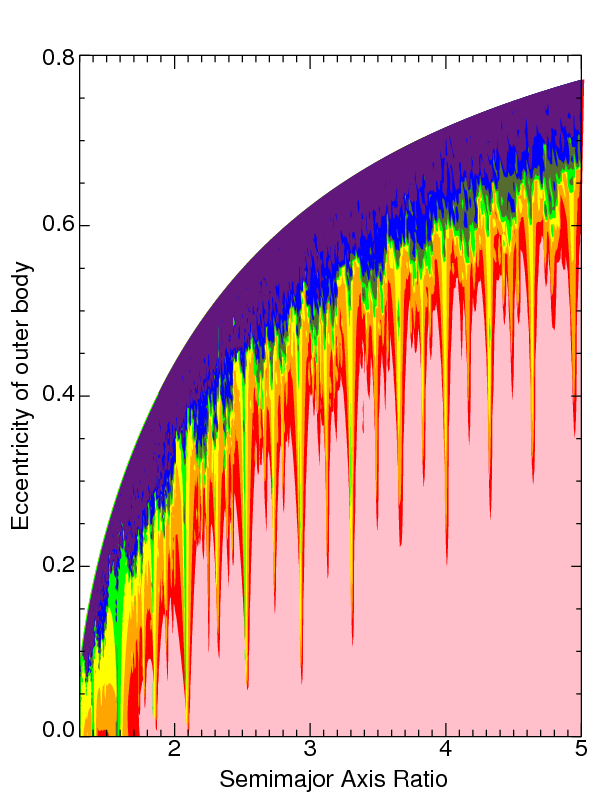}}&
    \multicolumn{3}{c}{\includegraphics[width=0.33\textwidth]{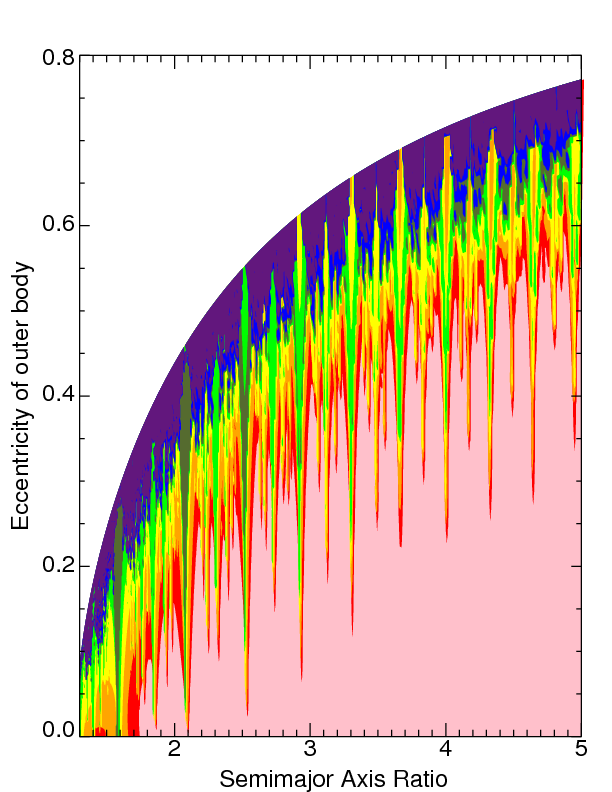}}&
    \multicolumn{3}{c}{\includegraphics[width=0.33\textwidth]{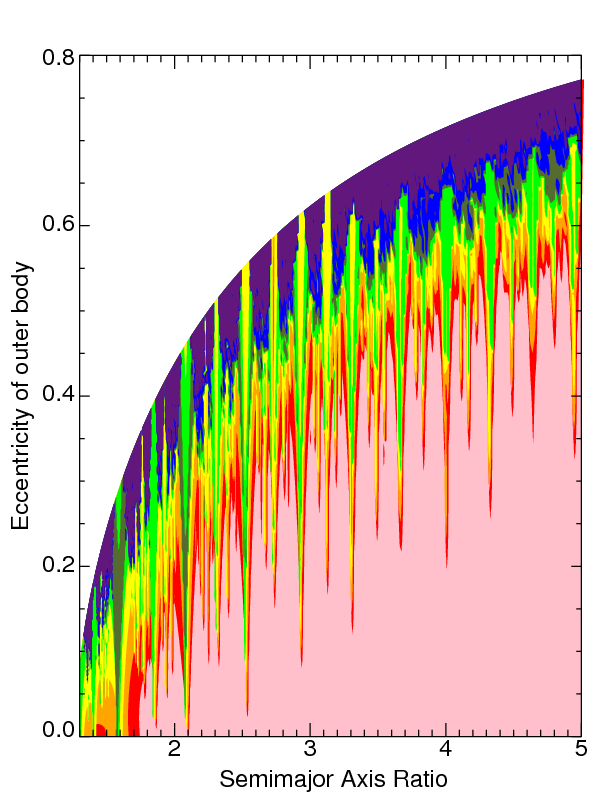}}\\[-10pt]
    \multicolumn{3}{c}{\includegraphics[width=0.33\textwidth]{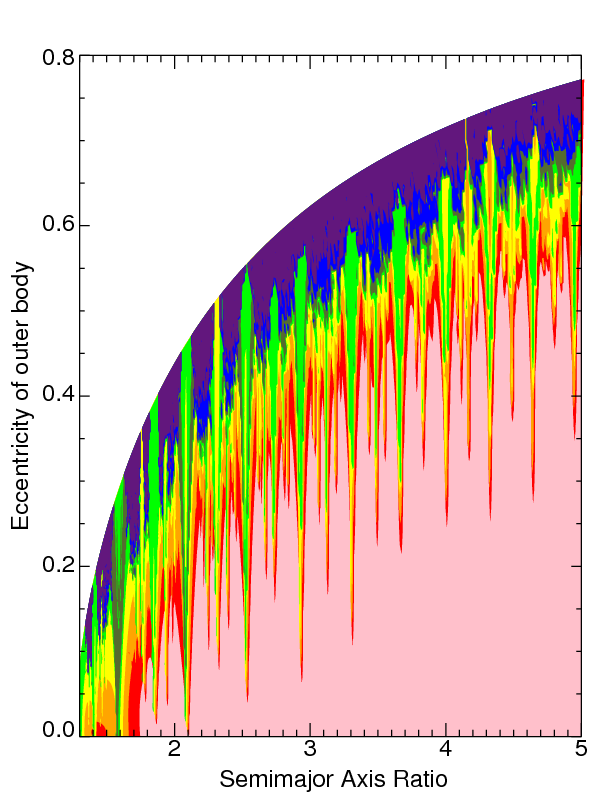}}&
    \multicolumn{3}{c}{\includegraphics[width=0.33\textwidth]{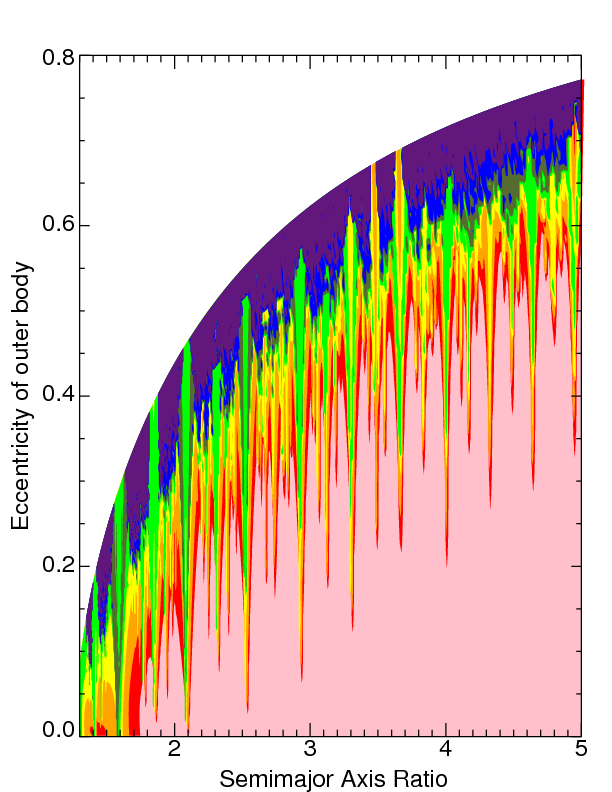}}&
    \multicolumn{3}{c}{\includegraphics[width=0.33\textwidth]{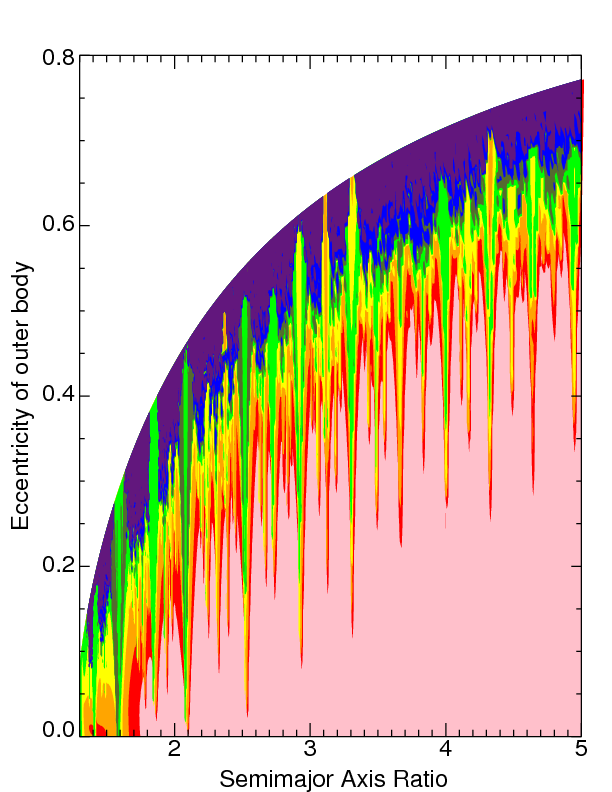}}\\[-10pt]
    \multicolumn{3}{c}{\includegraphics[width=0.33\textwidth]{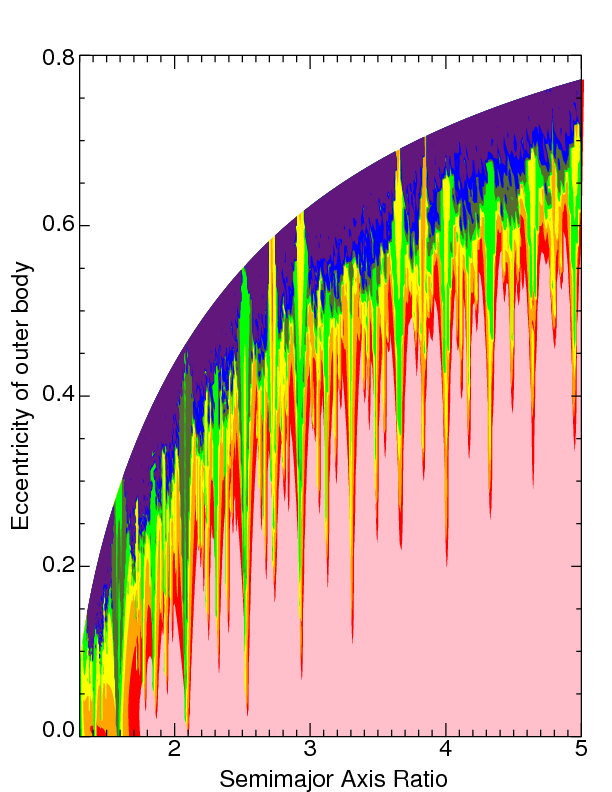}}&
    \multicolumn{3}{c}{\includegraphics[width=0.33\textwidth]{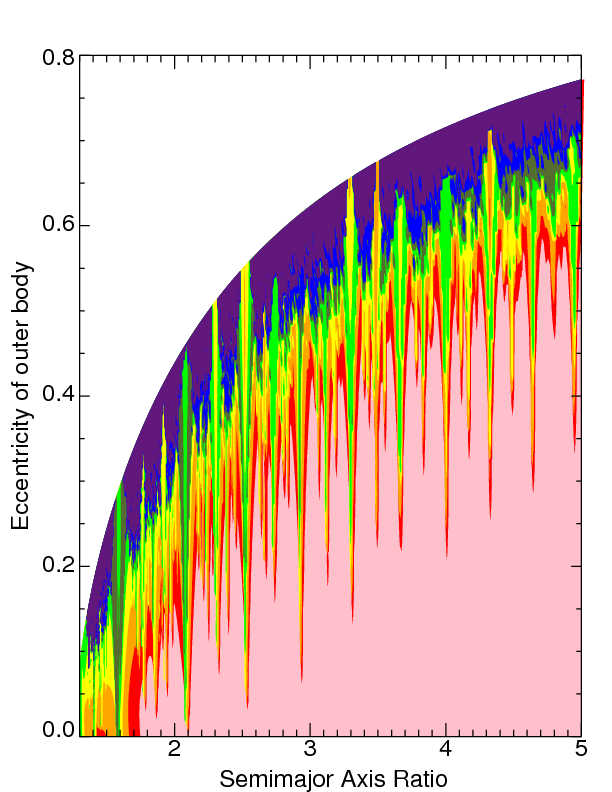}}&
    \multicolumn{3}{c}{\includegraphics[width=0.33\textwidth]{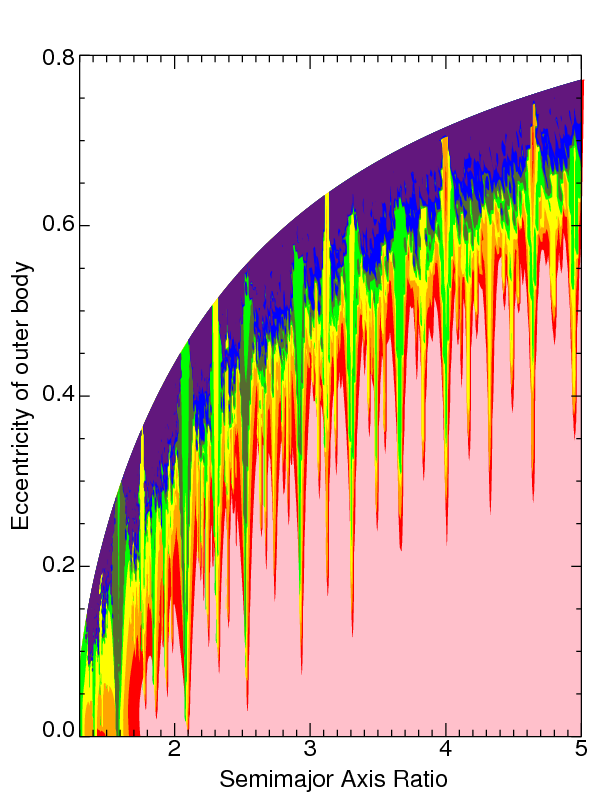}}\\[-10pt]
  \end{tabular}
}
%
     \put(-405,220) {\textbf{$\Pi_o =  0^{\circ}$}}
     \put(-261,220) {\textbf{$\Pi_o =  10^{\circ}$}}
     \put(-118, 220) {\textbf{$\Pi_o =  20^{\circ}$}}
     \put(-405,55) {\textbf{$\Pi_o =  30^{\circ}$}}
     \put(-261,55) {\textbf{$\Pi_o =  40^{\circ}$}}
     \put(-118, 55) {\textbf{$\Pi_o =  50^{\circ}$}}
     \put(-405,-110) {\textbf{$\Pi_o =  60^{\circ}$}}
     \put(-261,-110) {\textbf{$\Pi_o =  70^{\circ}$}}
     \put(-118,-110) {\textbf{$\Pi_o =  80^{\circ}$}}
%
%
  \caption{RMS TTV amplitude for $\Pi_i = \varpi_o = 0^{\circ}$ as $\Pi_o$ is increased in $10^{\circ}$ increments from $0^{\circ}$ to $80^{\circ}$ in the upper left, upper middle, upper right, middle left, center, middle right, lower left, lower middle and lower right panels, respectively.   
The contour levels are the same as in Figs. \ref{fiducial} and \ref{minmax}.  All systems are sampled for $N = 874$.  Note the different locations in which ``flames'' arise in the blue and purple regions in the plots.}
  \label{9angles}
\end{figure*}

Although such insight is difficult to discern when considering how the entire $(a_o, e_o)$ phase space varies with individual orbital angles, line plots for fixed $(a_o, e_o)$ values show more structure.  First, however, we consider ``flames'' plots as a function of $\Pi_o$.  In each plot of Fig. \ref{9angles}, the initial orbital angles are $\Pi_i = \varpi_o = 0^{\circ}$, with $\Pi_o = 0^{\circ}, 10^{\circ}, ... 80^{\circ}$.  Systems with planets initially at conjunction have high signal amplitude at high (near $e_H$) eccentricity, and, unlike in all other plots, saturate PCs at these eccentricities with $S({\bf Q}) > 10^3 $ s at $N = 874$.  For $\Pi_o = 10^{\circ} - 80^{\circ}$, in each plot, several PCs produce signals which are over two orders of magnitude lower than $10^3$ s at high $e_o$.  The ``flames" which produce these effects appear at different sets of PCs in each plot, and have no immediately recognizable pattern.  The lack of an apparent pattern in these plots, despite fixing all but one parameter and despite the high value of $N = 874$, expresses well the difficulty in trying to use transit timing variations to deduce the properties of unseen planets.

\begin{figure*}
  \centering
\resizebox{6in}{!} {
  \begin{tabular}{ccc ccc ccc}
    \multicolumn{3}{c}{\includegraphics[width=0.33\textwidth,height=0.37\textheight]{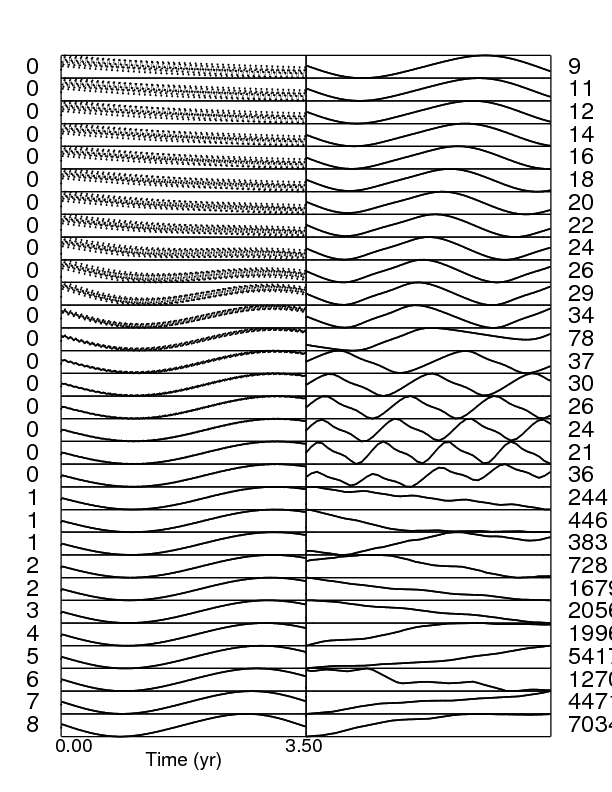}}&
    \multicolumn{3}{c}{\includegraphics[width=0.33\textwidth,height=0.37\textheight]{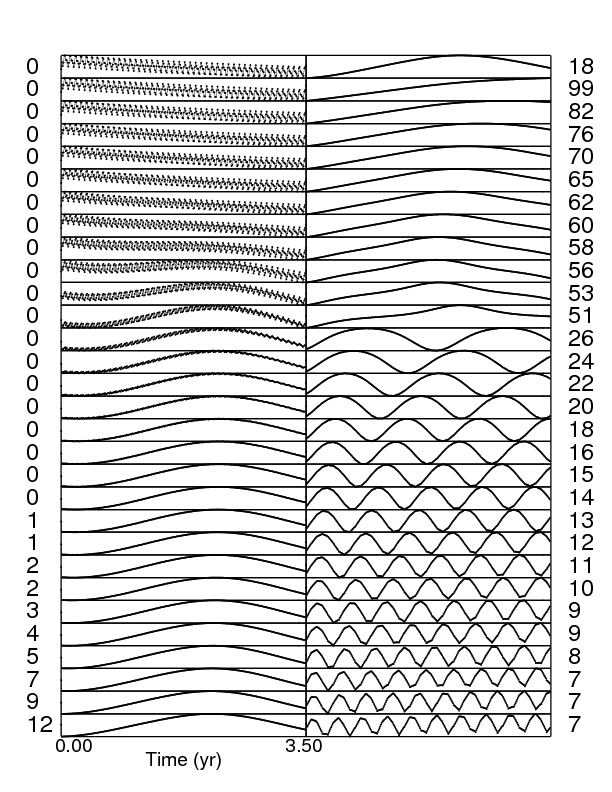}}&
    \multicolumn{3}{c}{\includegraphics[width=0.33\textwidth,height=0.37\textheight]{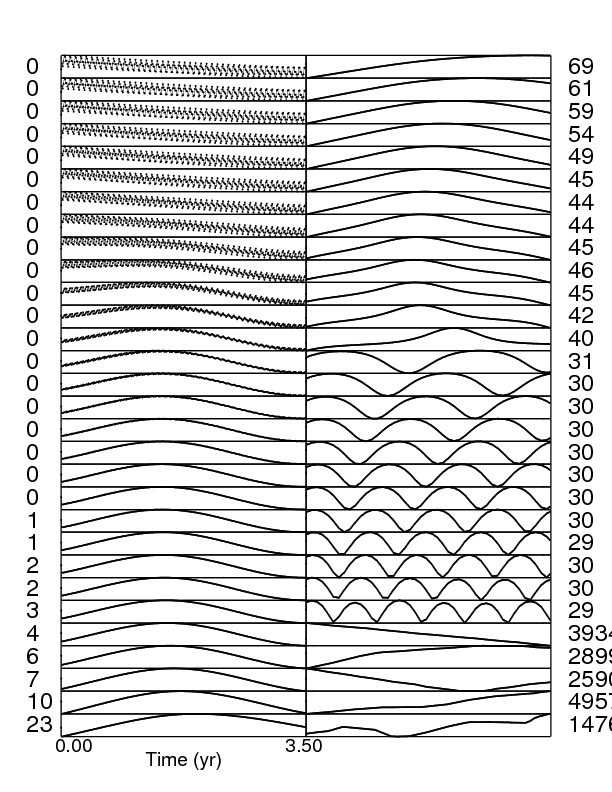}}\\[-10pt]
    \multicolumn{3}{c}{\includegraphics[width=0.33\textwidth,height=0.37\textheight]{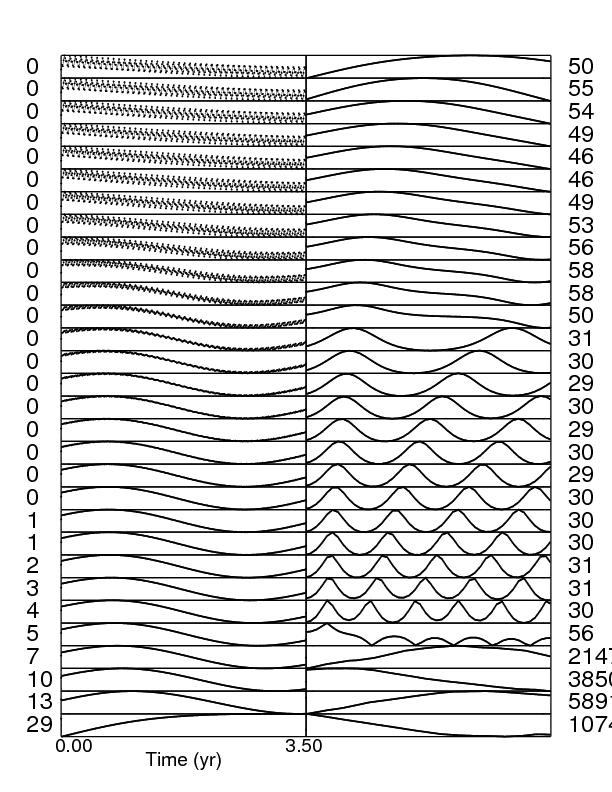}}&
    \multicolumn{3}{c}{\includegraphics[width=0.33\textwidth,height=0.37\textheight]{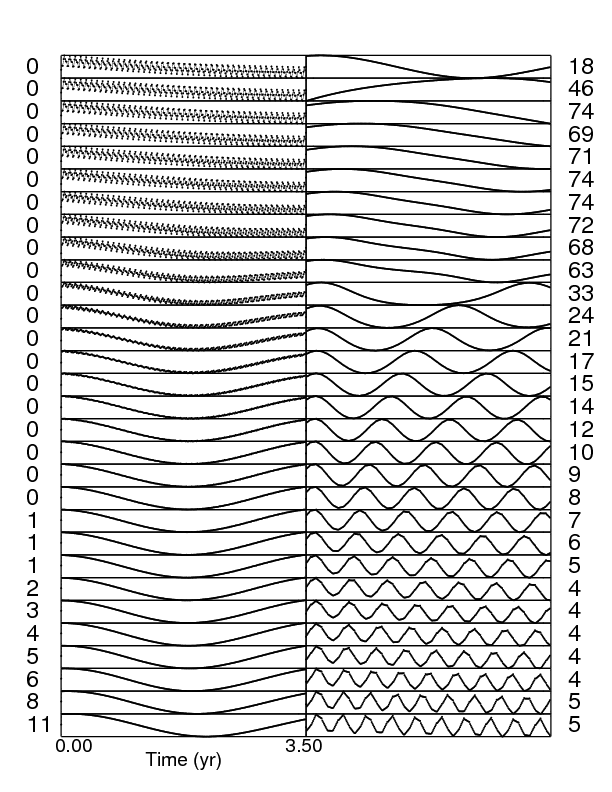}}&
    \multicolumn{3}{c}{\includegraphics[width=0.33\textwidth,height=0.37\textheight]{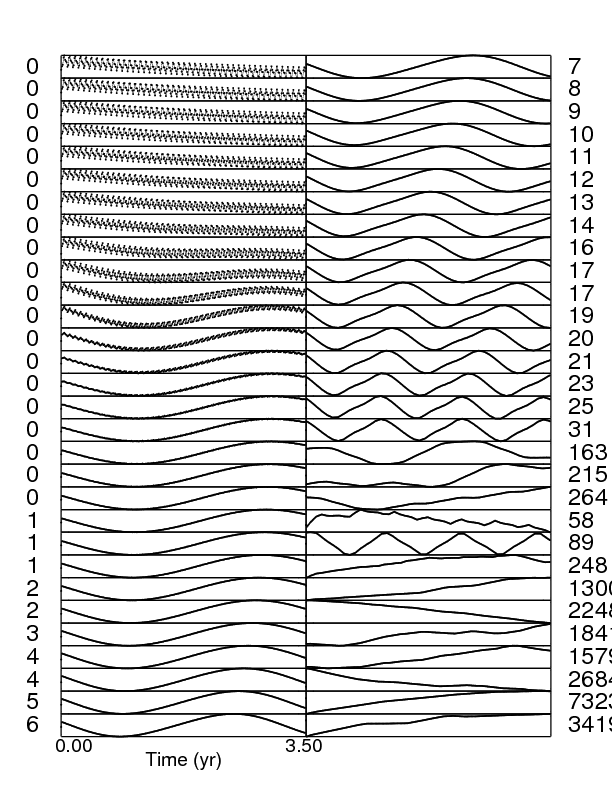}}\\[-10pt]
    \multicolumn{3}{c}{\includegraphics[width=0.33\textwidth,height=0.37\textheight]{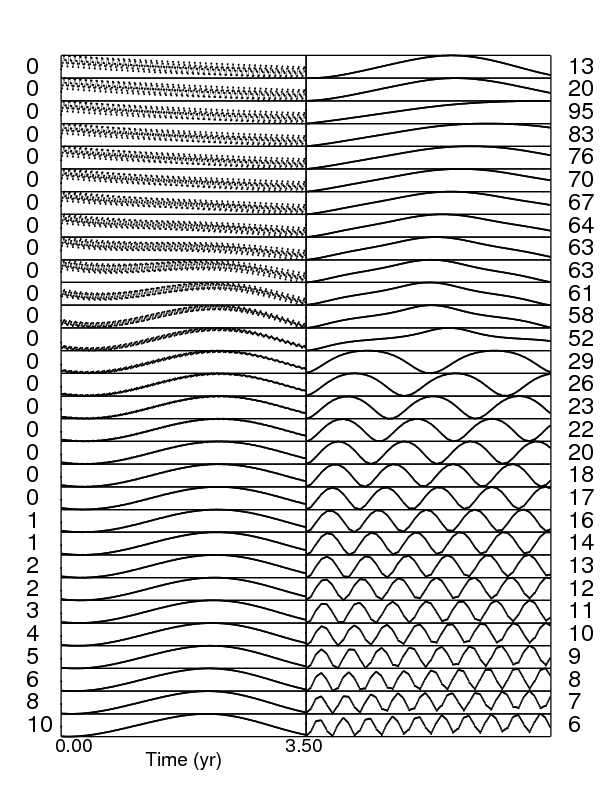}}&
    \multicolumn{3}{c}{\includegraphics[width=0.33\textwidth,height=0.37\textheight]{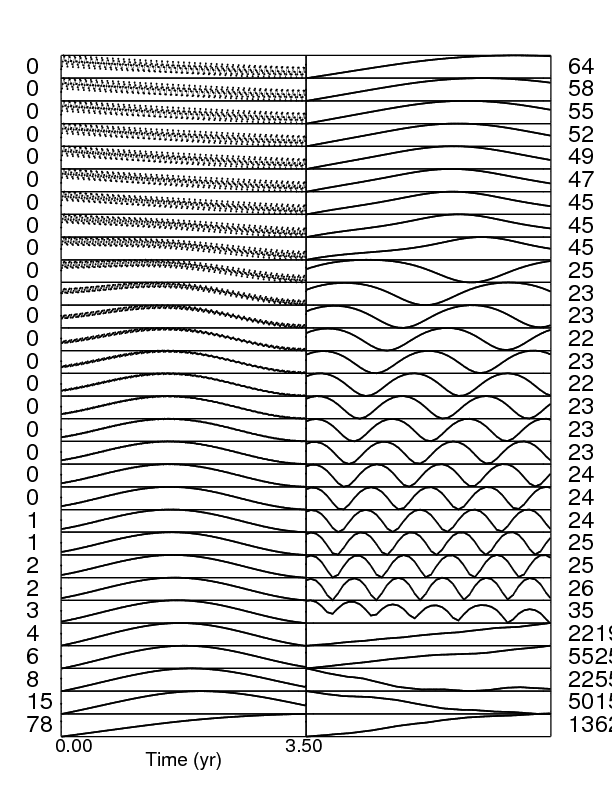}}&
    \multicolumn{3}{c}{\includegraphics[width=0.33\textwidth,height=0.37\textheight]{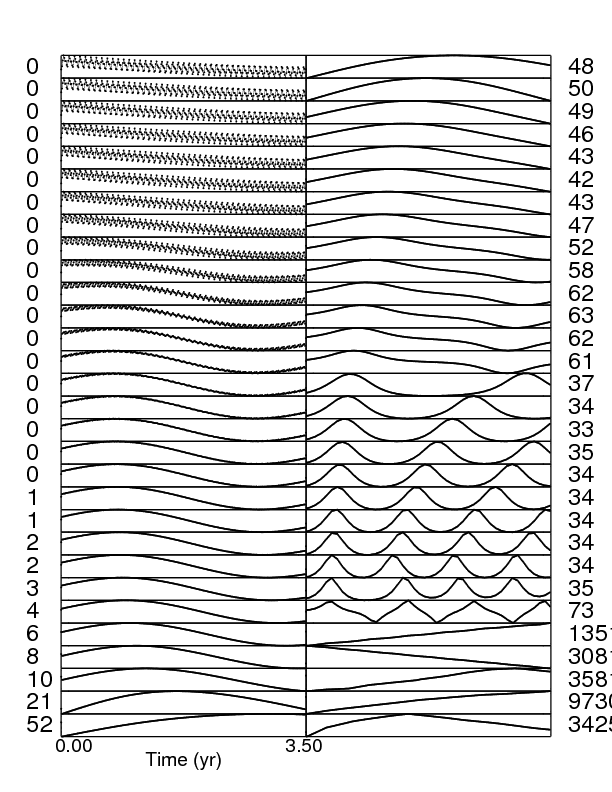}}\\[-10pt]
  \end{tabular}
}
%
     \put(-384,279){\large $\Pi_o =  0^{\circ}$}
     \put(-240,279){\large $\Pi_o =  10^{\circ}$}
     \put(-97, 279){\large $\Pi_o =  20^{\circ}$}
     \put(-384,85){\large $\Pi_o =  30^{\circ}$}
     \put(-240,85){\large $\Pi_o =  40^{\circ}$}
     \put(-97, 85){\large $\Pi_o =  50^{\circ}$}
     \put(-384,-102){\large $\Pi_o =  60^{\circ}$}
     \put(-240,-102){\large $\Pi_o =  70^{\circ}$}
     \put(-97,-102){\large $\Pi_o =  80^{\circ}$}
%
%
%
  \caption{Transit curves over $3.5$ yr close to the $7$:$1$ PC for $\Pi_i = \varpi_o = 0^{\circ}$ as $\Pi_{o}$ is increased in $10^{\circ}$ increments from $0^{\circ}$ to $80^{\circ}$ as indicated on top of the 9 panels. Within each panel, there are 60 systems sampled; $e_o$ increases from $0$ in the top left of each panel downward, first in the left column and then the right column, until $e_o = e_H$ in the bottom right of each panel. The vertical range of each plot is scaled, and we provide the value of $S({\bf Q})$ beside each curve.  Note the qualitatively different locations in eccentricity space of TTV signal forms in the right column of each panel.}
  \label{strips71}
\end{figure*}

An alternative and perhaps more demonstrative method of exhibiting this signal decrease at high eccentricities is through sequences of individual transit curves.  Fig. \ref{strips71}  displays 9 sequences (panels) of scaled curves, where each curve traces 3.5 yr, as $e_o$ is increased from $0$ to $e_H$ at the $7$:$1$ PC ($a_o/a_i = 3.66$).  In each panel, $e_o$ is increased from the top left curve to the bottom left, to the upper right to the bottom right.  The numbers accompanying each curve represents the RMS TTV amplitude of that curve in seconds.  The curves are scaled such that their extrema are the plot boundaries, so that one could better see detail and modulation.

The figure demonstrates that at this PC: 1) although low eccentricity ($e_o \sim 0$) curves are highly structured and modulated, their signal amplitude is negligible, 2) at mid-to-high ($e_o \sim 0.7 e_H$) eccentricity, the frequency of the curves suddenly increases, 3) in only some cases, eccentricities closest to $e_H$ will exhibit long-period ($> 3.5$) yr signals of $\ge 10^3$ s, 4) the transitions to these high signals is often sudden (caused by a difference of $e \lesssim 0.018$), and 5) in no case does the signal monotonically increase with increasing $e_o$.   In particular, for $\Pi_o = 10^{\circ}, 40^{\circ},$ and $60^{\circ}$, TTV curves exhibit low ($< 10$s) amplitudes at $e_H$.  For the other values of $\Pi_o$, TTV signals at these high eccentricities exhibit long ($> 3.5$ yr) period  variations and very high ($> 10^3$s) amplitudes.  This qualitative difference might vanish at high enough eccentricities, beyond the Hill Stability Limit.  Note additionally that the eccentricity range which allows for high-frequency (featuring several crests over 3.5 yr) TTV signals varies significantly (over a factor of 2) for different $\Pi_o$ values.  For $\Pi_o = 50^{\circ}$, the long-trend signals near $e_H$ are punctured by an orderly, periodic $89$ s signal amplitude, showcasing the unpredictability of TTV signals near MMR.

The above analysis keeps $\varpi_o$ fixed.  Although this angle refers to the outer planet's orbit and not the planet's location on that orbit, the value of $\varpi_o$ may play a crucial role in the dynamics.  In fact, systems with $\varpi_o = 90^{\circ}$ and $\varpi_o = 180^{\circ}$ (not plotted) do show completely different patterns of low amplitude ($< 100s$) ``flames" near the Hill Stability Limit from those shown in Fig. \ref{9angles}.  The starkest difference appears in the $\Pi_o = 0^{\circ}$ case, where these flames are absent for $\varpi_o = 0^{\circ}$, fully saturate the Hill Stability region for $\varpi_o = 90^{\circ}$, and appear only for $a_o/a_i \le 2.5$ when $\varpi_o = 180^{\circ}$.

The above analysis does not necessarily extend to other PCs.  As the coefficients of the terms in disturbing function, or the different shapes of the libration widths in Fig. \ref{fiducial} would indicate, the resonant structure of each PC is different.  Analysis of planets thought to be near or in a particular MMR hence would benefit from a high-resolution and complete exploration of the phase space at that location. 

\begin{figure*}
  \centering
  \begin{tabular}{cccccc}
    \multicolumn{2}{c}{\includegraphics[width=0.30\textwidth,height=0.41\textheight]{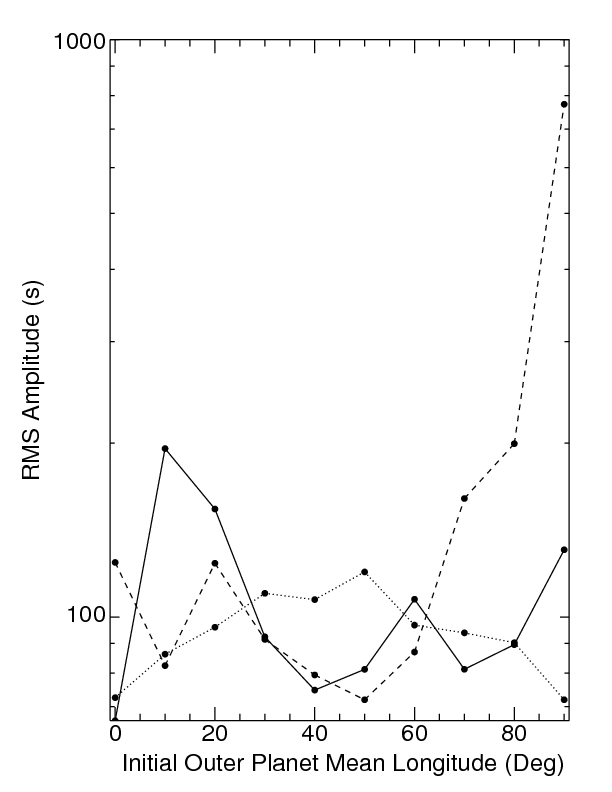}}&
    \multicolumn{2}{c}{\includegraphics[width=0.30\textwidth,height=0.41\textheight]{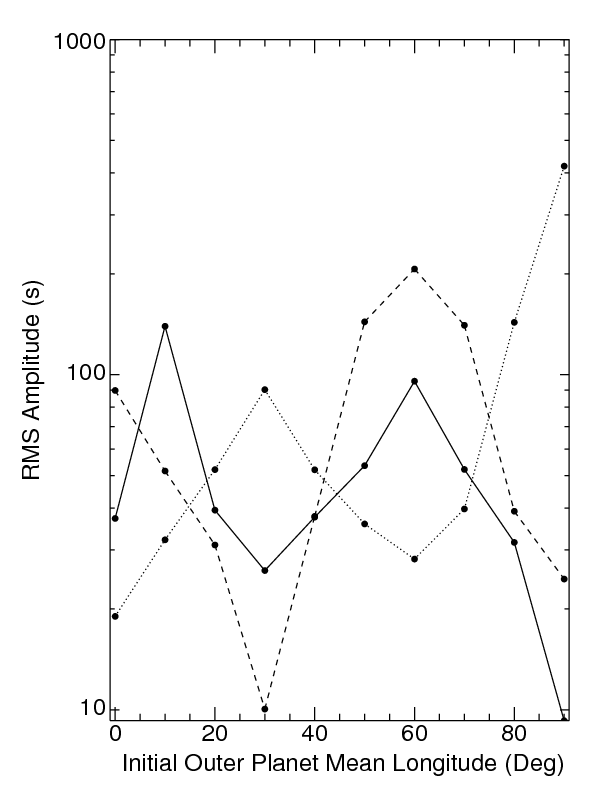}}&
    \multicolumn{2}{c}{\includegraphics[width=0.30\textwidth,height=0.41\textheight]{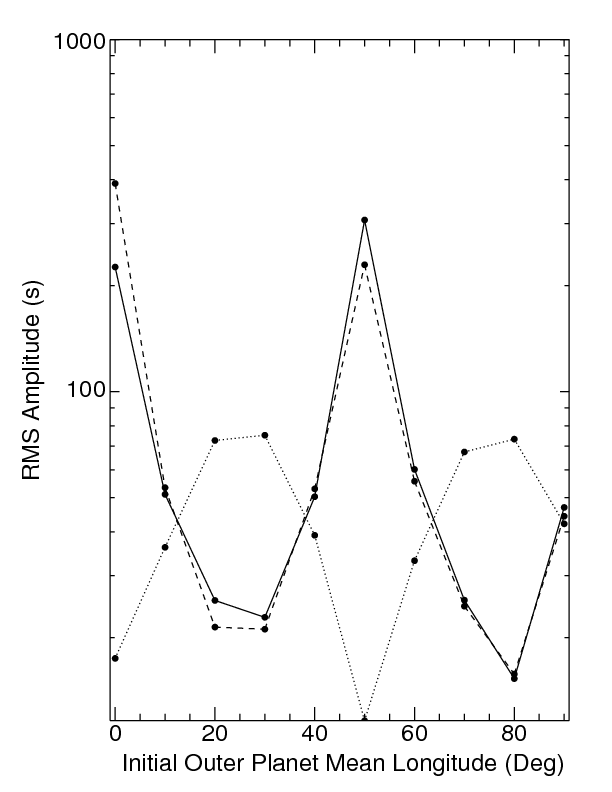}}\\
    \multicolumn{2}{c}{\includegraphics[width=0.30\textwidth,height=0.41\textheight]{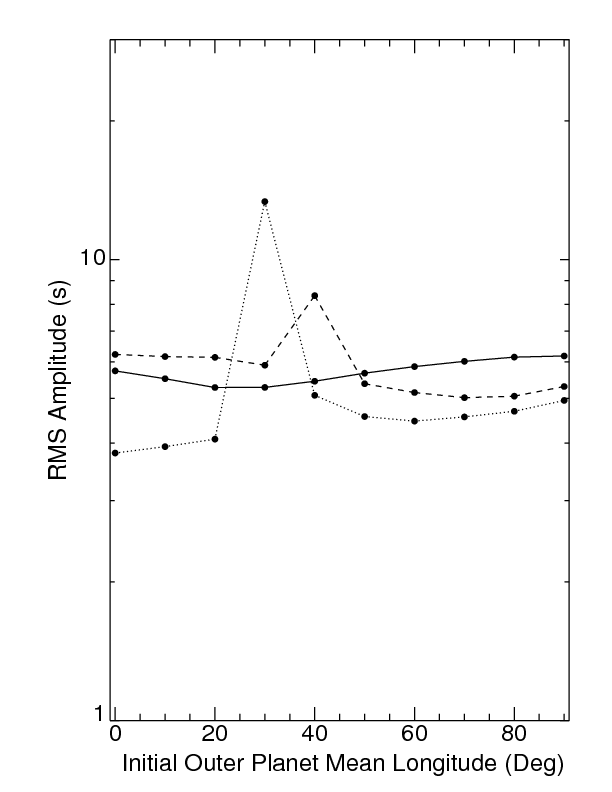}}&
    \multicolumn{2}{c}{\includegraphics[width=0.30\textwidth,height=0.41\textheight]{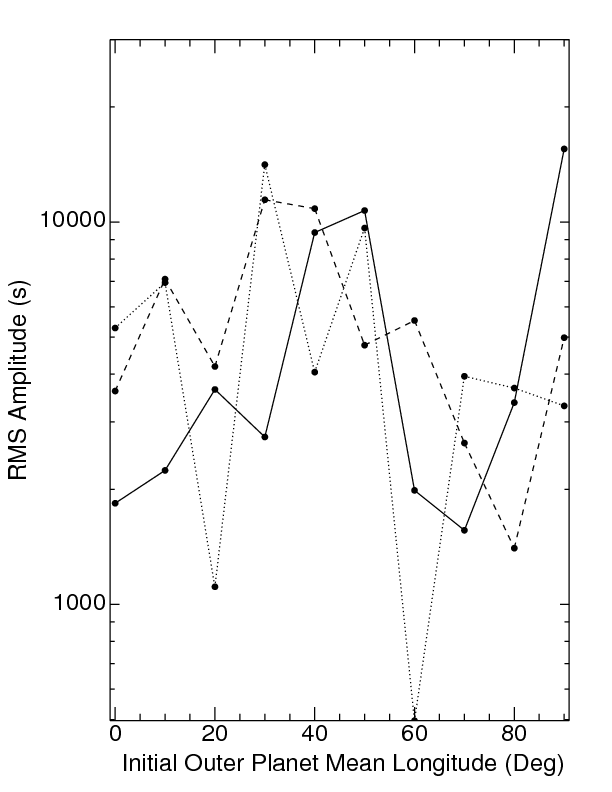}}&
    \multicolumn{2}{c}{\includegraphics[width=0.30\textwidth,height=0.41\textheight]{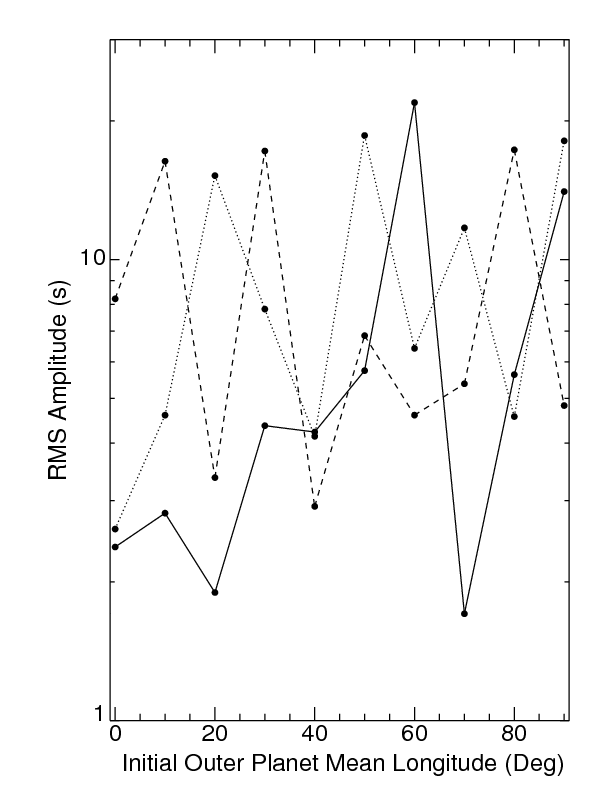}}
\\
  \end{tabular}
%
  \put(-410,212){$2$:$1$ PC}
  \put(-410,200){$e_o=0.20$}
  \put(-260,212){$3$:$1$ PC}
  \put(-260,200){$e_o=0.30$}
  \put(-110,212){$7$:$2$ PC}
  \put(-110,200){$e_o=0.36$}
  \put(-410,-20){$a_o/a_i=1.5$}
  \put(-410,-32){$e_o=0.035$}
  \put(-260,-20){$a_o/a_i=2.2$}
  \put(-260,-32){$e_o=0.45$}
  \put(-110,-20){$a_o/a_i=3.0$}
  \put(-110,-32){$e_o=0.455$}
%
%
  \caption{TTV RMS signal amplitude vs. $\Pi_o$ for systems at the $2$:$1$, $3$:$1$ and $7$:$2$ PC (upper panels, left to right) and for secular systems XVII, XX and XXII from Table \ref{regimes} (lower panels, left to right).  In each plot, the solid, dotted and dashed curves correspond to $\varpi_o = 0^{\circ}$, $\varpi_o = 90^{\circ}$ and $\varpi_o = 180^{\circ}$.  All signal amplitudes were sampled after 10 yr of continuous observations ($N=874$). Note the lack of a discernible correlation of $\Pi_o$ with RMS amplitude.}
  \label{lamlines}       
\end{figure*}
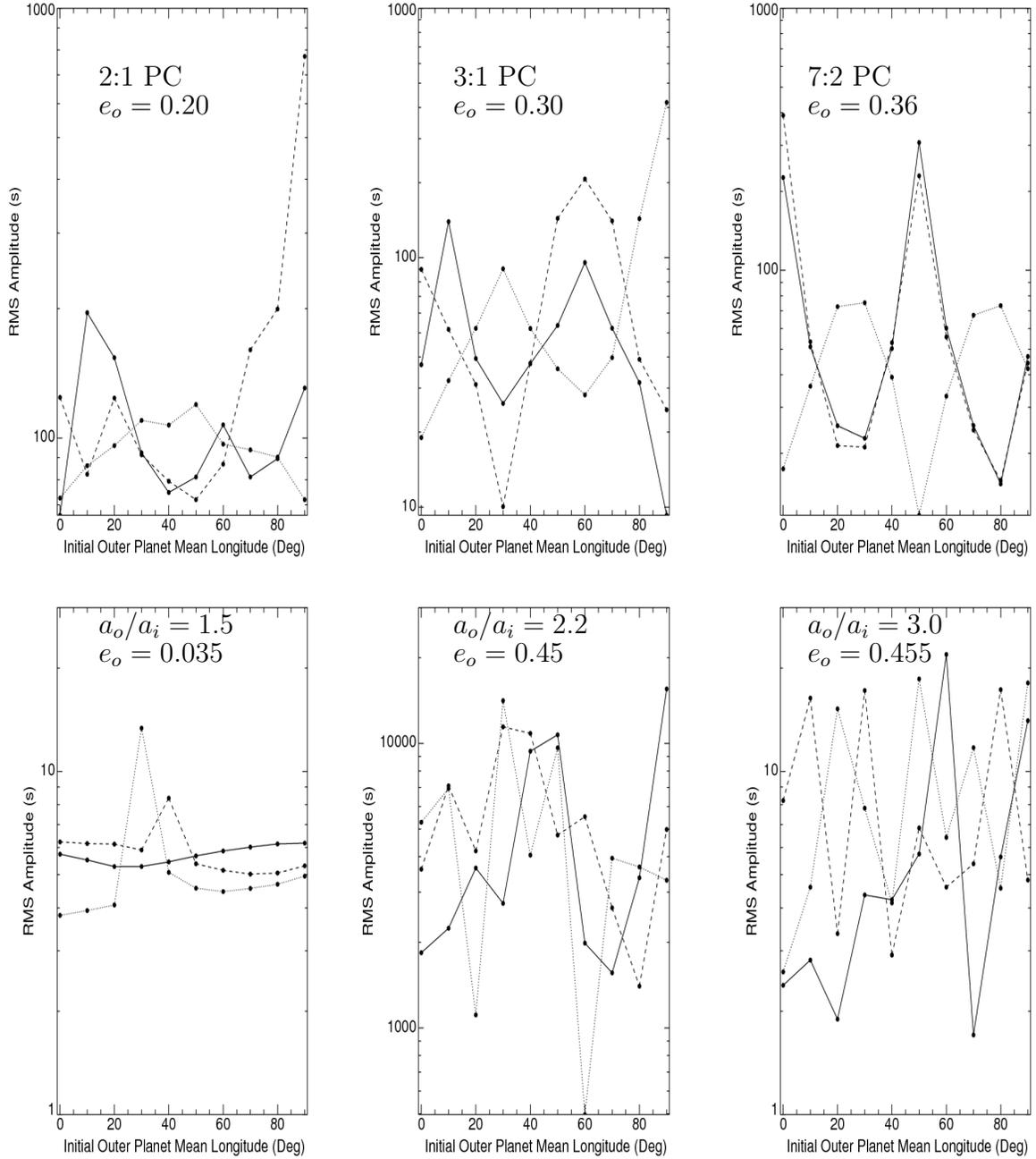

In order to better determine if the variations with $\Pi_o$ in Fig. \ref{9angles} have structure, we now consider the signal variations while fixing $(a_o, e_o)$.  We plot this variation for selected systems (I, II, VII, XVII, XX, XXII from Table \ref{regimes}) in Fig. \ref{lamlines}, where the solid, dotted and dashed curves in each plot correspond to $\varpi_o = 0^{\circ}, 90^{\circ},$ and $180^{\circ}$.  The y-axis of all plots are logarithmic.  At the coarse resolution of $10^{\circ}$ per data point, the curves in the upper panels are vaguely oscillatory.  For both the $2$:$1$ (I) and $3$:$1$ (II) systems, the $S({\bf Q})$ sharply (by several hundred seconds over $10^{\circ}$) peaks at $\Pi_o = 90^{\circ}$, but for a different value of $\varpi_o$ at each PC.  At the $7$:$2$ PC, this peak does not occur; curves for the three values of $\varpi_o$ sampled have $S({\bf Q})$ with $N=874$ values all within 10s of one another at $\Pi_o = 90^{\circ}$.  All three upper panel plots indicate an anticorrelation for the blue and olive curves, with the most pronounced difference (over $400$s) for the $3$:$1$ PC at $\Pi_o = 90^{\circ}$.  

The variation of $S({\bf Q})$ with $\Pi_o$ and $\varpi_o$ for the secular systems sampled (lower panels of Fig. \ref{lamlines}) qualitatively differ from those for the in-PC systems.  For the low eccentricity ($e_o = 0.14 e_H$) system XVII, the curves have fewer extrema, and have $S({\bf Q}) < 10$s except for the dotted and dashed curves peaking at $30^{\circ}$ and $40^{\circ}$.  The other two secular systems sampled are at high ($e_o \ge 0.73 e_H$) eccentricity, and both feature curves with multiple signal extrema, despite the 3-4 orders of magnitude difference in $S({\bf Q})$ between those two plots.  The greatest extrema occurs at $60^{\circ}$ for the dotted curve in system XX and at $70^{\circ}$ for the solid curve in system XXII.  The apparent anticorrelation between the solid and dotted curves from the in-PC systems appears to vanish for these secular systems.

\begin{figure*}
  \centering
  \begin{tabular}{ccc}
    \multicolumn{3}{c}{\includegraphics[width=0.55\textwidth,height=0.30\textheight]{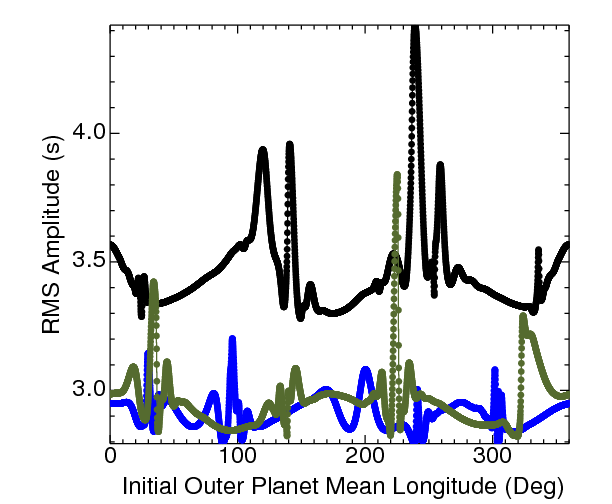}}\\
    \multicolumn{3}{c}{\includegraphics[width=0.55\textwidth,height=0.30\textheight]{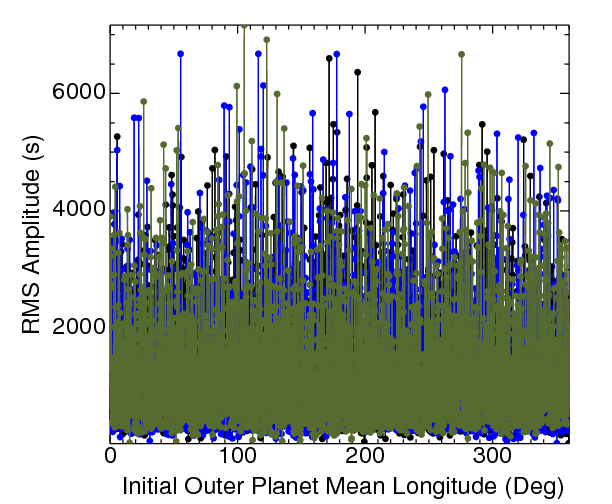}}\\
    \multicolumn{3}{c}{\includegraphics[width=0.55\textwidth,height=0.30\textheight]{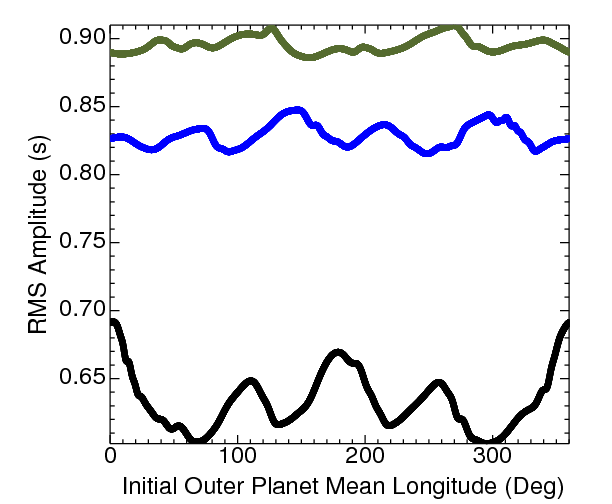}}\\
  \end{tabular}
%
  \put(-10,215){ { \huge $a_o/a_i=1.5$}}
  \put(-10,185){ { \huge$e_o=0.035$}}
  \put(-10,15){ { \huge$a_o/a_i=1.7$}}
  \put(-10,-15){ { \huge$e_o=0.28$}}
  \put(-10,-185){ { \huge$a_o/a_i=1.8$}}
  \put(-10,-215){ { \huge$e_o=0.10$}}
%
%
  \caption{High resolution plots of TTV RMS signal vs. $\Pi_o$.  Every $0.1^{\circ}$ of $\Pi_o$ is sampled for secular systems XVII (upper panel), XVIII (middle panel) and XIX (lower panel).  In each plot, the black, blue and green curves correspond to $\varpi_o = 0^{\circ}$, $\varpi_o = 90^{\circ}$ and $\varpi_o = 180^{\circ}$.  All signal amplitudes were sampled after 10 yr of continuous observations ($N=874$). Note the wide variation in sensitivity of signal amplitude to $\Pi_o$.}
  \label{highres}
\end{figure*}

In order to observe the variation of $S({\bf Q})$ with $\Pi_o$ at a higher resolution of $\Pi_o$, we performed additional simulations.  For the secular systems XVII, XVIII and XIX, we varied $\Pi_o$ over the entire $\left[ 0^{\circ}, 360^{\circ} \right]$ range by sampling the angle at intervals of $0.1^{\circ}$, and report the results in Fig. \ref{highres}.  The three panels show qualitatively different behavior.  The bottom panel (with $S({\bf Q}) < 1$ s) feature smooth curves, the top panel curves (with $2.8$ s $<  S({\bf Q}) < 4.4$ s) show spikes and dips which could be hidden in broader resolution studies, and the middle panel (with $S({\bf Q})$ up to $7 \times 10^3$ s) shows little structure whatsoever.  The figure demonstrates that TTV amplitude can vary by seconds due to changes of a few degrees in $\Pi_o$ in some regimes, and vary by thousands of seconds due to changes of a fraction of a degree in $\Pi_o$ in others.

We plot these curves primarily to demonstrate the difficulties inherent in solving for the mass and orbital parameters of the unseen planet.  Because of the broad parameter space, one is hard-pressed to determine ``representative" systems in each regime.  Additionally, because of the  semimajor axis resolution of our simulations, the typical difference between nominal MMR $a_o$ values and those from our simulations is $\sim 10^{-3} - 10^{-4}$.  This difference may be significant due to the sensitivity of TTV signal profiles on the planetary semimajor axis ratio.  Thus, we conclude that any TTV signal will need individual analysis, as opposed to being characterized by a few summary statistics.  Further, limits on the mass of the planets possible for a given TTV dataset must be mindful of the possibility that a putative planet could have orbital elements that result in a signal much smaller than is typical for a planet of a given mass, semimajor axis and eccentricity.

\section{Correlations with Masses}

Thus far, we have fixed the masses of the star and both planets in all systems studied.  Because these three masses are hierarchical ($M_{\odot} \gg M_J \gg M_{\oplus}$), varying any one of them by a factor of a few won't significantly alter the contour phase space structure of Fig. \ref{fiducial}.  However, mass variation might qualitatively affect individual systems at the edge of a secular, near-PC, or in-PC regime, or with an outer planet at a moderate-to-high ($> 0.5 e_H$) eccentricity.  

\begin{figure*}
  \centering
\vskip -2em
\resizebox{6in}{4.3in} {
  \begin{tabular}{cccc cccc cccc cccc cccc}
    \multicolumn{5}{c}{\includegraphics[trim = 28mm 20mm 0mm 0mm, clip, angle=0, width=0.30\textwidth]{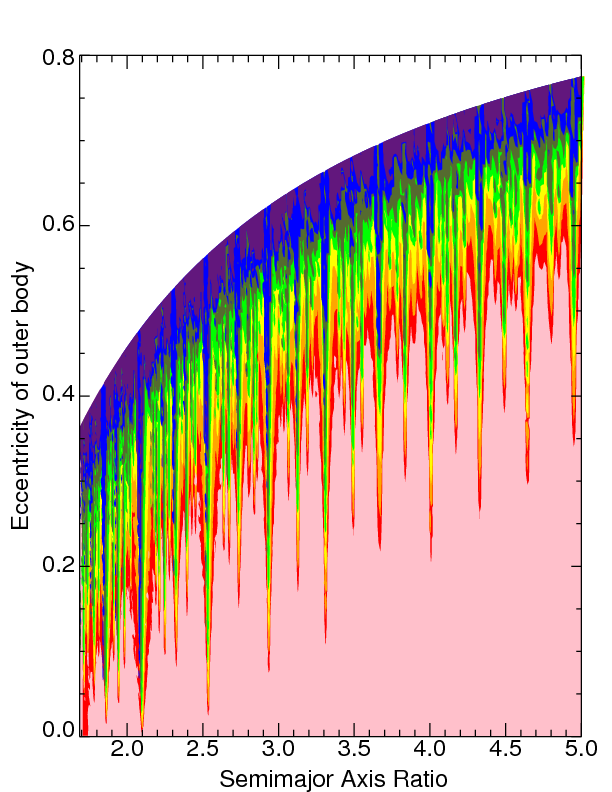}}&
    \multicolumn{5}{c}{\includegraphics[trim = 28mm 20mm 0mm 0mm, clip, angle=0, width=0.30\textwidth]{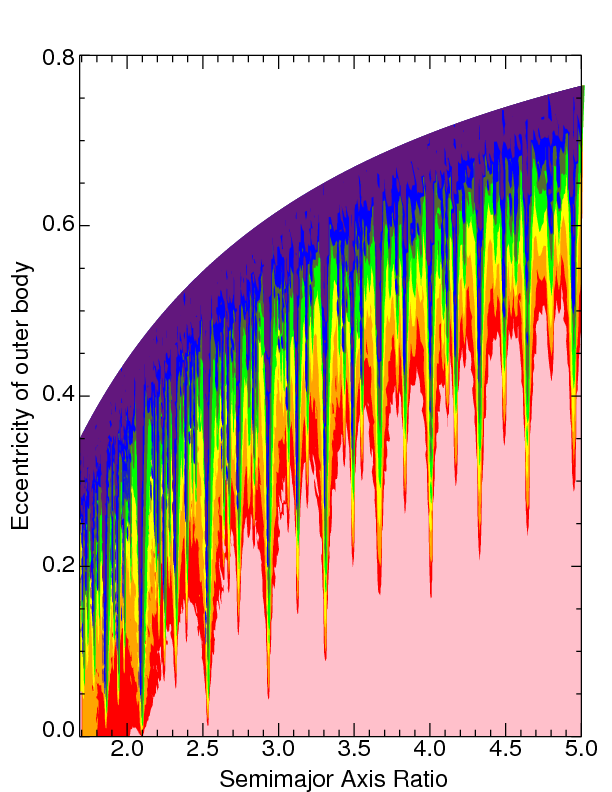}}&
    \multicolumn{5}{c}{\includegraphics[trim = 28mm 20mm 0mm 0mm, clip, angle=0, width=0.30\textwidth]{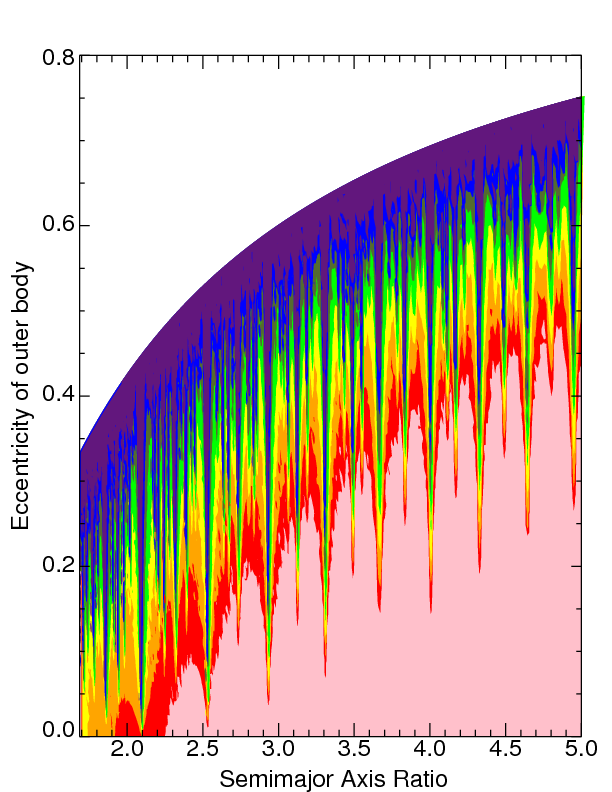}}&
    \multicolumn{5}{c}{\includegraphics[trim = 28mm 20mm 0mm 0mm, clip, angle=0, width=0.30\textwidth]{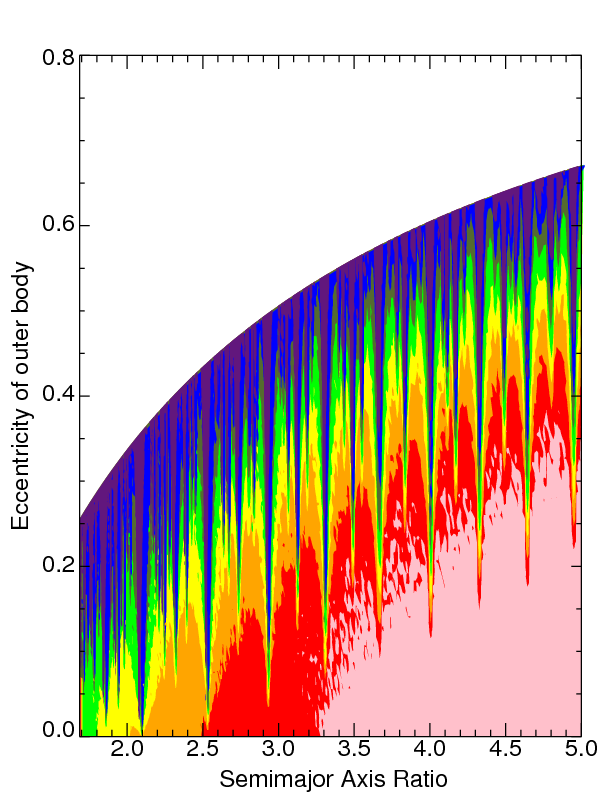}}\\[-10pt]
    \multicolumn{5}{c}{\includegraphics[trim = 28mm 20mm 0mm 0mm, clip, angle=0, width=0.30\textwidth]{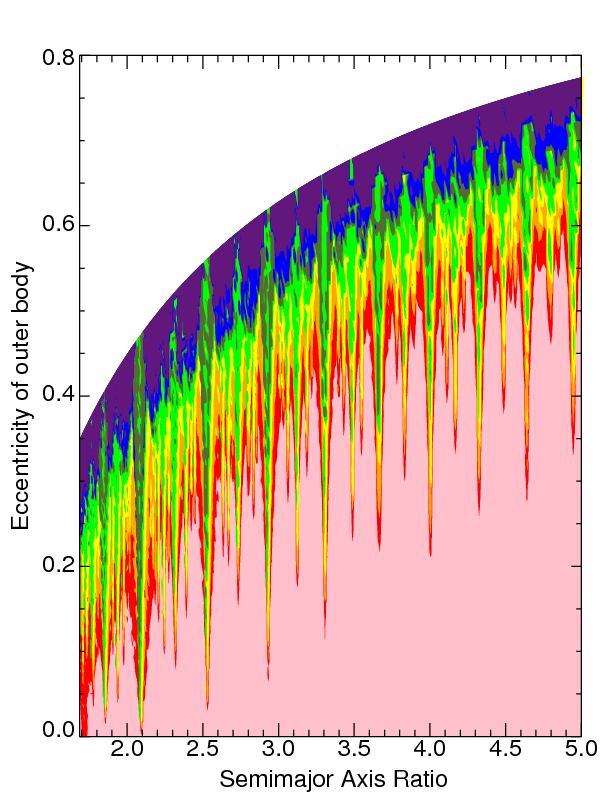}}&
    \multicolumn{5}{c}{\includegraphics[trim = 28mm 20mm 0mm 0mm, clip, angle=0, width=0.30\textwidth]{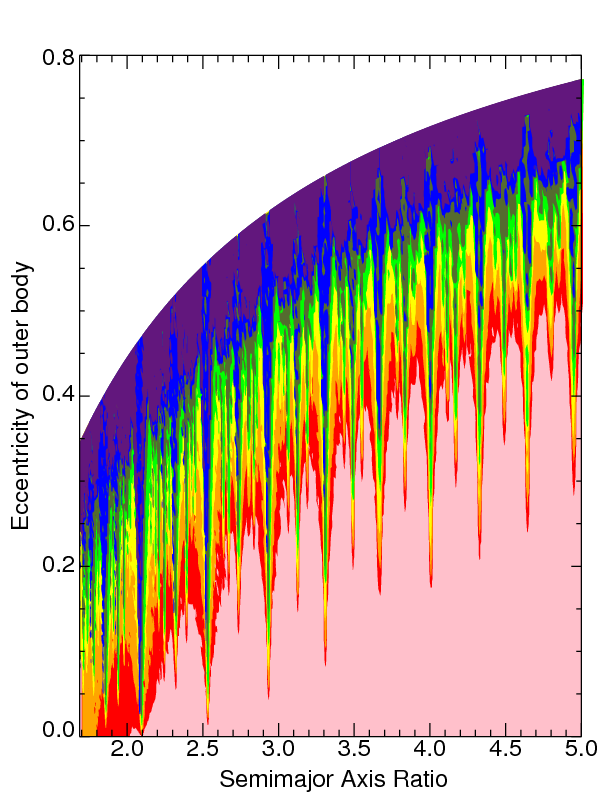}}&
    \multicolumn{5}{c}{\includegraphics[trim = 28mm 20mm 0mm 0mm, clip, angle=0, width=0.30\textwidth]{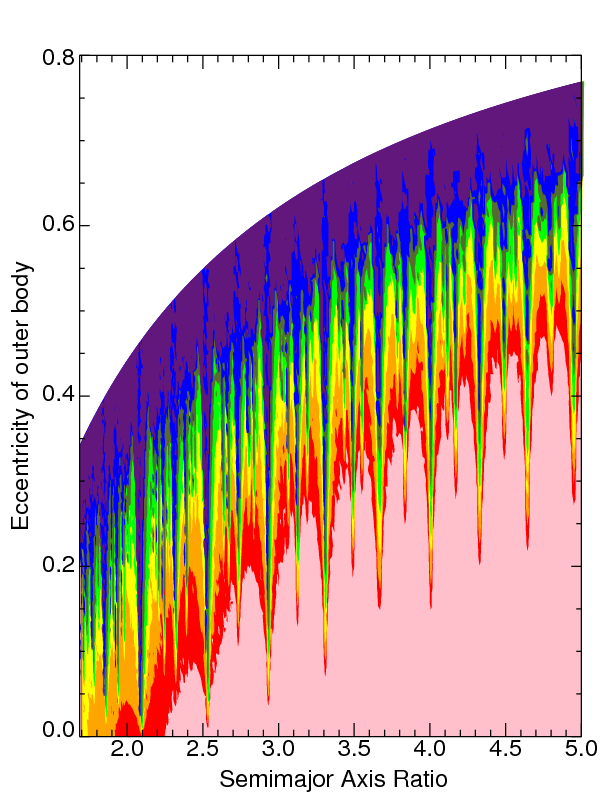}}&
    \multicolumn{5}{c}{\includegraphics[trim = 28mm 20mm 0mm 0mm, clip, angle=0, width=0.30\textwidth]{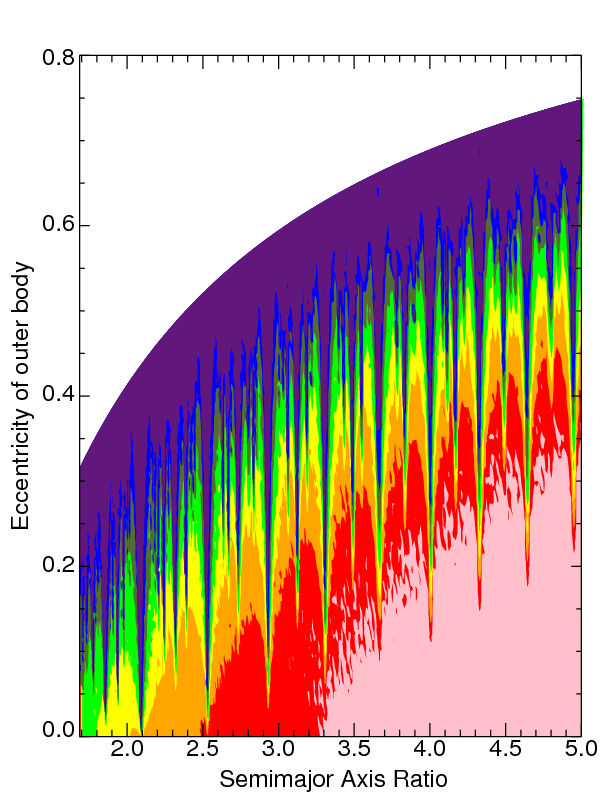}}\\[-10pt]
    \multicolumn{5}{c}{\includegraphics[trim = 28mm 20mm 0mm 0mm, clip, angle=0, width=0.30\textwidth]{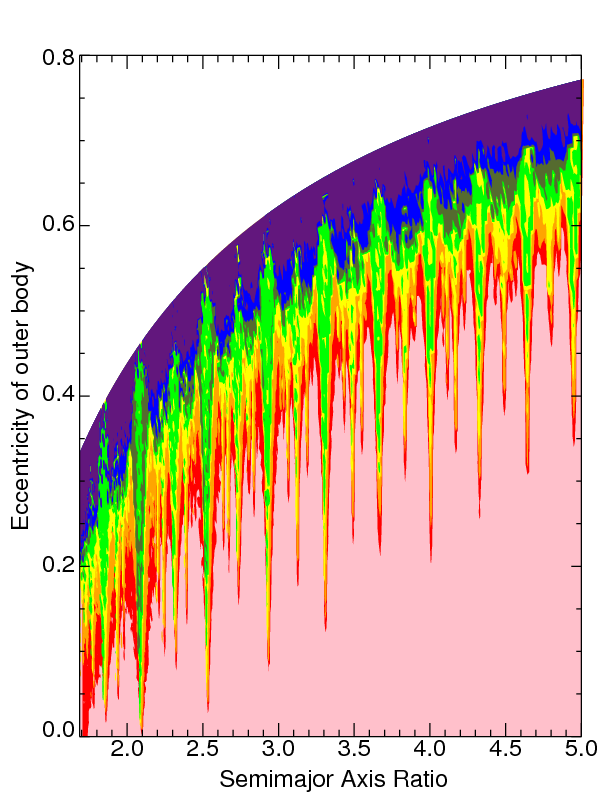}}&
    \multicolumn{5}{c}{\includegraphics[trim = 28mm 20mm 0mm 0mm, clip, angle=0, width=0.30\textwidth]{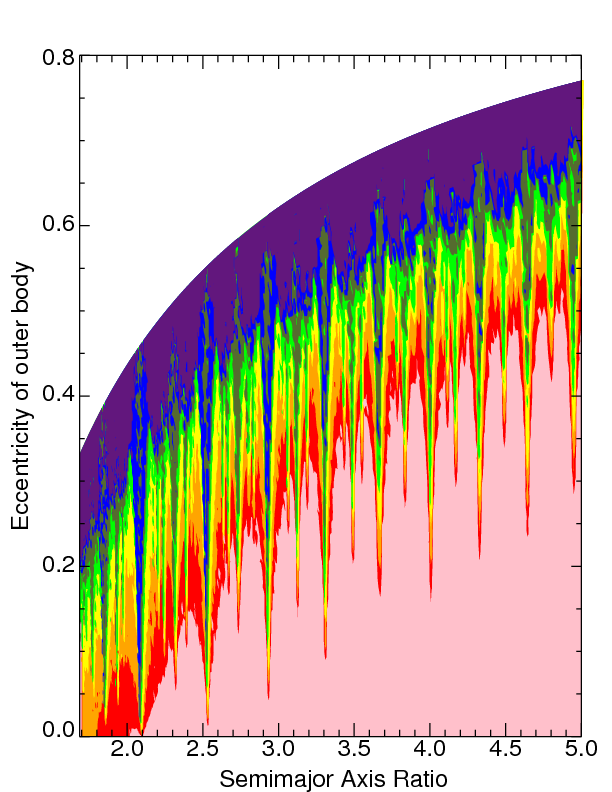}}&
    \multicolumn{5}{c}{\includegraphics[trim = 28mm 20mm 0mm 0mm, clip, angle=0, width=0.30\textwidth]{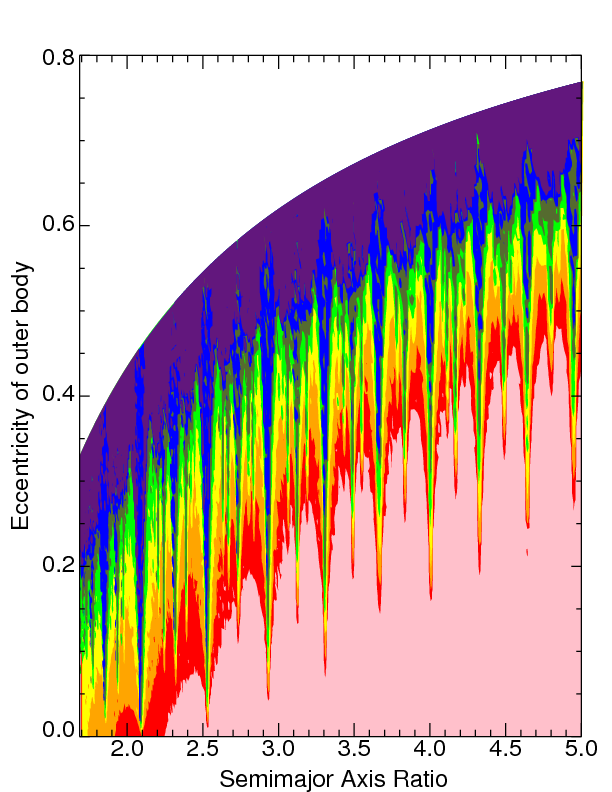}}&
    \multicolumn{5}{c}{\includegraphics[trim = 28mm 20mm 0mm 0mm, clip, angle=0, width=0.30\textwidth]{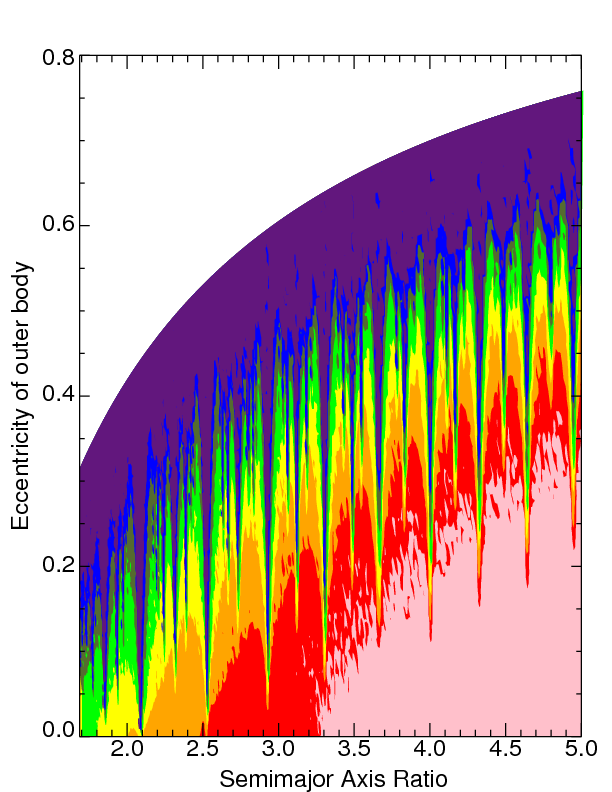}}\\[-10pt]
    \multicolumn{5}{c}{\includegraphics[trim = 28mm 20mm 0mm 0mm, clip, angle=0, width=0.30\textwidth]{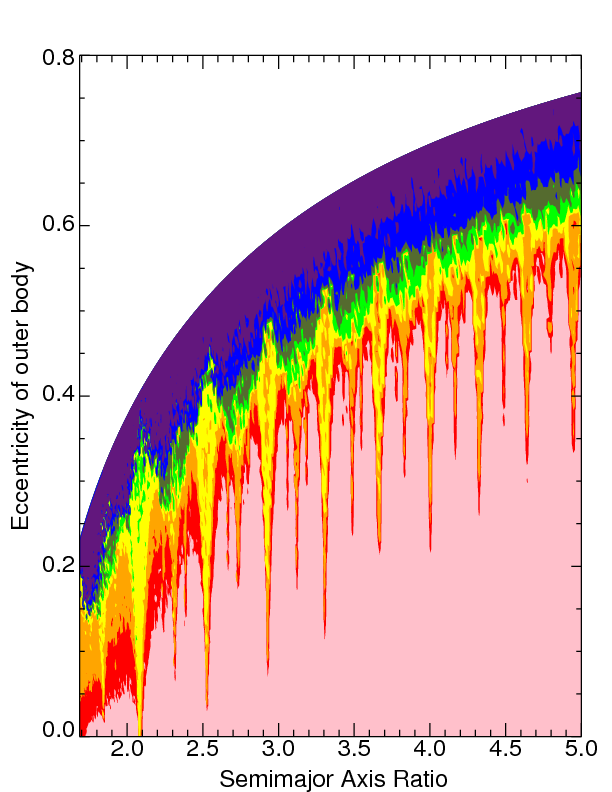}}&
    \multicolumn{5}{c}{\includegraphics[trim = 28mm 20mm 0mm 0mm, clip, angle=0, width=0.30\textwidth]{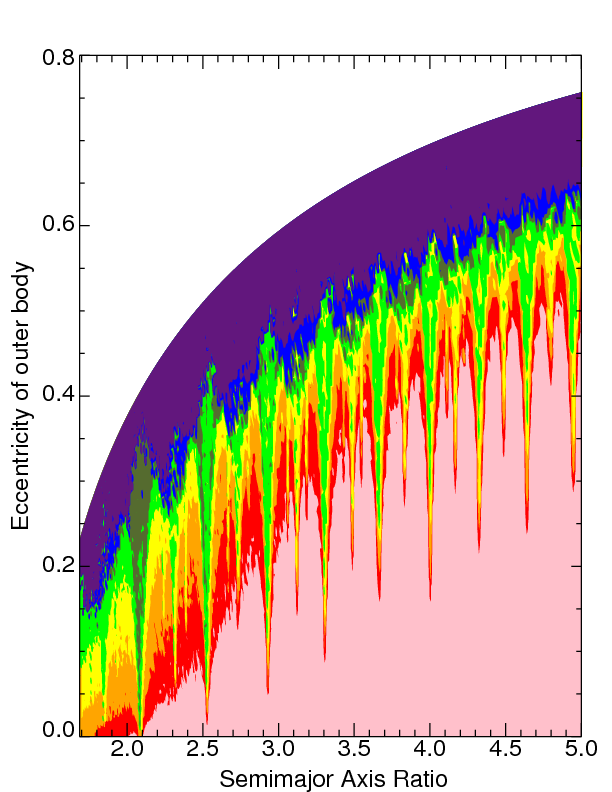}}&
    \multicolumn{5}{c}{\includegraphics[trim = 28mm 20mm 0mm 0mm, clip, angle=0, width=0.30\textwidth]{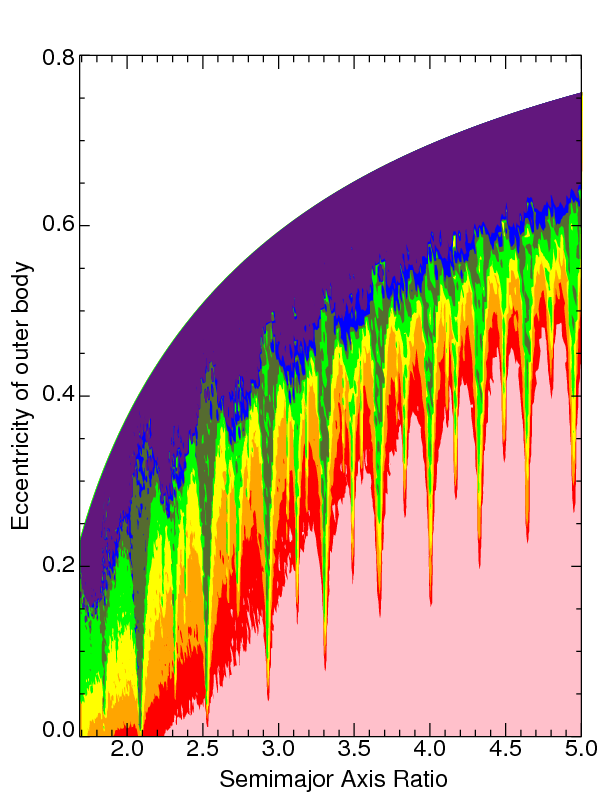}}&
    \multicolumn{5}{c}{\includegraphics[trim = 28mm 20mm 0mm 0mm, clip, angle=0, width=0.30\textwidth]{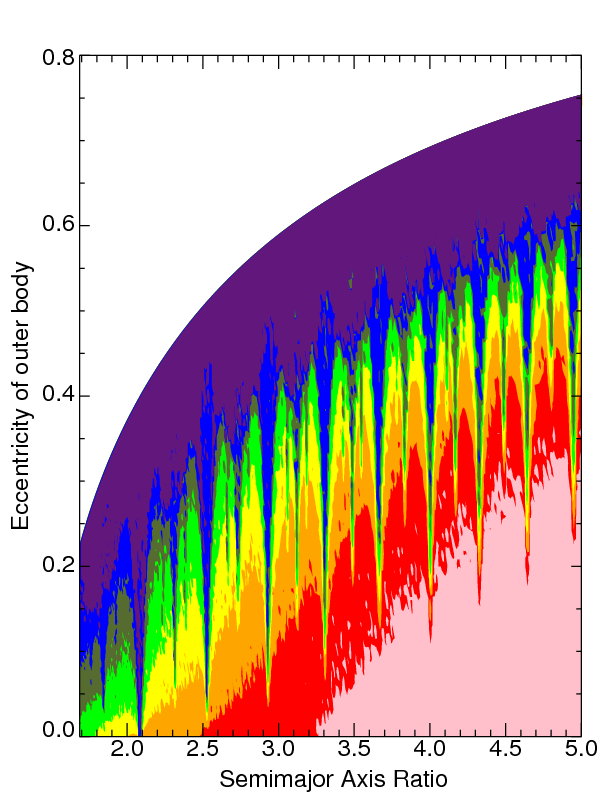}}\\[-10pt]
    \multicolumn{5}{c}{\includegraphics[trim = 28mm 20mm 0mm 0mm, clip, angle=0, width=0.30\textwidth]{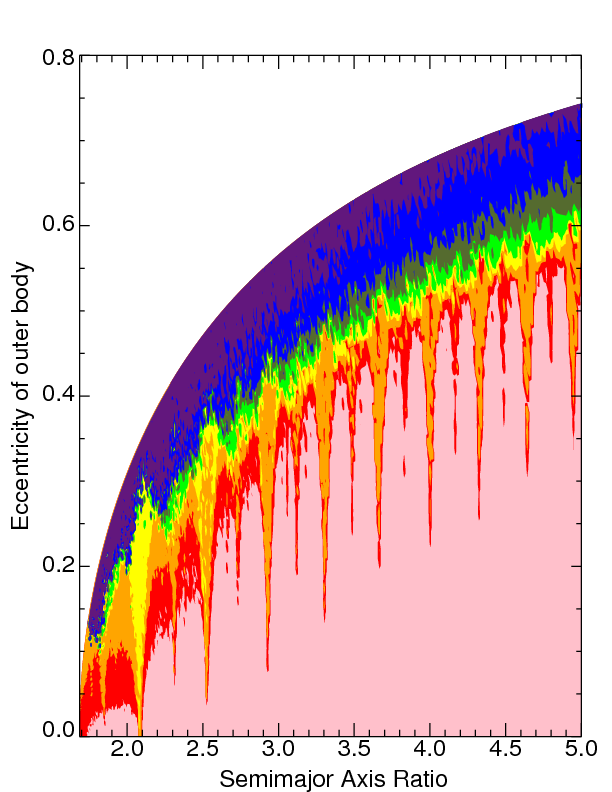}}&
    \multicolumn{5}{c}{\includegraphics[trim = 28mm 20mm 0mm 0mm, clip, angle=0, width=0.30\textwidth]{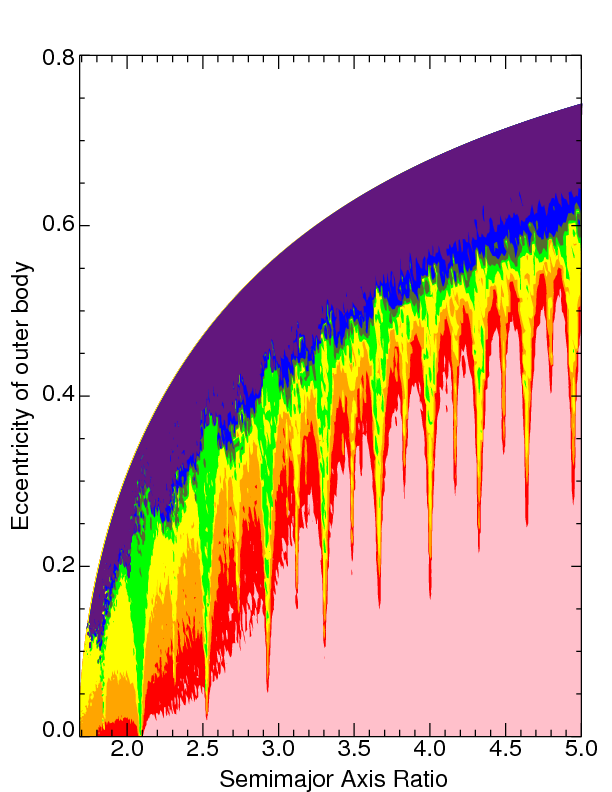}}&
    \multicolumn{5}{c}{\includegraphics[trim = 28mm 20mm 0mm 0mm, clip, angle=0, width=0.30\textwidth]{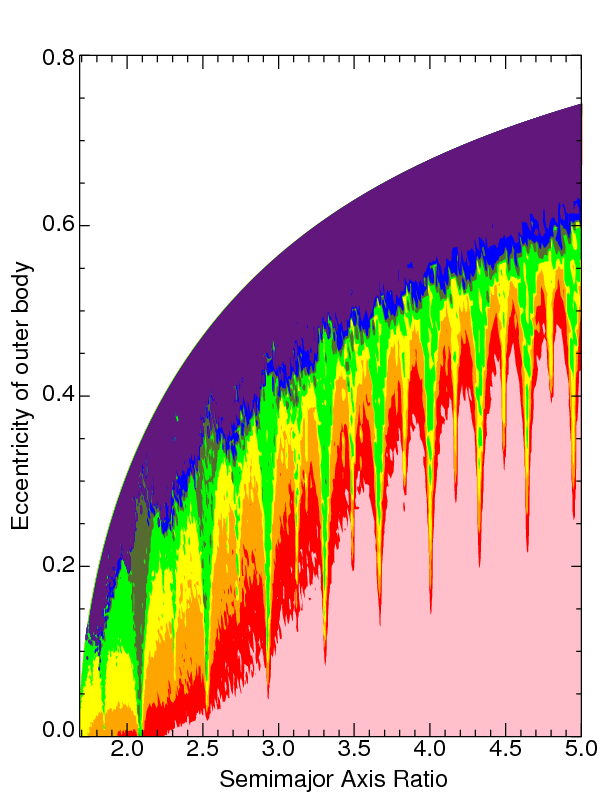}}&
    \multicolumn{5}{c}{\includegraphics[trim = 28mm 20mm 0mm 0mm, clip, angle=0, width=0.30\textwidth]{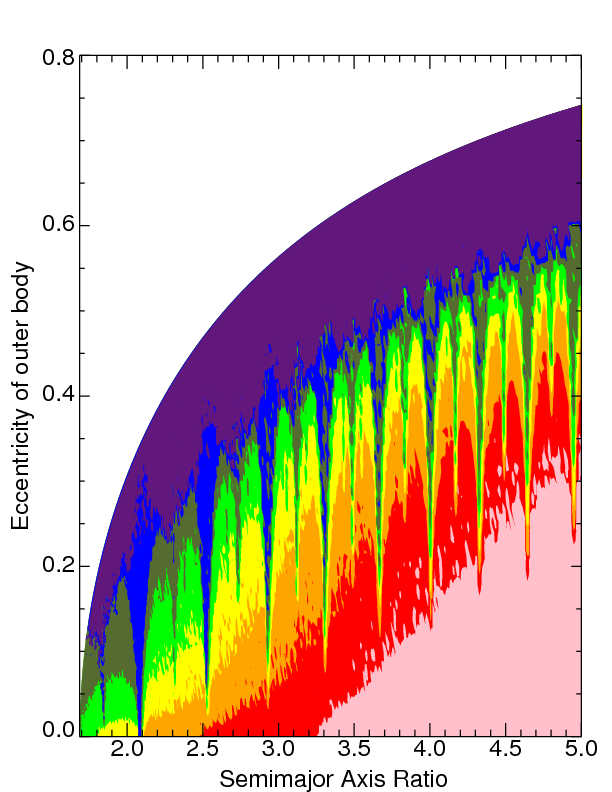}}\\[-10pt]
  \end{tabular}
}
%
  \put(-490,305){{\huge $M_i$}}
  \put(-506,234){{\huge $0.1 M_J$}}
  \put(-506,111){{\huge $0.5 M_J $}}
  \put(-506,-12){{\huge $1.0 M_J $}}
  \put(-506,-135){{\huge $5.0 M_J $}}
  \put(-506,-258){{\huge $10.0 M_J $}}
  \put(-460,330){{\huge $M_o$}}
  \put(-415,330){{\huge $1.0 M_{\oplus}$}}
  \put(-310,330){{\huge $5.0 M_{\oplus}$}}
  \put(-205,330){{\huge $10.0 M_{\oplus}$}}
  \put(-100,330){{\huge $50.0 M_{\oplus}$}}
%
%
%
  \caption{ {The median RMS TTV amplitude for 5 different initial orbital configurations for (horizontally) 1, 5, 10 and 50 $M_{\oplus}$ external perturbers and (vertically) 0.1, 0.5, 1, 5 and 10 $M_{J}$ transiting planets.  The contour levels are the same as in Figs. \ref{fiducial} and \ref{minmax}.  All systems are sampled for $N = 874$.  Note that increasing $M_o$ increases the TTV amplitude in most cases, while increasing $M_i$ restricts the number of PCs which are discernible.}}
  \label{masscont}
\end{figure*}

We consider four different external perturber masses ($1 M_{\oplus}, 5 M_{\oplus}, 10 M_{\oplus},  50 M_{\oplus}$) and five different transiting planet masses ($0.1 M_J, 0.5 M_J, 1 M_J,  5 M_J, 10 M_J$).  Because Hill Stability is a function of these masses, transiting inner planets more massive than $1 M_J$ will most significantly restrict the semimajor axis-eccentricity space in which the system is guaranteed to be stable.  Therefore, in Fig. \ref{masscont}, the phase space plotted is smaller than that of Fig. \ref{fiducial} for all 20 combinations of planetary masses. 

Increasing the mass of the outer planet increases the TTV signal in almost all areas of phase space.  More subtly, increasing mass causes the signal amplitude at resonance to be higher than that from near-resonance, in contrast to Fig. \ref{N3to1}.  Fewer PCs signatures are discernible at higher-mass transiting planets; for $M_i = 10 M_J$, only PCs with $q = 1$ or $2$ can be identified by inspection in the figure, independent of the value of $M_o$.  Correspondingly, one can distinguish the greatest number of PCs for the lowest $M_i$ values.  Additionally, for the most massive interior and exterior planets, the highest-signal stable (purple) region is greatest in extent.

\begin{figure*}
  \centering
\resizebox{6in}{!} {
  \begin{tabular}{cc cc cc}
    \multicolumn{3}{c}{\includegraphics[height=3in,width=0.50\textwidth]{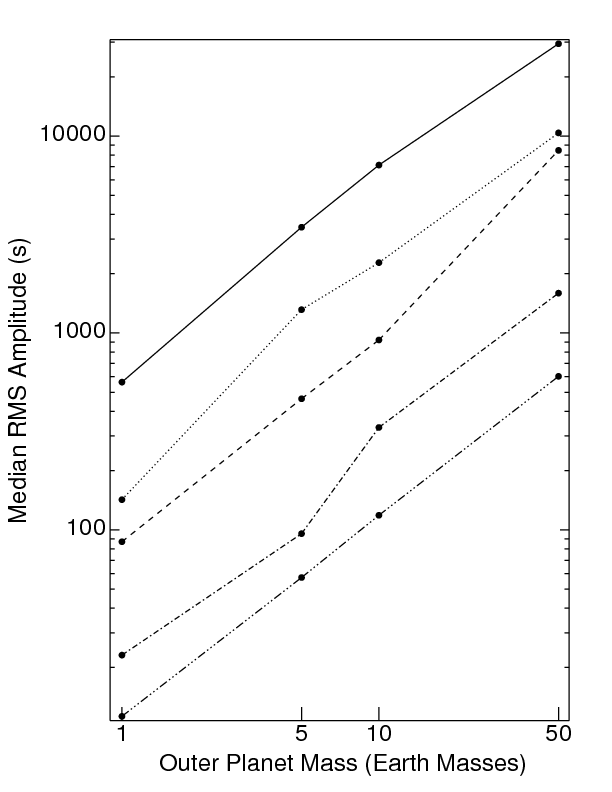}}&
    \multicolumn{3}{c}{\includegraphics[height=3in,width=0.50\textwidth]{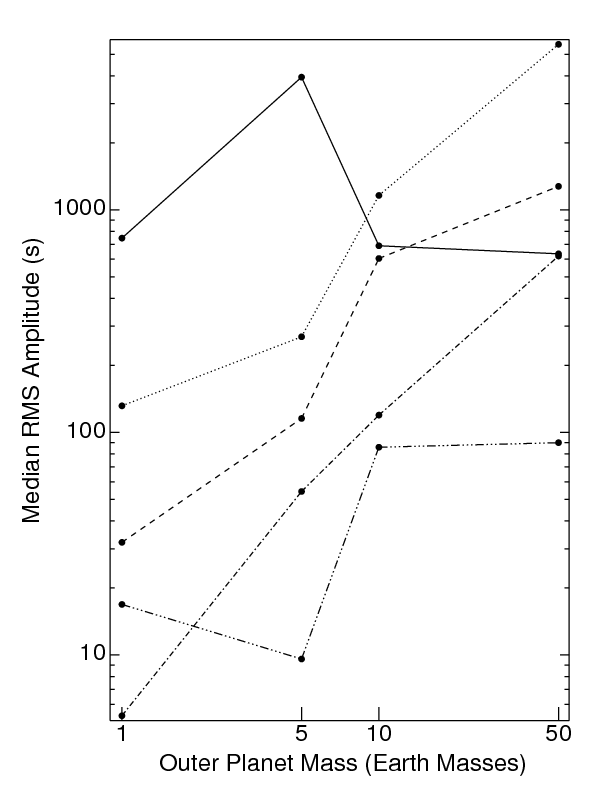}}\\[-10pt]
    \multicolumn{3}{c}{\includegraphics[height=3in,width=0.50\textwidth]{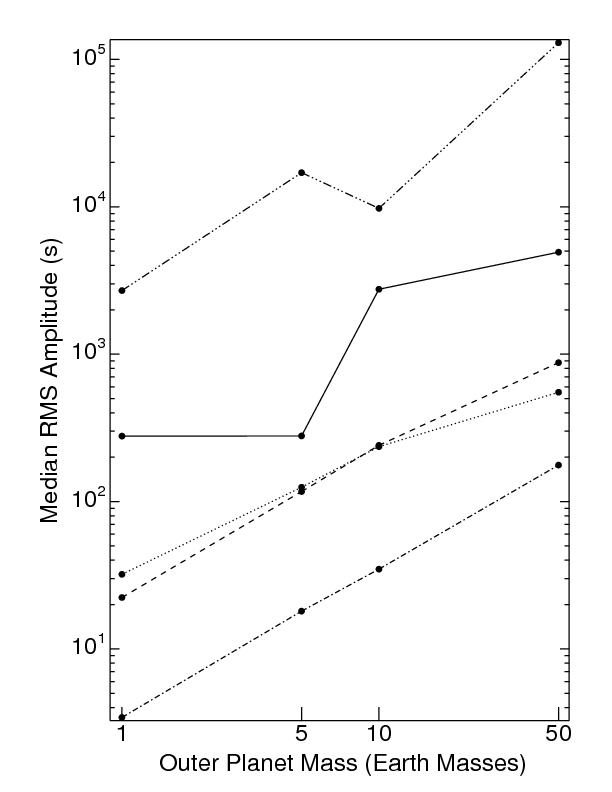}}&
    \multicolumn{3}{c}{\includegraphics[height=3in,width=0.50\textwidth]{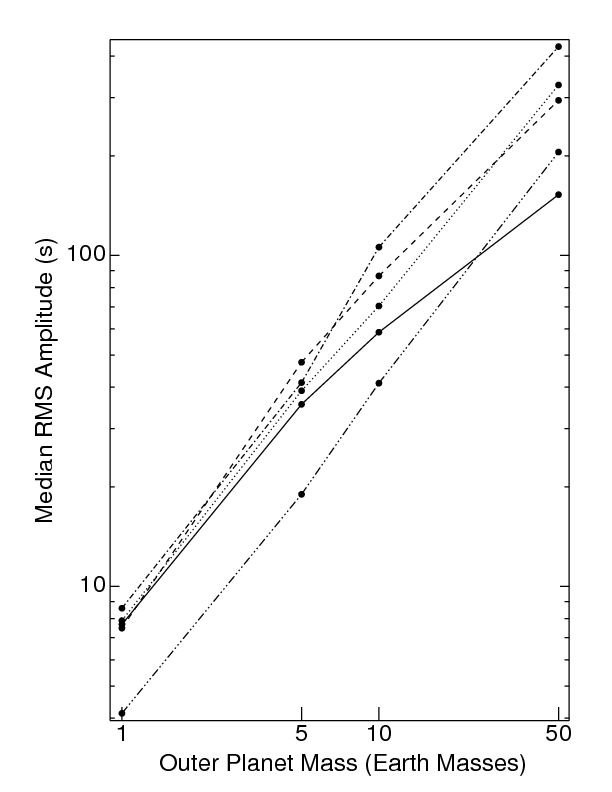}}\\[-10pt]
    \multicolumn{3}{c}{\includegraphics[height=3in,width=0.50\textwidth]{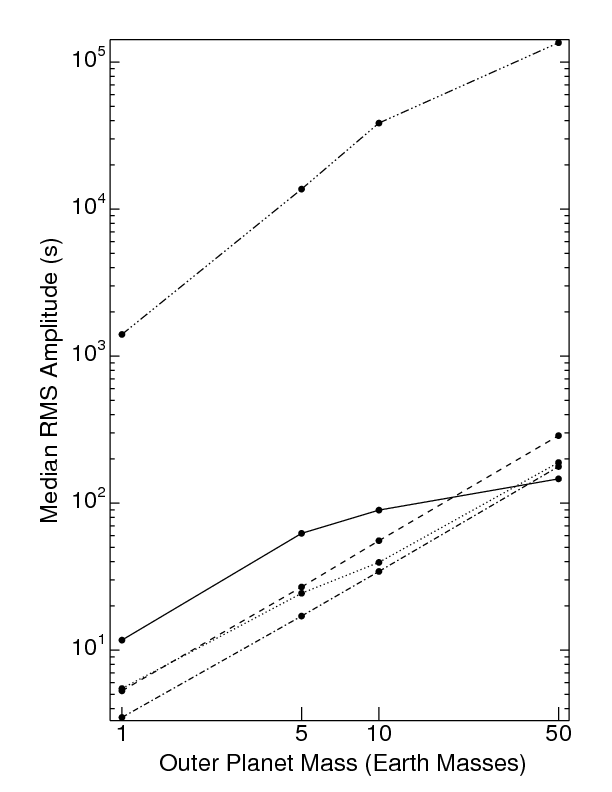}}&
    \multicolumn{3}{c}{\includegraphics[height=3in,width=0.50\textwidth]{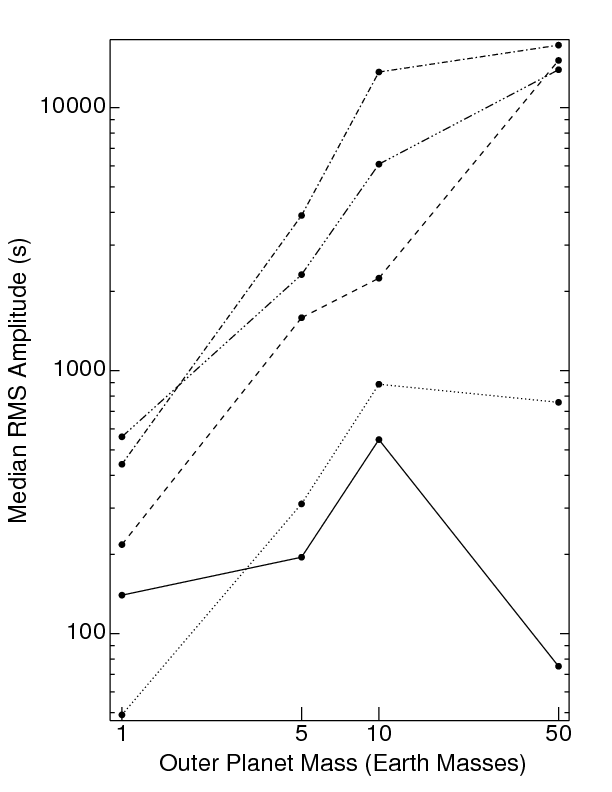}}\\[-10pt]
  \end{tabular}
}
  \put(-490,183){{\large $2$:$1$ PC}}
  \put(-490,171){{\large $e_o=0.20$}}
  \put(-10,183){{\large $8$:$1$ PC}}
  \put(-10,171){{\large $e_o=0.57$}}
  \put(-490,6){{\large Near $2$:$1$ PC}}
  \put(-490,-6){{\large $e_o=0.11$}}
  \put(-10,6){{\large Near $5$:$2$ PC}}
  \put(-10,-6){{\large $e_o=0.20$}}
  \put(-490,-171){{\large $a_o/a_i=1.5$}}
  \put(-490,-183){{\large $e_o=0.035$}}
  \put(-10,-171){{\large $a_o/a_i=4.405$}}
  \put(-10,-183){{\large $e_o=0.665$}}
%
%
  \caption{TTV RMS signal vs. $M_o$ for systems at the $2$:$1$ and $8$:$1$ PC (upper panels, left to right), systems near the $2$:$1$ and $5$:$2$ PC (middle panels, left to right), and secular XVII and XXIII systems (lower panels, left to right) from Table \ref{regimes}.  In each plot, the solid, dotted, dashed, dot-dashed and triple dot-dashed curves correspond to $M_i = 0.1 M_J, 0.5 M_J, 1 M_J, 5 M_J$, and $10 M_J$.  All signal amplitudes represent the median values from sampling 5 randomly chosen sets of initial orbital angles after 10 yr of continuous observations.  Note that although there is a power-law-like dependence of RMS amplitude on $M_o$ (unlike for $M_i$), this dependence breaks down for some orbital architectures.}
  \label{massline}
\end{figure*}

Unlike the correlation between TTV signal and orbital angles, the correlation between TTV signal and planetary mass is robustly exponential in most stable regimes.  Plotted in Fig. \ref{massline} are 6 systems (I, IV, IX, XIV, XVII, XVIII) from Table \ref{regimes}, with the solid, dotted, dashed, dot-dashed and triple dot-dashed curves corresponding to $M_i = 0.1 M_J, 0.5 M_J, 1 M_J, 5 M_J$, and $10 M_J$.   All six systems illustrate explicitly a roughly exponential increase in $S({\bf Q})$ as a function of $M_o$.  The upper panel systems, which correspond to systems in the $2$:$1$ and $8$:$1$ PCs, demonstrate that $S({\bf Q})$ decreases with increasing $M_i$, unlike in the panels below, and that for $8$:$1$, all curves monotonically increase, whereas the curves for $2$:$1$ do not.  The middle panel of the figure displays $2$:$1$ and $5$:$2$ near-PC systems with $e_o = 0.44 e_H$  and $e_o = 0.49 e_H$, respectively. The difference in $S({\bf Q})$ upon varying the planetary masses is striking: 6 orders of magnitude for the $2$:$1$ near-PC system, and just 2 orders of magnitude for the $5$:$2$ near-PC system. The secular systems in the lowest panel show two anomalous features:  1) for system XVII, $S({\bf Q})$ for the $M_i = 10 M_J$ curve is 2 orders of magnitude higher than any of the other $M_i$ curves, 2) system XXIII joins the in-PC $8$:$1$ system as the only ones which contain a curve (solid; $M_i = 0.1 M_J$) where $S({\bf Q})$ decreases as $M_o$ is increased from $1 M_{\oplus}$ to $50 M_{\oplus}$.

\section{Correlations with Orbital Parameter Evolution} 

Over the course of a multi-year observing campaign, the semimajor axis and eccentricity of the transiting planet might vary noticeably.  N-body simulations show that an exterior terrestrial-planet could induce semimajor axis and eccentricity variations of $\sim 0.003$ AU and $\sim 0.005$, respectively, of the massive transiting Jupiter.  If this variation is observed over time, then in principle these profiles could supplement TTVs as a method for identifying or confining the parameters of unseen planets. 

Away from PC, although various formulations \citep{verarm2007} of Laplace-Lagrange secular theory (see \citealt*{murder2000}) may be used to provide approximations of $e_i(t)$ profiles which don't exceed low-to-moderate eccentricities $\sim 0.2-0.4 e_H$, these secular timescales (often $10^{4-8}$yr) exceed the duration of optimistic observing campaigns ($10$ yr) by many orders of magnitude.  One should instead consider evolution on orbital timescales.  One could apply perturbation theory without averaging over mean longitude in order to model these systems.  Alternatively, suites of N-body simulations are prudent since the CPU time required to evolve two-planet systems on the timescale of 10 yr is on the order of $\mu s$.  


At PC, and especially in MMR, theory may yield an additional useful constraint.  Because of the mass hierarchy in our system, and the initially circular orbit of $M_i$, the resonant state of the systems studied here is likely to be dominated by a single resonant angle of the form in Eq. (\ref{resangle}).  In this situation, constants of motion exist which relate just two orbital parameters to each other through conservation of angular momentum and energy.  For any eccentricity-type of resonance, one of these constants, $C_1$, may be expressed \citep{veras2007} in terms of $a_i(t)$ and $e_i(t)$ only (recall that $e_i \equiv e_i(t=0)$ and  $a_i \equiv a_i(t=0)$):

\begin{eqnarray}
C_1 & \equiv & \sqrt{a_i} \left[ p \left( \sqrt{1 - e_{i}^2} - 1  \right) - \left(p-q \right)  \right]
\label{ck1}
\\
& = & \sqrt{a_{i}(t)} \left[ p \left(  \sqrt{1 - e_{i}(t)^2} - 1 \right)  - \left(p-q \right)  \right]
,
\label{ck2}
\end{eqnarray}

\noindent{}Another constant, $C_2$, relates $a_i(t)$ to $a_o(t)$:

\begin{eqnarray}
C_2 & \equiv & \sqrt{a_o} \left(  \frac{q M_o}{p M_i} \right) 
\frac{M_{\star} \sqrt{M_{\star} + M_o + M_i} }
{ \left(  M_{\star} + M_1 \right)^{3/2} }
-
\sqrt{a_i}
\label{ck3}
\\
& = & \sqrt{a_o(t)} \left(  \frac{q M_o}{p M_i} \right) 
\frac{M_{\star} \sqrt{M_{\star} + M_o + M_i } }
{ \left(  M_{\star} + M_o \right)^{3/2} }
-
\sqrt{a_i(t)}
\equiv
\sqrt{a_{o}(t)} Y - \sqrt{a_{i}(t)}
\label{ck4}
\end{eqnarray}

We rewrite Eqs. ({\ref{ck1})-(\ref{ck4}) as

\begin{equation}
\sqrt{1 - e_{i}(t)^2}
=
1
+ \frac{1}{q}
\left[
\left(  Y \left[ \frac{\sqrt{a_o(t)} - \sqrt{a_{o}}}
               {\sqrt{a_{i}}} \right] + 1 \right)^{-1}
\left( 
q
\left( 
\sqrt{1 - e_{i}^2} - 1
\right)
- \left( p-q \right)
\right)
+
\left( p-q \right)
\right]
,
\label{big1}
\end{equation}

\noindent{which} illustrates how a transiting planet's eccentricity is predicted to change as a function of the variation in the external perturber's semimajor axis for a given MMR.  However, this expression does require knowledge of the initial semimajor axes and eccentricities of both planets.

Assuming that the outer planet is much more massive than the inner planet, the unseen outer planet will experience greater variations of its orbital elements over time than the transiting inner planet.  Here we consider only the semimajor axis ratio range of 1.3-3.2.  For $e_o = 0$, the variation of $e_o(t)$ typically does not exceed $0.01$, but may reach $0.1$ for systems near the 2:1 PC.  The $e_o(t)$ profiles in this regime include sinusoidal, lightly modulated, and heavily modulated curves with characteristic periods ranging from a few days to a few years.  For $e_o = 0.5 e_H$, typically $0.01 \le e_o(t) \le 0.1$, with a greater proportion of sinusoidal-like curves.  For $e_o = 0.75 e_H$, typically $e_o(t) \ge 0.1$, with variations up to $\approx 0.6$.  Many $e_o(t)$ profiles have little apparent structure, and are periodic only on long ($> 10 $yr) timescales.   The outer planets with the smallest eccentricity variation, at $e_o(t) \approx 0.01$, have corresponding TTV RMS amplitudes of just $10$s.  RMS signal amplitudes of $\sim 10^3$s may correspond to $e_o$ variations of $<0.10$ or $>0.40$.  For $e_o = e_H$, $e_o(t)$ profiles vary chaotically and have typical amplitudes $> 10^4$s.  For $a_o/a_i > 3.2$, $e_o(t)$ profiles are generally flat except for those corresponding to $e_o \approx e_H$, in which they vary chaotically. 

Transiting systems at or near PCs is perhaps of the greatest observational interest.  Therefore, we trace the time evolution of the resonant angle in each system to determine if and over what timescale, and with what amplitude, this angle librates.  We consider the strongest (lowest-order) PCs for $q = 1, 2$ and $3$, and fix $\Pi_i = \varpi_o = 0^{\circ}$ and  $\Pi_o = 60^{\circ}$.  For each PC, we sample the resonant argument evolution for all simulations with the three values of $a_o/a_i$ simulated which are closest to the value of $\left( p/q \right)^{2/3}$. Here, we describe the results qualitatively for this narrow region of resonant phase space.

All these systems are near a PC, and at eccentricities near $e_H$ are all within the libration width of the corresponding commensurability.  However, only a fraction of these systems actually exhibit a librating resonant angle, and hence are ``in" MMR.  The resonant angle may librate for different timespans; throughout an observational campaign or periodically during that campaign.  The importance of a system actually in MMR versus just having a period ratio close to that commensurability is arguable, but a topic we now address.

First, we consider a commensurability that is saturated by resonant systems.  We plot the systems sampled for the $2$:$1$ PC in Fig. \ref{MMR2to1}.  The upper, middle and lower panels represent systems at the three different initial semimajor axis ratios ($1.5812$, $1.5865$ and $1.5919$), and the left and right panels represent the ranges $0 > e_o > 0.5 e_H$ and $0.5 e_H > e_o > e_H$, with $e_o$ decreasing downward.  Every row of each panel represents a different system; the left column represents the TTV signal, with corresponding amplitude in seconds to the left; the middle column represents the ``apsidal angle" ($\equiv \varpi_o - \varpi_i$); the right column represents the resonant angle ($2\lambda_o - \lambda_i - \varpi_o$) in red and another angle ($2\lambda_o - \lambda_i - \varpi_i$) in green.  The middle and right columns have $0^{\circ} - 180^{\circ}$ plot ranges, whereas the left column's range is scaled according to $S({\bf Q})$.  The curves are shown for a time evolution of about 1.1 yr (corresponding to $N \approx 100$).

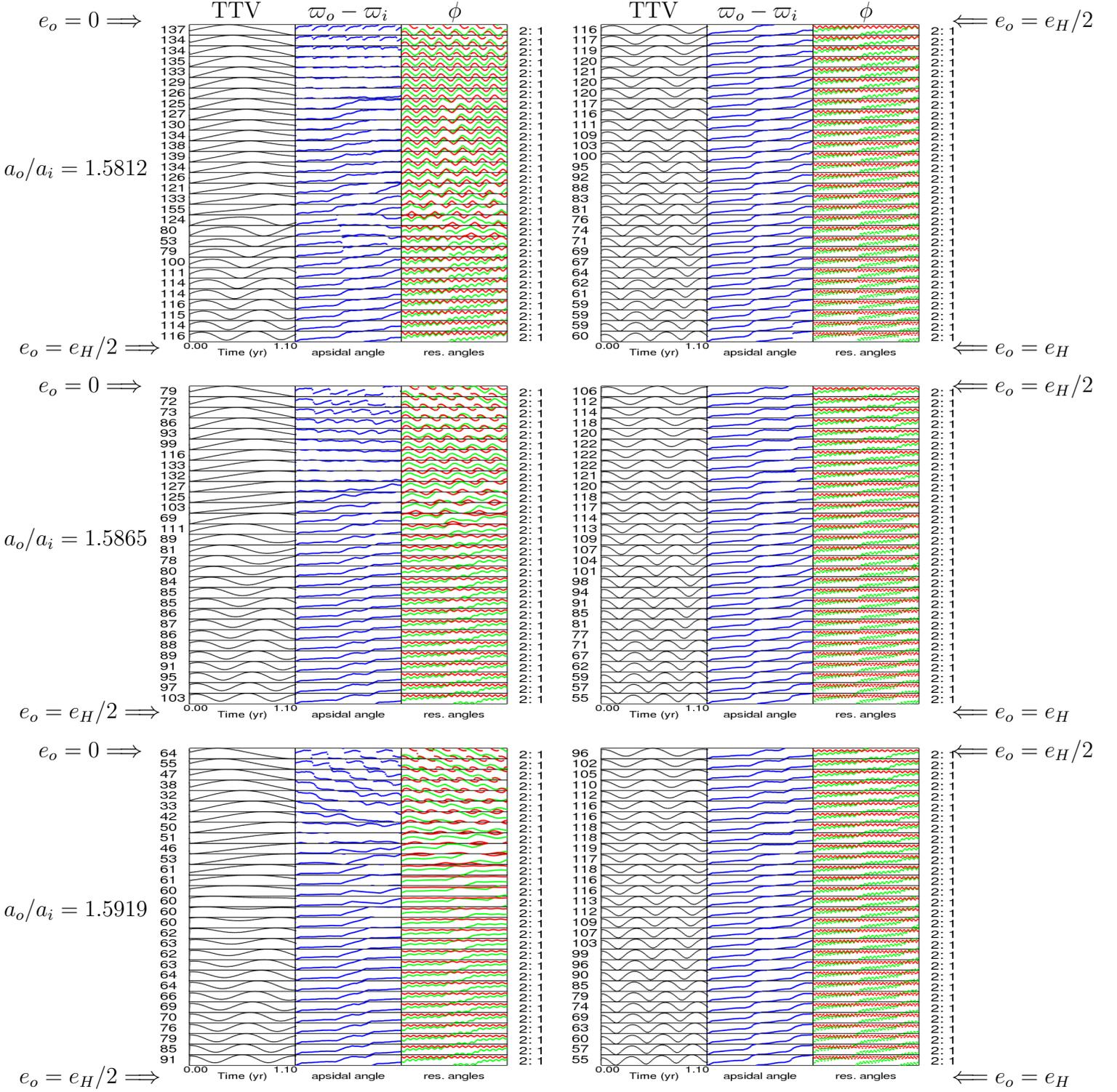
\begin{figure*}
  \centering
\resizebox{6in}{!} {
  \begin{tabular}{cc cc cc}
    \multicolumn{3}{c}{\includegraphics[height=4.0in,width=4.3in]{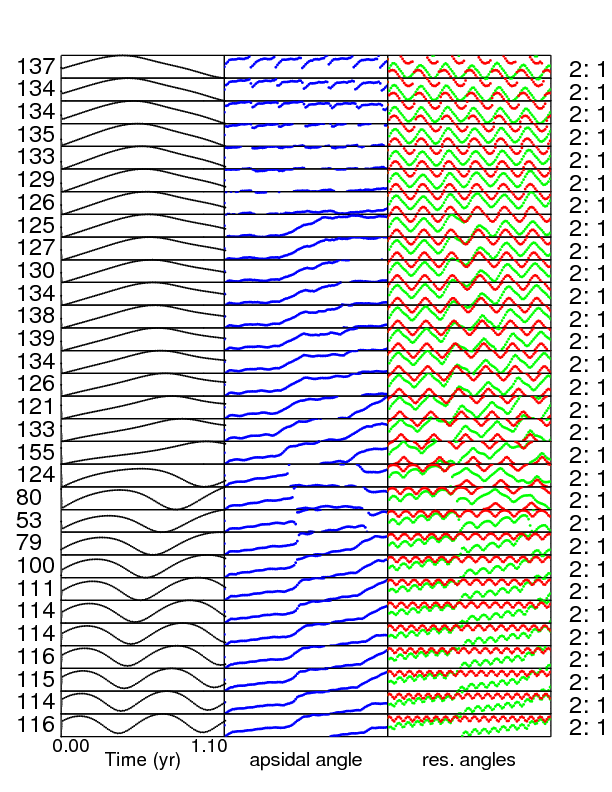}}&
    \multicolumn{3}{c}{\includegraphics[height=4.0in,width=4.3in]{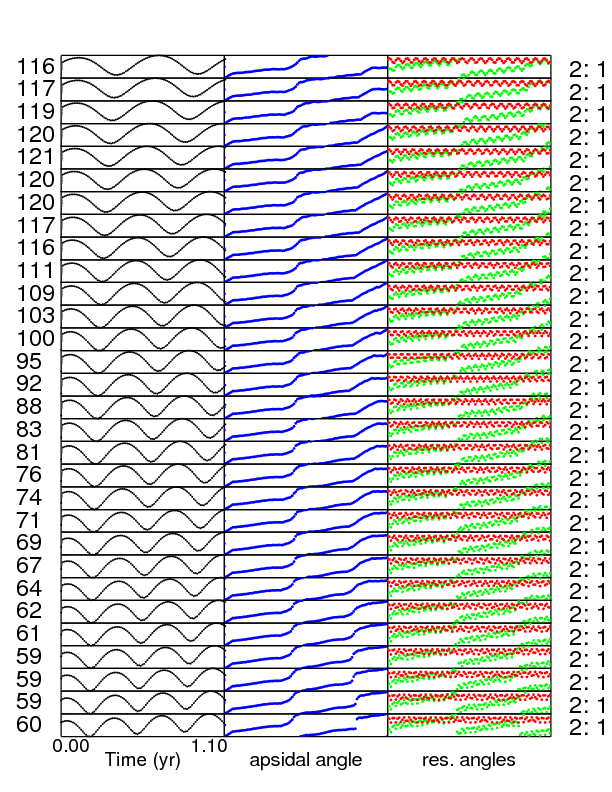}}\\[-10pt]
    \multicolumn{3}{c}{\includegraphics[height=4.0in,width=4.3in]{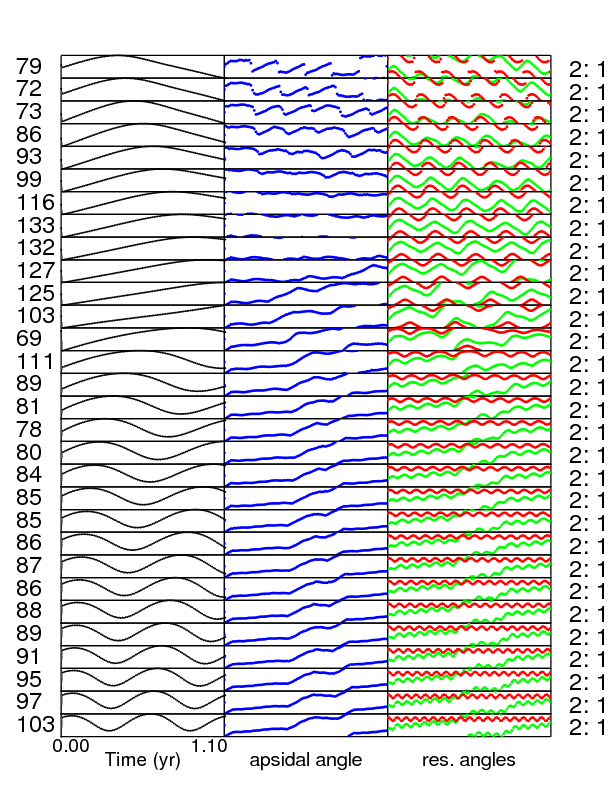}}&
    \multicolumn{3}{c}{\includegraphics[height=4.0in,width=4.3in]{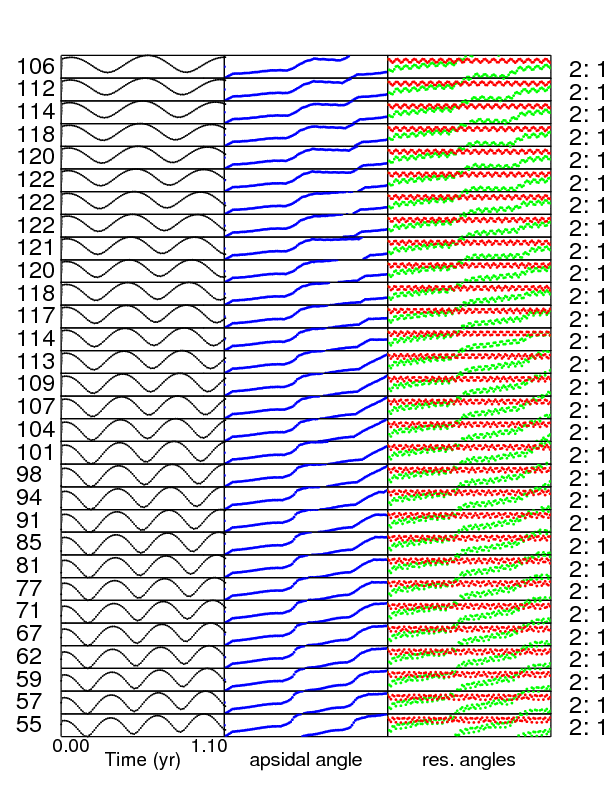}}\\[-10pt]
    \multicolumn{3}{c}{\includegraphics[height=4.0in,width=4.3in]{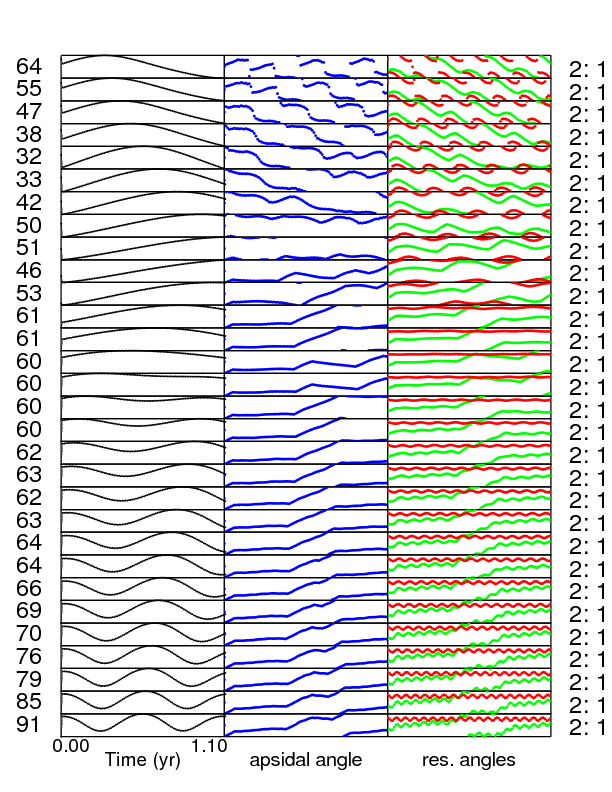}}&
    \multicolumn{3}{c}{\includegraphics[height=4.0in,width=4.3in]{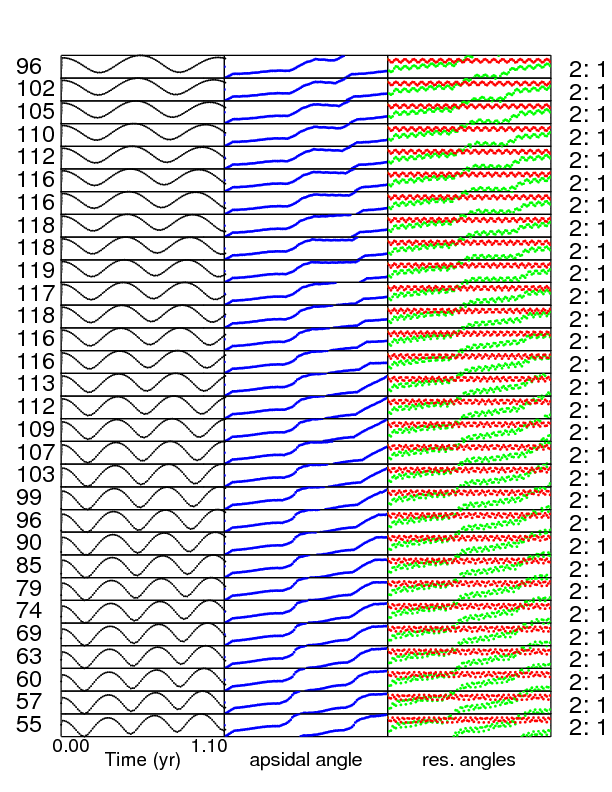}}\\[-10pt]
  \end{tabular}
}
%
  \put(-503,193){$a_o/a_i = 1.5812$}
  \put(-503,0){$a_o/a_i = 1.5865$}
  \put(-503,-193){$a_o/a_i = 1.5919$}
  \put(-485,270){$e_o = 0 \Longrightarrow$}
  \put(-495,100){$e_o = e_H/2 \Longrightarrow$}
  \put(-485,80){$e_o = 0 \Longrightarrow$}
  \put(-495,-90){$e_o = e_H/2 \Longrightarrow$}
  \put(-485,-110){$e_o = 0 \Longrightarrow$}
  \put(-495,-280){$e_o = e_H/2 \Longrightarrow$}
  \put(-10,270){$\Longleftarrow e_o = e_H/2$}
  \put(-10,100){$\Longleftarrow e_o = e_H$}
  \put(-10,80){$\Longleftarrow e_o = e_H/2$}
  \put(-10,-90){$\Longleftarrow e_o = e_H$}
  \put(-10,-110){$\Longleftarrow e_o = e_H/2$}
  \put(-10,-280){$\Longleftarrow e_o = e_H$}
  \put(-395,275){TTV}
  \put(-345,275){$\varpi_o - \varpi_i$}
  \put(-272,275){$\phi$}
  \put(-178,275){TTV}
  \put(-132,275){$\varpi_o - \varpi_i$}
  \put(-58,275){$\phi$}
%
%
  \caption{Transit curves, apsidal angle evolution, and resonant angle ($\phi$) evolution over 1.1 yr ($N \approx 100)$ close to the $2$:$1$ PC for $\Pi_i = \varpi_i = \varpi_o = 0^{\circ}$ and $\Pi_o = 60^{\circ}$.  All x-axes are in time.  The values of $a_o/a_i$ for the upper, middle and lower panels are $1.5812$, $1.5865$ and $1.5919$, respectively. In a given pair of horizontal panels, $e_o$ is incremented uniformly from $0$ to $e_H$ top to bottom, left to right.  The RMS amplitude is given in seconds besides each transit plot (whose range is appropriately scaled), and the range for the angle plots are [$0,180^{\circ}$]. The red curve plots Eq. (\ref{resangle}) and the green the other $2$:$1$ PC angle.  The blue curve plots the apsidal angle.  Note the difference in the upper, middle and lower panels of the TTV amplitude, which can vary by several factors despite the small ($< 0.006$ AU) difference between $a_o$ values, and the regions where blue, red and green curves all librate (for ACR).}
  \label{MMR2to1}
\end{figure*}

The red curve librates for all 180 systems in the figure, demonstrating that all these planets are locked in MMR, at least for about 1 yr.  At the lowest values of $e_o$, this libration occurs about $0^{\circ}$, and is hence said to be ``symmetric'' and ``aligned''.  For other values of $e_o$, the libration is about $45^{\circ}$ and is hence said to be ``antisymmetric''.  The green and blue curves librate for just a few systems at each $a_o$ value.  The libration of these angles means that the system is in Apsidal Corotation Resonance \citep[ACR;][]{beamic2003,feretal2003,lee2004,kleetal2005,beaetal2006,micetal2006a,voyhad2006,sanetal2007,micetal2008a,micetal2008b}.  The location of these systems occurs at $\approx 0.2 e_H$, but does vary sensitively on the value of $a_o$.  The $2$:$1$ MMR is the only resonance studied which features a librating apsidal angle.

The primary difference between the upper, middle and lower panels are the amplitudes of the TTV signals, which can vary by several factors despite the small ($< 0.006$ AU) difference between $a_o$ values. The highest signals (with $> 100$s) appear in two eccentricity groupings in the upper and middle panels, and just one eccentricity grouping in the lower panels.  The first group of high signals in the upper and middle panels ends when the libration of the resonant angle becomes antisymmetric and the TTV signal qualitatively changes character.  The gap between groups of high signals for the upper panels is five times smaller than that for the middle panels.  For $0.5 < e_o < e_H$, the libration curves, TTV signal profiles, and TTV amplitudes are relatively insensitive (varying by a few percent) to changes in $a_o$, unlike for $0 < e_o < 0.5 e_H$.  In this low-eccentricity regime, the highest amplitudes seen in the middle panels occurs at ACR; the same is not true for the lower or upper panels.  In none of those panels does the TTV signal profile noticeably change at ACR.

Unlike for the $2$:$1$ commensurability, other $p$:$q$ commensurabilities feature systems that are not in resonance, a situation corresponding to a circulation of the red angle.  For all $q=1$, and $p=2, ... 11$, only the $6$:$1$ PC contains no MMRs.  Systems also appear in MMRs for the $3$:$2$, $5$:$2$, $7$:$2$, $9$:$2$, $7$:$3$, $8$:$3$, $10$:$3$, and $11$:$3$ PCs, but none for the $5$:$3$ PC.  However, recall importantly that these results hold for a narrow region of phase space where $\Pi_i = \varpi_o = 0^{\circ}$  and $\Pi_o = 60^{\circ}$ and for just one set of masses.

As $e_o$ increases, the transition between non-resonant and resonant systems produces a sudden change in TTV signal amplitude (by a factor of at least several) for all commensurabilities where this transition can be seen.  Therefore, TTV signals could clearly distinguish resonant from non-resonant systems, even if both are within the libration width of a particular commensurability.  Further, the transition between symmetric and antisymmetric libration corresponds to a sudden (by a factor of at least 2) change in TTV signal in several cases around the $2$:$1$, $3$:$1$, $4$:$1$, $5$:$1$, $7$:$1$, $8$:$1$ and $10$:$1$ PCs.  The signal jump caused by this shift in libration center is typically not as great as the jump caused by the transition to resonance from circulation.  For $0.75 e_H \le e_o \le e_H$, only the $2$:$1$, $5$:$1$, $7$:$1$, $3$:$2$ and $9$:$2$ PC systems are in MMR.  At the other commensurabilities, the resonant angle varies in a chaotic fashion, a qualitatively different behavior from the smooth circulation (featuring a continuous curve) found at non-resonant lower-eccentricity regimes.

\section{Correlation with TTV Signal Shape}

In previous sections, we have focused our investigations on TTV signal amplitudes.  Their importance stems from the ability of observers to distinguish a TTV curve from measurement uncertainty.  As we have demonstrated, the TTV signal amplitude is highly sensitive to the orbital parameters of both planets, and is of limited utility when attempting to identify an unseen planet.  In order to help break the degeneracy, one can employ the {\it shape} of the TTV curve.  As seen in Fig. \ref{sample}, TTV curves may take on a variety of forms.  By incorporating this form into a TTV analysis, one may be able to better pinpoint the mass and orbital parameters of the unseen planet.

A quantifiable way of obtaining TTV shape data is to consider the autocorrelation function $\mathcal{A}$ at a given (time) lag $L$:

\begin{equation}
\mathcal{A}_L = \frac{ \sum_{k=1}^{N-L} x_k  x_{k+L}  }
                     { \sum_{k=1}^{N}  x_{k}^2  }
\end{equation}

\noindent{where} $x_k$ represents the deviation of the $k$th transit time from the constant period model (to be distinguished from $S({\bf Q})$ at $N=k$, which represents the RMS amplitude from the first $k$th observations).  The function $\mathcal{A}_L$ is bounded by $[-1,1]$ and can be computed solely from the transit observation data alone.  In principle, $\mathcal{A}_L$ provides $N$ constraints, in addition to $S({\bf Q})$, on the mass and orbital parameters of the unseen planet.  

We find that the autocorrelation function may provide a useful summary of the TTV signal shape (see Figs. \ref{LagCont}-\ref{LagAngSec}).  Figure \ref{LagCont} plots the autocorrelation as contours for 9 different values of $L$ (2, 3, 4, 5, 6, 20, 50, 100, 200) for $N = 313$ ($\approx 3.5$ yrs) and $\Pi_o = \Pi_i = \varpi_o = 0^{\circ}$.  One notes immediately that although some ``flame''-like features from Fig. \ref{fiducial} are apparent for low ($L \le 6$) and high ($L = 200$) lags, the contours generally sculpt the phase space differently than does $S({\bf Q})$.  This difference is most apparent in the low eccentricity ($e_o \le 0.2$) regimes, which demonstrate rich structure in autocorrelation space.  For $L=100$, sharp ``flames'' puncture this regime for initial semimajor axis ratios of up to $\approx 4$.  This low-eccentricity regime is transformed drastically as one increases $L$ from $L=2$ to $L=6$.  Note importantly that a purple flume (with $\mathcal{A}_L$ close to unity) is almost precisely centered on the $L$:$1$ PC for $L =2, 3, 4, 5$, and $6$ (corresponding to $a_o/a_i = 1.59, 2.08, 2.52, 2.92, 3.30$ and $3.66$).  Such a correlation is a promising sign of the utility of $\mathcal{A}_L$ in helping to diagnose plausible integrations for future study. 

\begin{figure*}
  \centering
\resizebox{6in}{!} {
  \begin{tabular}{ccc ccc ccc}
    \multicolumn{3}{c}{\includegraphics[height=0.35\textheight,width=0.33\textwidth]{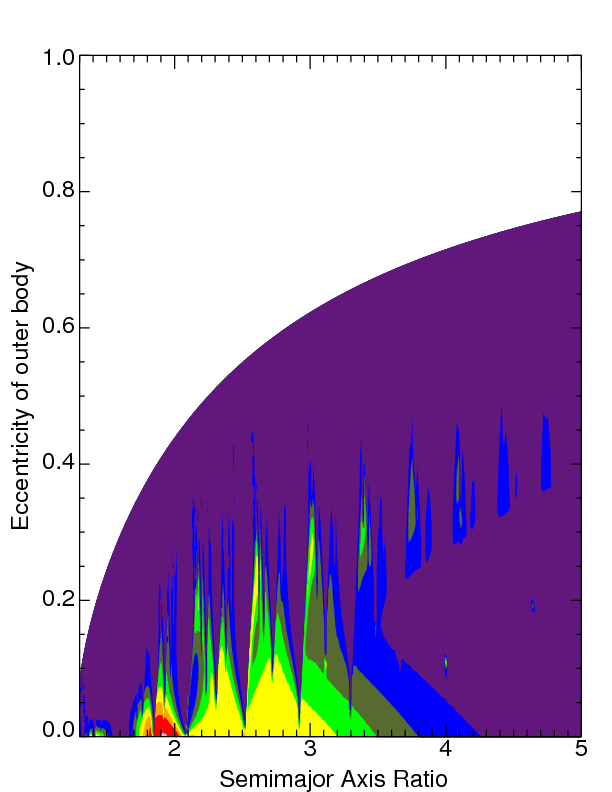}}&
    \multicolumn{3}{c}{\includegraphics[height=0.35\textheight,width=0.33\textwidth]{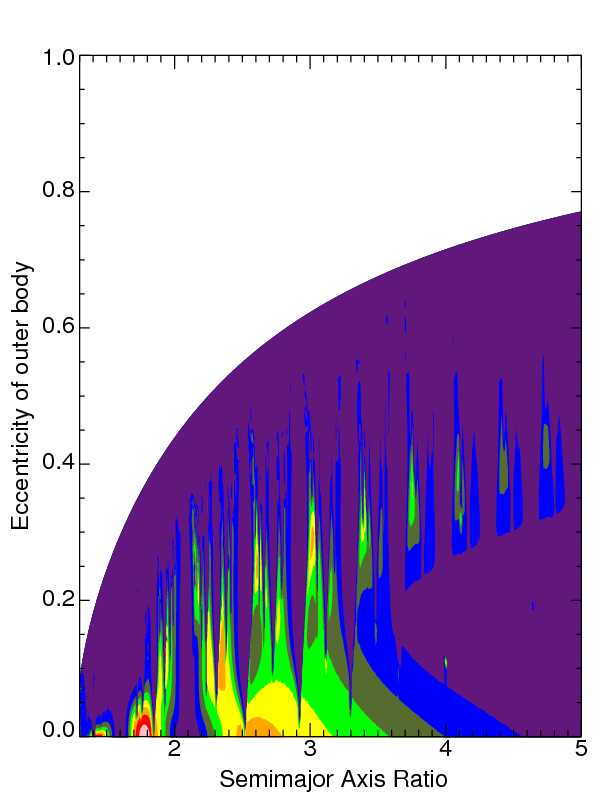}}&
    \multicolumn{3}{c}{\includegraphics[height=0.35\textheight,width=0.33\textwidth]{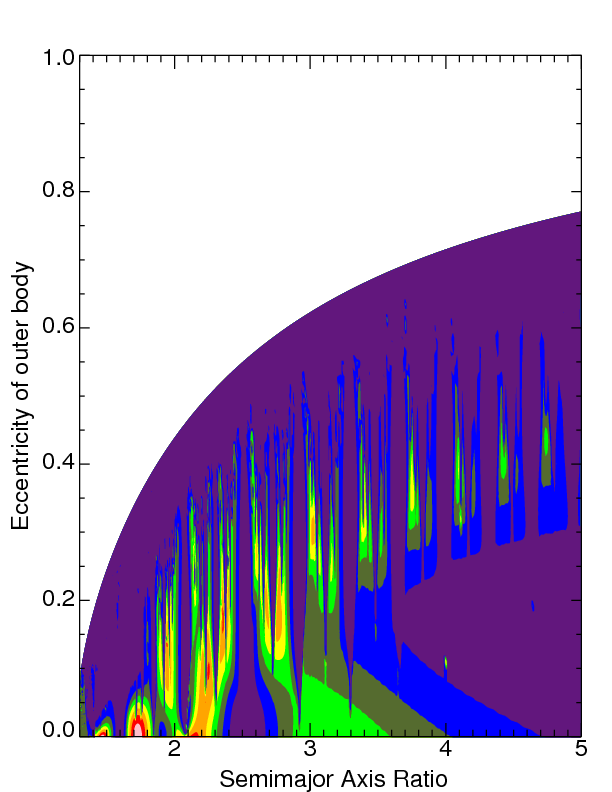}}\\[-10pt]
    \multicolumn{3}{c}{\includegraphics[height=0.35\textheight,width=0.33\textwidth]{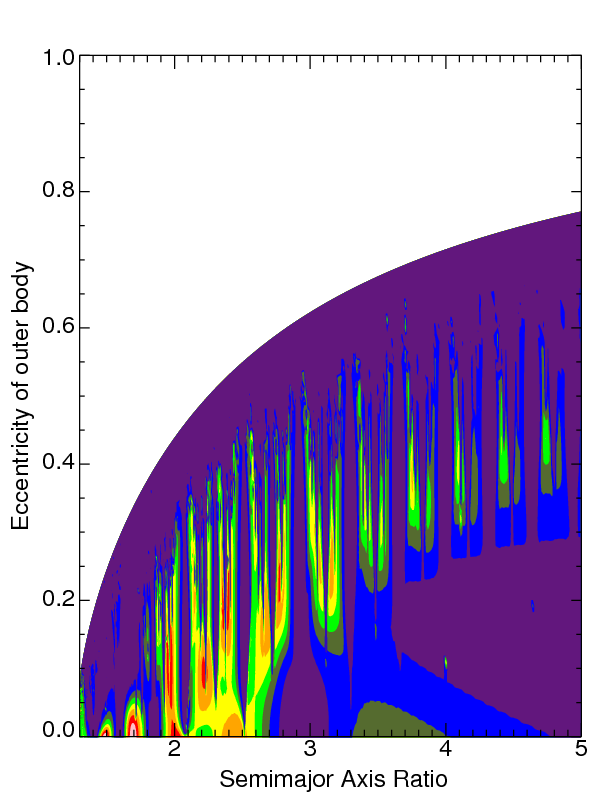}}&
    \multicolumn{3}{c}{\includegraphics[height=0.35\textheight,width=0.33\textwidth]{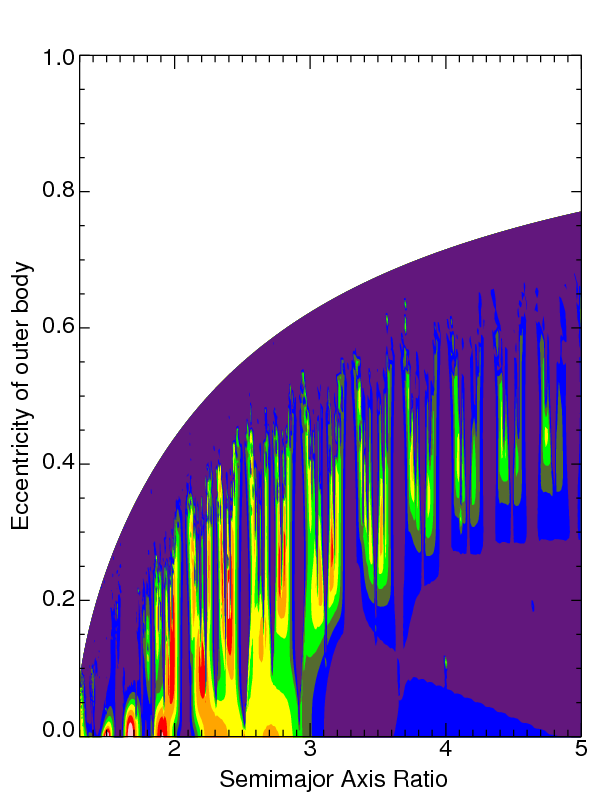}}&
    \multicolumn{3}{c}{\includegraphics[height=0.35\textheight,width=0.33\textwidth]{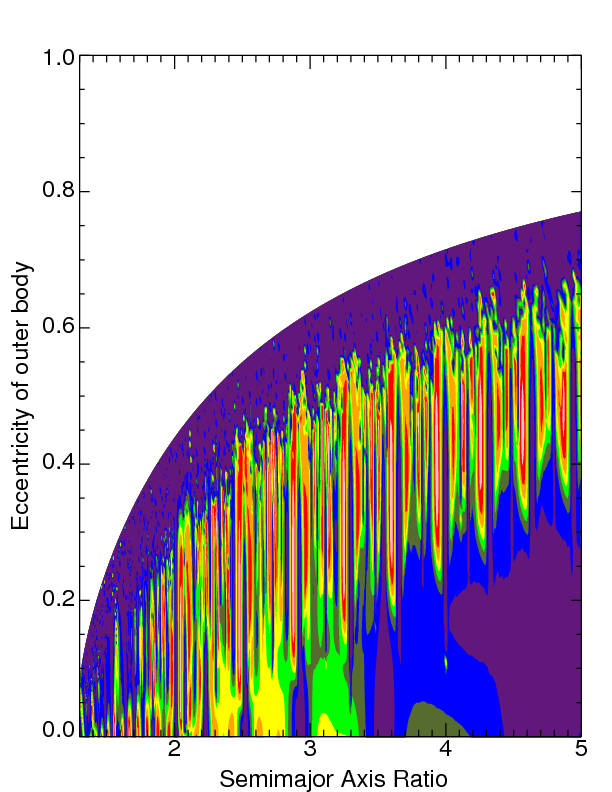}}\\[-10pt]
    \multicolumn{3}{c}{\includegraphics[height=0.35\textheight,width=0.33\textwidth]{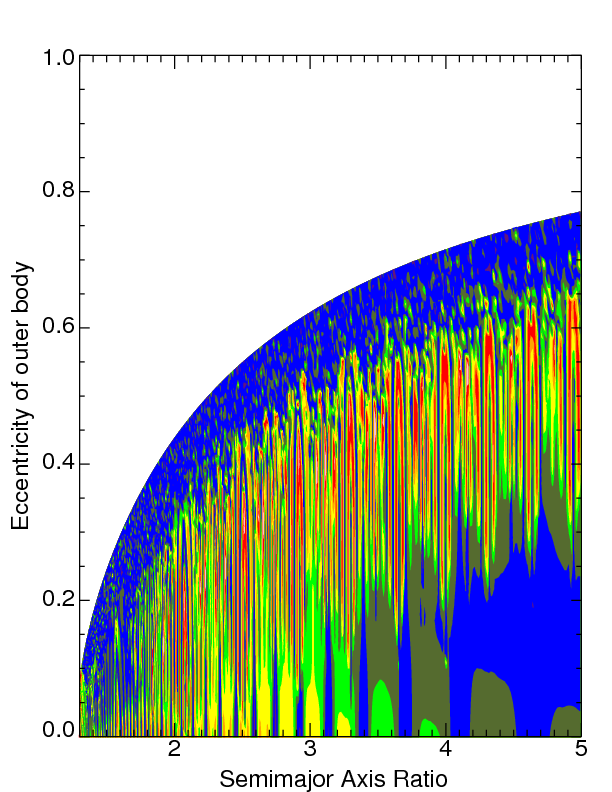}}&
    \multicolumn{3}{c}{\includegraphics[height=0.35\textheight,width=0.33\textwidth]{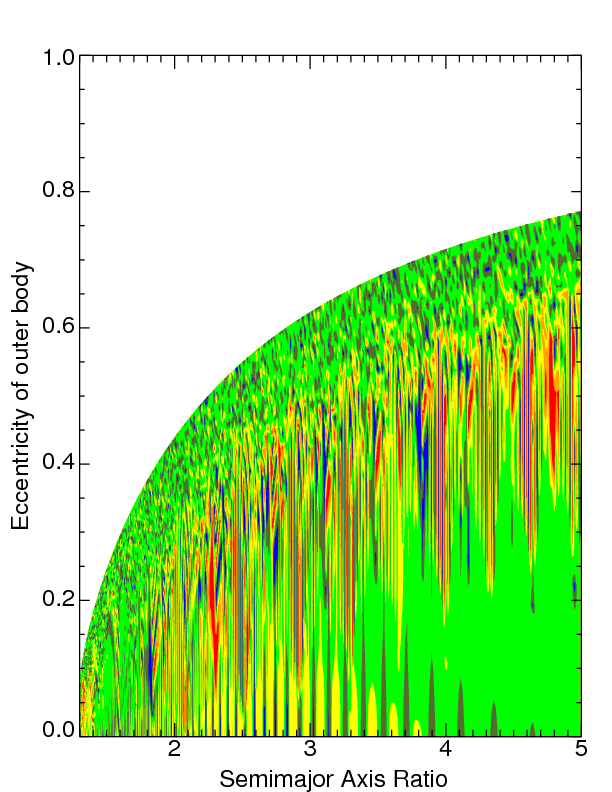}}&
    \multicolumn{3}{c}{\includegraphics[height=0.35\textheight,width=0.33\textwidth]{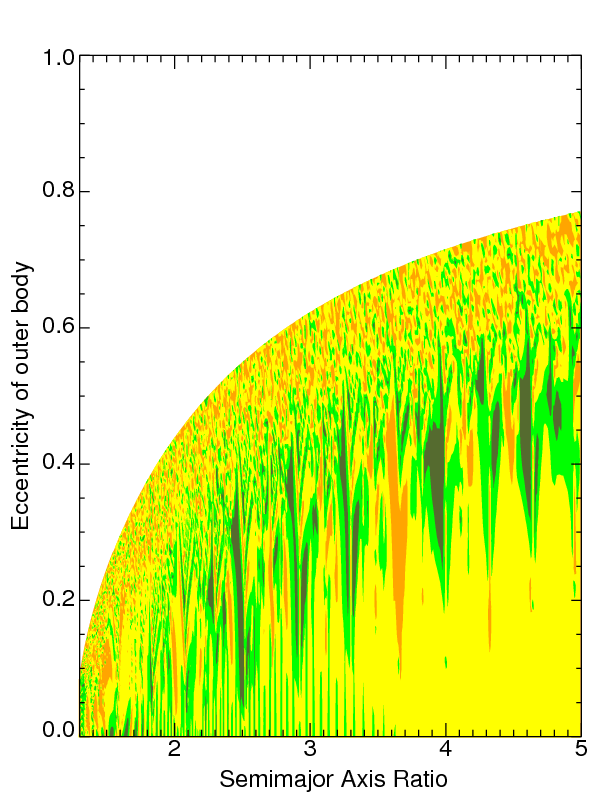}}\\[-10pt]
  \end{tabular}
}
  \put(-520,210){\includegraphics[trim = 0mm 0mm 0mm 0mm, clip, width=0.23\textwidth]{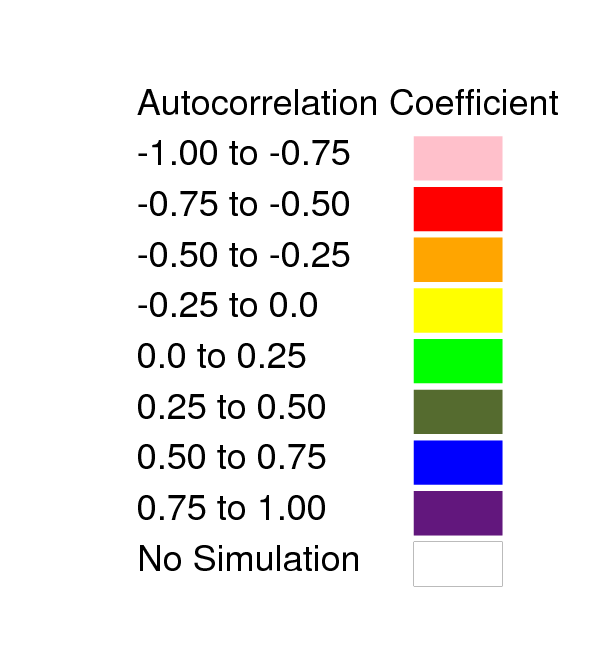}}
     \put(-350,238) {\textbf{\large L=2}}
     \put(-206,238) {\textbf{\large L=3}}
     \put(-63, 238) {\textbf{\large L=4}}
     \put(-350,60) {\textbf{\large L=5}}
     \put(-206,60) {\textbf{\large L=6}}
     \put(-63, 60) {\textbf{\large L=20}}
     \put(-350,-118) {\textbf{\large L=50}}
     \put(-206,-118) {\textbf{\large L=100}}
     \put(-63,-118) {\textbf{\large L=200}}
%
%
  \caption{The autocorrelation function $\mathcal{A}_L$ at lags corresponding to $L = 2, 3, 4, 5, 6, 20, 50, 100$ and $200$ from left to right, top to bottom, for $\Pi_i = \Pi_o = \varpi_i = \varpi_o = 0^{\circ}$ and $N = 313$  ($\approx 3.5$ yrs).  The contours are pink ($-1.0$ to $-0.75$), red ($-0.75$ to $-0.50$), orange ($-0.5$ to $-0.25$), yellow ($-0.25$ to $0$), light green ($0$ to $0.25$), olive ($0.25$ to $0.5$), blue ($0.5$ to $0.75$), and purple ($0.75$ to $1.0$).  Note that each plot represents an independent measure on the TTV signal, and the first few lags trace out the locations of PCs.}
  \label{LagCont}
\end{figure*}

The above contours fixed the initial orbital angles, which we have shown to sensitively affect the TTV signal amplitude. When we instead fix the lag, e.g. at $L=3$, and vary $\Pi_o$ as in Fig. \ref{9angles}, then at the contour resolution of Fig. \ref{LagCont}, the change in initial angles appear to have no effect on the value of $\mathcal{A}_L$.  However, at finer resolutions, there is a variation.  For some systems, this variation in $\mathcal{A}_L$ strongly mimics the variation in $S({\bf Q})$ as $\Pi_o$ is changed.  Fig. \ref{LagAngSec} displays three secular systems: the system (XVII from Table \ref{regimes}) in the upper panel with $L=3$ shows remarkable agreement in the shapes and relative extrema of the solid, dotted and dashed curves (corresponding to $\varpi_o = 0^{\circ}$, $\varpi_o = 90^{\circ}$ and $\varpi_o = 180^{\circ}$) to the signal from the bottom left panel of Fig. \ref{lamlines}.  In contrast, the middle panels, which feature secular system XVIII, does not share the same agreement with the corresponding signal (not shown).  In between these two extremes, the bottom panels, featuring secular system XIX, illustrate different levels of correlation to the signal variation (not shown): strong for the dashed curve, moderate for the dotted curve, and none for the solid curve.  Therefore, the RMS amplitude and autocorrelation value might be strongly or weakly correlated depending on the particular orbital configuration.  The right panels of Fig. \ref{LagAngSec} demonstrate how the shape and magnitude of the $\mathcal{A}_L$ curves can qualitatively change for two different (arbitrarily chosen) values of $L$.

\begin{figure*}
  \centering
\resizebox{6in}{!} {
  \begin{tabular}{cc cc cc}
    \multicolumn{3}{c}{\includegraphics[height=3.0in,width=0.5\textwidth]{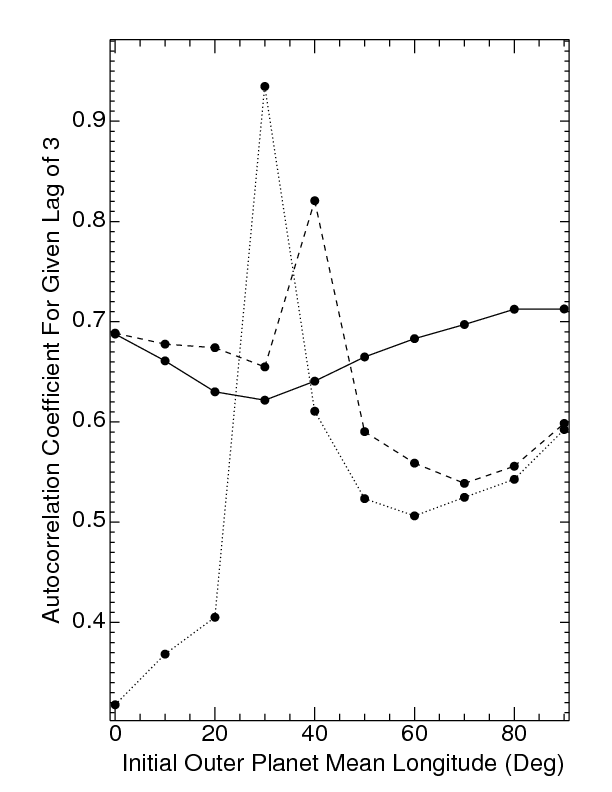}}&
    \multicolumn{3}{c}{\includegraphics[height=3.0in,width=0.5\textwidth]{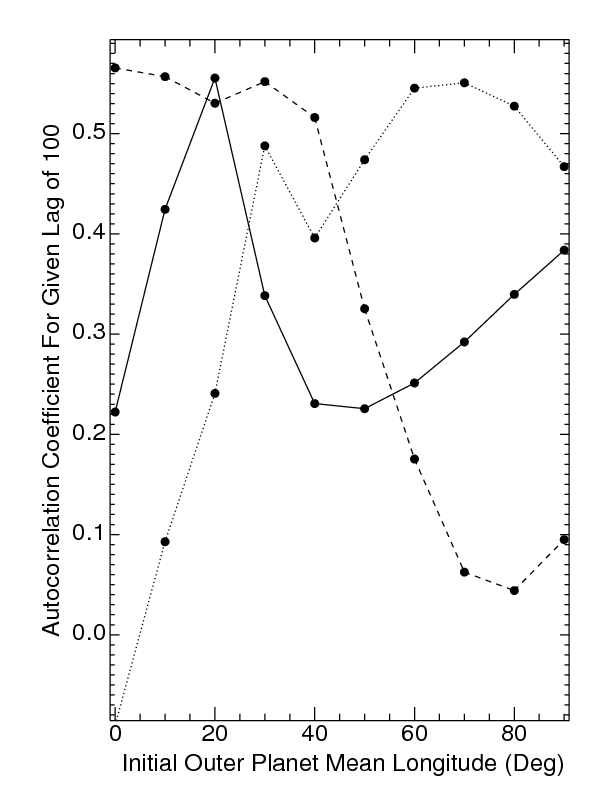}}\\[-10pt]
    \multicolumn{3}{c}{\includegraphics[height=3.0in,width=0.5\textwidth]{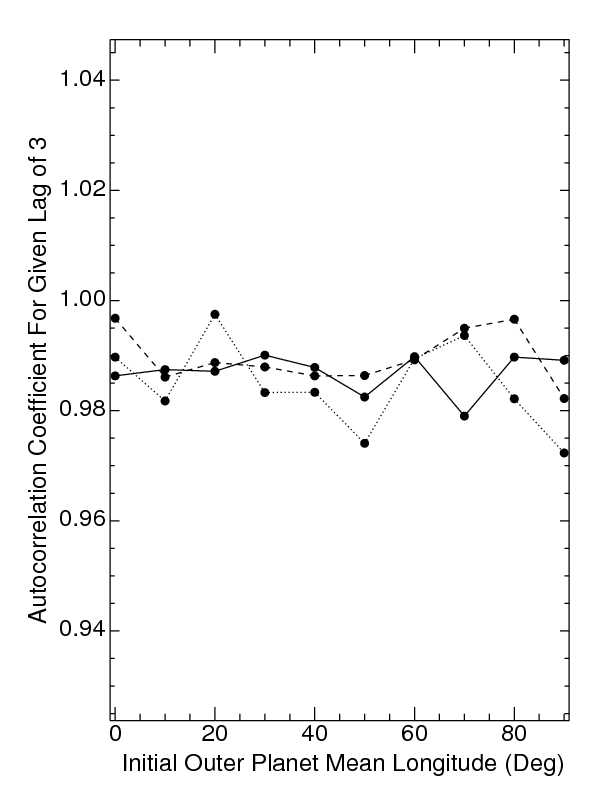}}&
    \multicolumn{3}{c}{\includegraphics[height=3.0in,width=0.5\textwidth]{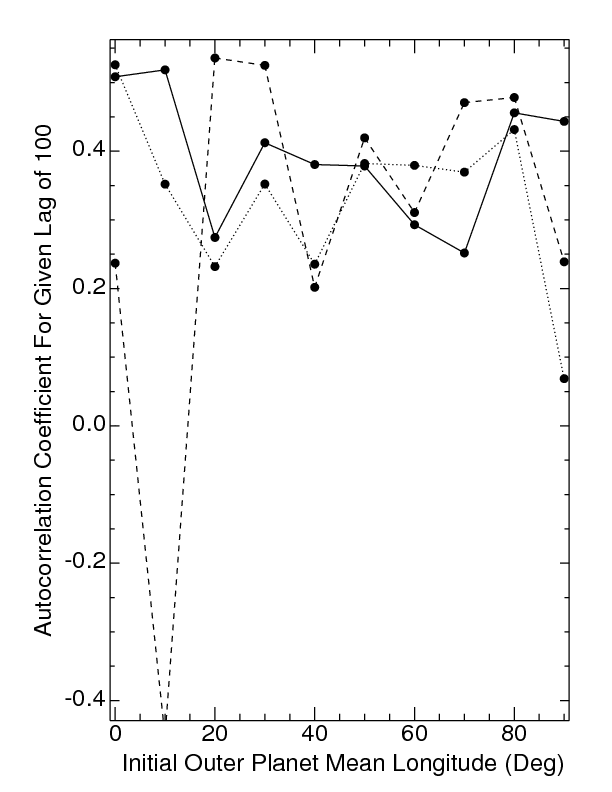}}\\[-10pt]
    \multicolumn{3}{c}{\includegraphics[height=3.0in,width=0.5\textwidth]{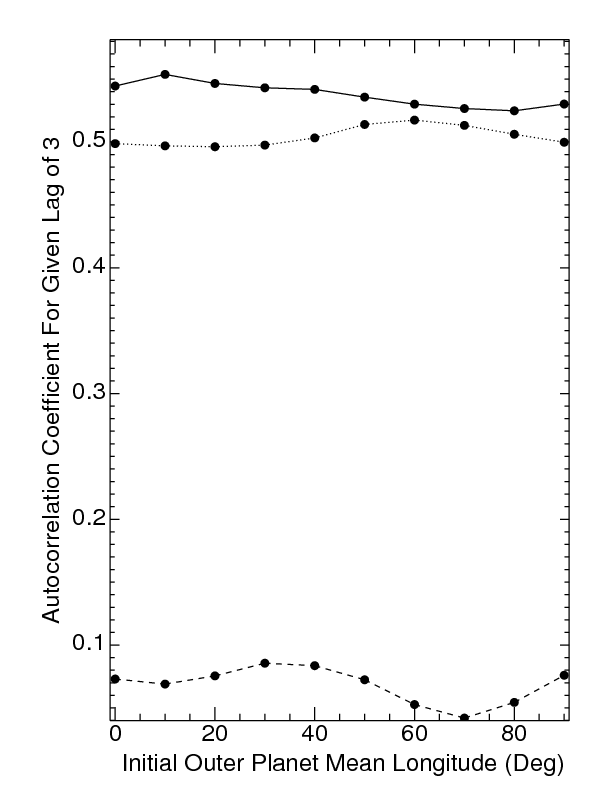}}&
    \multicolumn{3}{c}{\includegraphics[height=3.0in,width=0.5\textwidth]{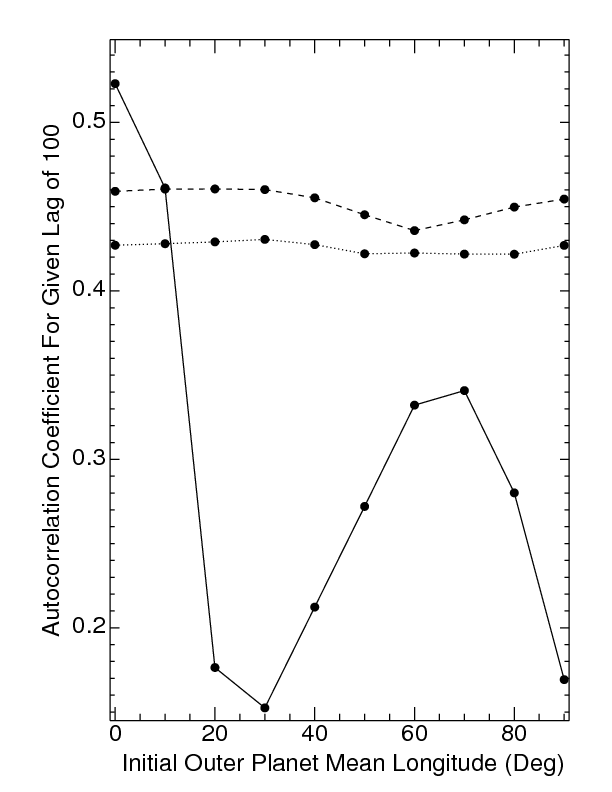}}\\[-10pt]
  \end{tabular}
}
%
  \put(-330,140){{\large $a_o/a_i=1.5$}}
  \put(-330,128){{\large $e_o=0.035$}}
  \put(-150,140){{\large $a_o/a_i=1.7$}}
  \put(-150,128){{\large $e_o=0.28$}}
  \put(-330,-30){{\large $a_o/a_i=1.8$}}
  \put(-330,-42){{\large $e_o=0.10$}}
  \put(-120,-30){{\large $a_o/a_i=1.5$}}
  \put(-120,-42){{\large $e_o=0.035$}}
  \put(-330,-200){{\large $a_o/a_i=1.7$}}
  \put(-330,-212){{\large $e_o=0.28$}}
  \put(-140,-172){{\large $a_o/a_i=1.8$}}
  \put(-140,-184){{\large $e_o=0.10$}}
%
%
  \caption{Autocorrelation $\mathcal{A}_3$ (left panels) and $\mathcal{A}_{100}$ (right panels) for the three secular systems XVII (upper panels), XVIII (middle panels) and XIX (lower panels) from Table \ref{regimes}.  In each plot, the solid, dotted and dashed curves correspond to $\varpi_o = 0^{\circ}$, $\varpi_o = 90^{\circ}$ and $\varpi_o = 180^{\circ}$.  Here, $N=874$. Note the similarity in form of the $\mathcal{A}_3$ for system XVII and its RMS amplitude (bottom left curve of Fig. \ref{lamlines}), and that when $L$ is increased to $100$, $\mathcal{A}_L$ changes both form and magnitude for most secular systems.
 }
\label{LagAngSec}
\end{figure*}


One may consider how $\mathcal{A}_L$ varies with planetary mass as well as initial orbital angles.  In many cases, we do find a correlation between the variation of $\mathcal{A}_L$ and $S({\bf Q})$ with $M_o$, even if the agreement fails to come close to that from some orbital angle regimes.  The variations induced by changing $M_o$ for a given $L$ tend to be greater by an order of magnitude than those induced from orbital angles alone.  In a sense, the variation caused by the angles is a modulation to those caused by the masses.


These preliminary results suggest a promising avenue in which to pursue the inverse TTV problem, particularly if additional effects discussed in Section \ref{trandisc} are coupled with amplitude and shape data.  Future work will entail identifying the critical lags which produce the autocorrelation extrema, quantifying how low ($L < 10$) lags track $L:1$ PCs, and incorporating $S({\bf Q})$ values and $\mathcal{A}_L$ for $N-1$ values of $L$ all as independent parameters to help remove the degeneracy of the inverse TTV problem.

\section{Discussion} \label{trandisc}

\subsection{Lagrange Unstable Systems}

The highest signature TTV amplitudes ($S({\bf Q}) \gtrsim 1000$s), like those generated from the curve in the fourth panel of Fig. \ref{sample}, might be the result of erratic changes in the orbital parameters of the planets.  First, we confirmed that the highest eccentricity systems simulated for this work are provably Hill stable.  In principle, such systems might be manifestly unstable even though they satisfy Hill Stability.  Although the planetary orbits will never cross in a Hill Stable system, the outer planet might get ejected because the outer planet is not bounded, or is Lagrange Unstable.  If this is the case, then the timescale for this ejection is expected to be orders of magnitude less than the lifetime of the system, so observing planets in such configurations would be rare.

We attempt to place an upper limit on the fraction of systems which exhibit this manifestly unstable behavior by investigating configurations which produce sudden time variations of orbital elements such as the semimajor axis, periastron and apastron, mean longitude and longitude of pericenter.  Such estimates may be used as a Lagrange stability filter in additional statistical studies of TTVs.  We find that comparing the maximum semimajor axis difference or ratio of both planets over the length of the simulation with their initial values provides a useful diagnostic.  In particular, we define

\begin{equation}
\chi
\equiv
\frac{{\rm max}\left(a_o(t) - a_i(t)\right)}
{a_o - a_i}
.
\label{filter}
\end{equation}

\noindent{}Therefore we compute $\chi$ for every one of the 120,000 simulations studied in our fiducial configuration of Figure. \ref{fiducial}, plus in selected configurations from Figure \ref{masscont} representing extreme values of the transiting and exterior planet masses.   We report the percent of all systems with $\chi$ less than given values from 1.1 to 10.0 in the left panel of Figure \ref{chivalues}.  The figure demonstrates that the large majority of Hill Stable systems have planets which do not vary their semimajor axis difference by more than 10\%; in these relatively quiescent systems, the outer planet is unlikely to be ejected.  All but a few systems simulated with a transiting planet of $0.1 M_J$ exhibit this quiescent behavior.  Alternatively, systems with a more massive transiting planet are the most likely to exhibit instability.  Nearly 20\% of all systems sampled with a $10 M_J$ transiting planet exhibit changes in their semimajor axis ratio approaching 1000\%.  Such systems could become unstable (in the sense of the outer planet being ejected) on timescales much shorter than the system lifetime. 

In order to help determine what Hill-stable initial conditions most likely generate this suggested Lagrange instability, one can consider the initial eccentricity of the outer planet.  We discover that highly eccentric outer planets close to the Hill Stability limit produce the sudden changes in semimajor axis characterized by $\chi \gtrsim 2.0$.  In order to quantify this finding, we plot in the right panel of Figure \ref{chivalues} the median (dots) and mean (squares) values of $e_o$ as a percent of $e_H$ which cause $\chi$ to be less than fixed values for the same mass configurations as in the left panel.  Note that these eccentricities correspond to the highest TTV signals ($S({\bf Q}) \gtrsim 1000$s; purple and blue regions in ``flames'' plots) in the fiducial case.  We caution that these results are statistical in nature.  The evolution of a particular systems is highly dependent on the initial orbital element configuration and will likely require long term ($\gg 10 $ yr) study in order to determine whether they will individually tend towards instability.  We find this sensitive dependency by noting that for $e_o = e_H$ and for given (initial) sets of $a_o/a_i$, altering initial sets of orbital angles can cause $\chi$ to vary from $\approx 1.1$ to $\approx 10.0$. 

\begin{figure*}
  \centering
  \begin{tabular}{cc}
    \multicolumn{1}{c}{\includegraphics[height=4.0in,width=0.46\textwidth]{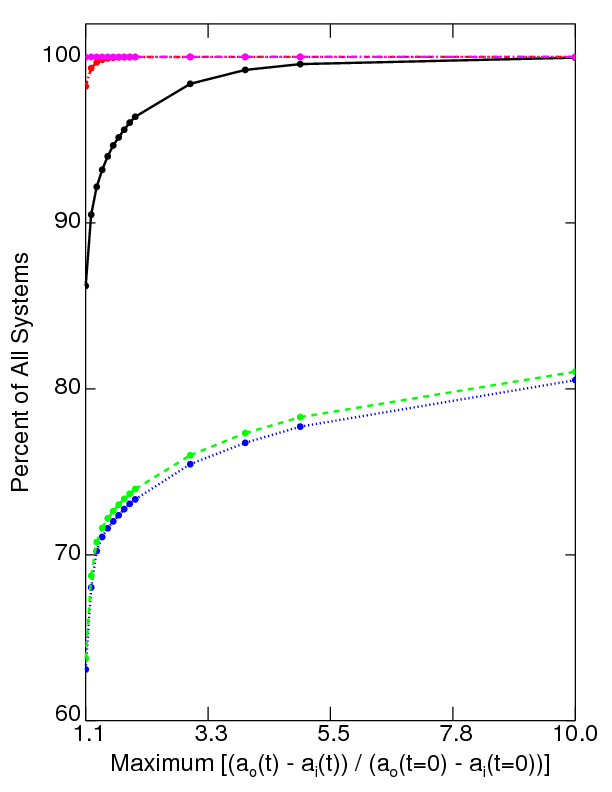}}&
    \multicolumn{1}{c}{\includegraphics[height=4.0in,width=0.46\textwidth]{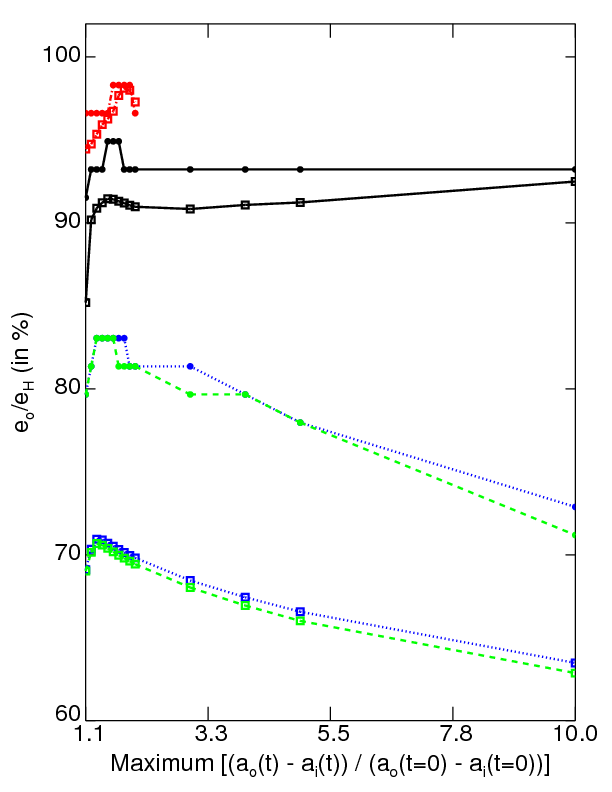}}
\\
  \end{tabular}
  \caption{Possible tracers of Lagrange Instability in Hill Stable systems.  The left panel displays the fraction of all 120,000 systems from Fig. \ref{fiducial} (solid black curves) as well as from the configurations in Fig. \ref{masscont} corresponding to ($M_i = 10 M_J, M_o = 1 M_{\oplus}$ ; blue dotted), ($M_i = 10 M_J, M_o = 50 M_{\oplus}$ ; green dashed), ($M_i = 0.1 M_J, M_o = 1 M_{\oplus}$ ; red dot dashed) and ($M_i = 0.1 M_J, M_o = 50 M_{\oplus}$ ; magenta triple-dot-dashed) whose value of $\chi \equiv {\rm max}\left(a_o(t) - a_i(t)\right)/{(a_o - a_i)}$ is less than the values on the X-axis.  The right panel shows similarly colored curves but for the median (dots) and mean (squares) values of $e_o/e_H$, expressed as a percent, at which $\chi$ exceeds the value on the X-axis.  The figure demonstrates that the fraction of Lagrange Unstable systems is likely to be at the few percent level for our fiducial simulations, and correspond to the purple and blue regions of the ``flames'' plots (Fig. \ref{fiducial}).
}
  \label{chivalues}
\end{figure*}

\subsection{Radial Velocity Follow-Up}

Ideally, the existence of unseen planetary companions suggested from TTVs can be confirmed with radial-velocity observations.  However, the low mass and orbital separation of such unseen perturbers may preclude them from radial-velocity detection (as well as observational limitations due to stellar properties).  In this section, we estimate the maximum magnitude of the radial-velocity semiamplitude produced by an external pertuber in a guaranteed-to-be-stable multi-planet exosystem with a hot Jupiter.  Because radial velocity semiamplitude $ \equiv K_o \propto 1/\sqrt{1 - e_{o}^2}$, a planet could most easily be detected by radial velocity surveys when its eccentricity is highest.  At the Hill Stability limit \citep{gladman1993},

\begin{eqnarray}
& &
\left( \mu_o + \mu_o \mu_i + \mu_i \right)^{-3}
\left(  \mu_o + \frac{\mu_i}{\alpha} \right)
\left(  \mu_i \sqrt{\alpha} + \mu_o \sqrt{1 - e_{o}^2} \right)^2
=
\nonumber
\\
& & 
1 + \frac{3^{4/3} \mu_o \mu_i}
{\left( \mu_o + \mu_i \right)^{4/3}}
-
\frac{\mu_o \mu_i \left( 11 \mu_i + 7 \mu_o \right)}
{ 3 \left( \mu_o + \mu_i \right)^2 }
.
\end{eqnarray}

\noindent{where} $\alpha \equiv a_i/a_o$, 
$\mu_i \equiv M_i/M_{\star}$ and $\mu_o \equiv M_o/M_{\star}  $.
Further, the radial velocity semiamplitude of the outer planet
in the coplanar edge-on case is \citep{leepea2003}:

\begin{equation}
K_{o,max} = \left( \frac{2 \pi G}{P_o} \right)^{1/3}
\frac{M_o}{\left( M_{\star} + M_o + M_i \right)^{2/3} }
\frac{1}{\sqrt{1 - e_{o}^2}},
\end{equation}

\begin{equation}
P_o = \frac{2 \pi a_{o}^{3/2}}
{G \left( M_{\star} + M_i + M_o  \right)}
.
\end{equation}

Therefore, the maximum semiamplitude of the external perturber, assuming $M_{\star} \gg M_i, M_o$, is

\begin{equation}
K_o \approx
\frac{\sqrt{\frac{G M_{\star}}{a_o}} \mu_{o}^2}
{
\left( \mu_i + \mu_o \right)^{3/2}
\left( \frac{\mu_i}{\alpha} + \mu_o \right)^{-1/2}
- \mu_i \sqrt{\alpha}
}
.
\label{radveleq}
\end{equation}

This limiting equation,  plotted in Fig. \ref{rv}, assumes that 1) the system is seen edge-on and 2) $e_o = e_H$.  The figure, which displays curves for given masses of $1 M_{\oplus}, 5 M_{\oplus}, 10 M_{\oplus}, 20 M_{\oplus}$, $50 M_{\oplus}$ and $1 M_{{\rm Saturn}}$ (in ascending order of the curves on the plot), demonstrates that given a hot Jupiter at $0.05$ AU around a Solar-like star, $K_o$ is a stronger function of $M_o$ than of $\alpha$. A single Earth-Mass external perturber is not yet detectable by radial velocity measurements, but a Super-Earth may be detectable by this means, independent of planetary separation.  Once detected by radial velocity measurements, those observations can be combined with TTVs in order to help characterize the  architecture of the system \citep{marlau2010}.

\begin{figure*}
  \centering
  \begin{tabular}{c}
    \multicolumn{1}{c}{\includegraphics[width=1.00\textwidth]{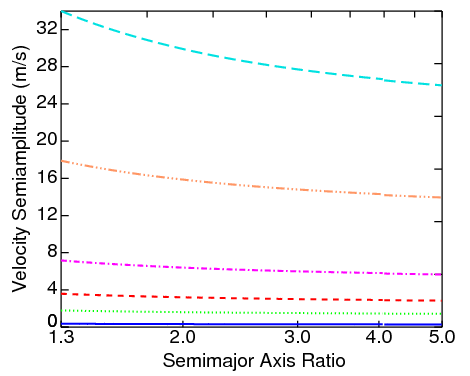}}
\\
  \end{tabular}
%
%
%
  \caption{The maximum semiamplitude of the radial velocity signature of external perturbers of masses equaling
$1 M_{\oplus}, 5 M_{\oplus}, 10 M_{\oplus}, 20 M_{\oplus}$ and $50 M_{\oplus}$ and $1 M_{{\rm Saturn}}$, corresponding to the solid/blue, dotted/green, dashed/red, dot-dashed/magenta, triple dot-dashed/orange and long dashed/aqua curves, respectively.  Note that, in the TTV regime studied in this work, a single Earth-Mass external perturber is not yet detectable by radial velocity measurements, but a Super-Earth may be detectable by this means.}
  \label{rv}
\end{figure*}

\subsection{Light Travel Time}

The variation in Light Travel Time (LTT) during each transit may exceed the previously cited detectability threshold of 10 s depending on the planetary masses and orbital configuration. Therefore, TTV models might need to incorporate leading-order LTT effects in order to correctly describe the motion.  All our simulations incorporate this effect.  Here we show where the effect might be important.  In order to help determine when LTT effects may be neglected, we plot this contribution in Fig. \ref{LTT}.  Note the domain of the plot is three times as wide as those from Figs. \ref{fiducial} and \ref{minmax}, but still sampled with 400 points along the X-axis.  Figure \ref{LTT} quantifies the contribution of LTT effects for $N=874$ (left panels) and $N=50$ (right panel).  The contours plot the quantity $\log{(S({\bf Q})/T({\bf Q}))}$, where $T({\bf Q})$ is the TTV RMS amplitude produced by LTT effects alone.  The figure demonstrates that the LTT's contribution for $M_o = M_{\oplus}$ for an idealized $10$ yrs of coverage can be at the few percent to tens of percent level, but does not dominate the TTV signal.  However, for a more massive external perturber, LTT can dominate the signal, especially for hierarchical (widely separated) systems with an outer planet eccentricity up to $0.5 e_H$.  

\begin{figure*}
  \centering
  \begin{tabular}{cccc}
    \multicolumn{2}{c}{\includegraphics[width=0.5\textwidth,height=0.36\textheight]{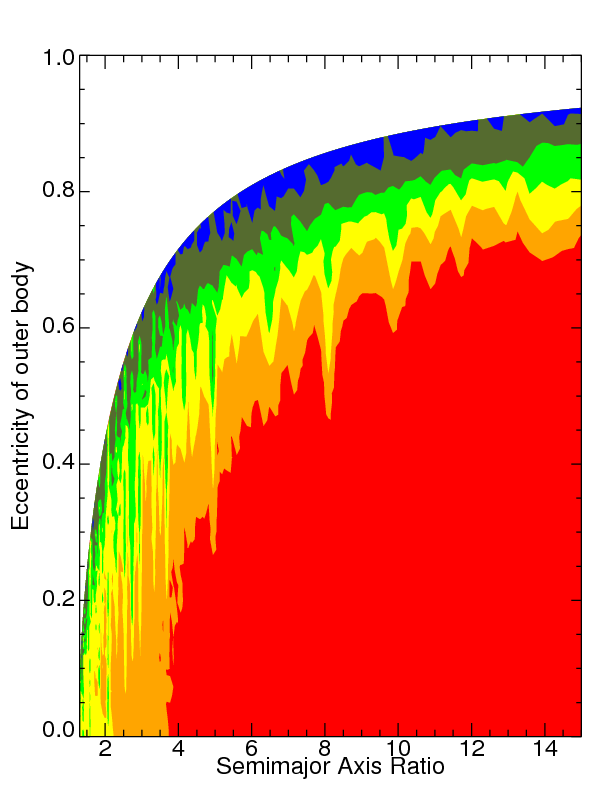}}&
    \multicolumn{2}{c}{\includegraphics[width=0.5\textwidth,height=0.36\textheight]{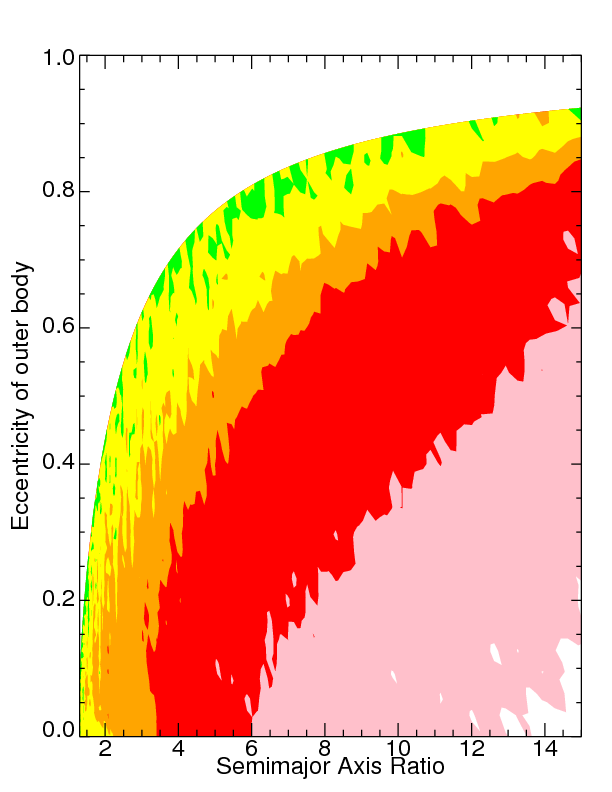}}\\
    \multicolumn{2}{c}{\includegraphics[width=0.5\textwidth,height=0.36\textheight]{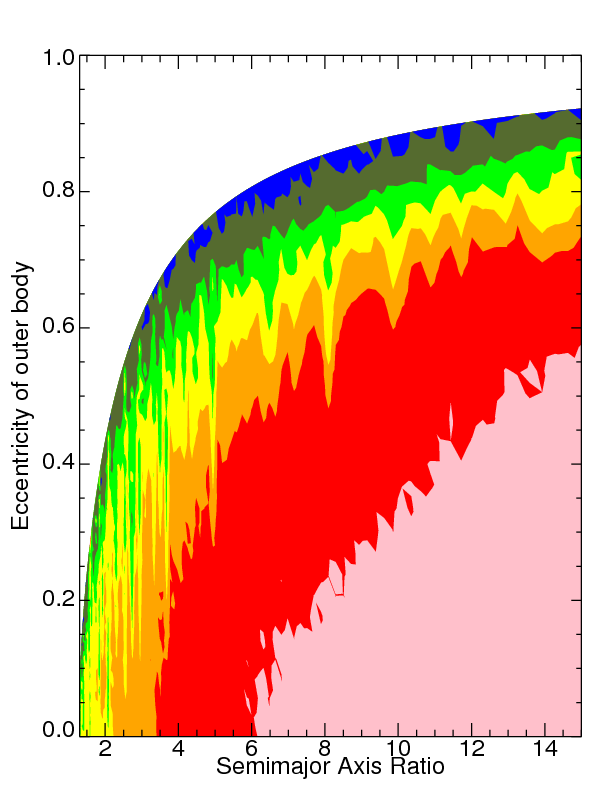}}&
    \multicolumn{2}{c}{\includegraphics[width=0.5\textwidth,height=0.36\textheight]{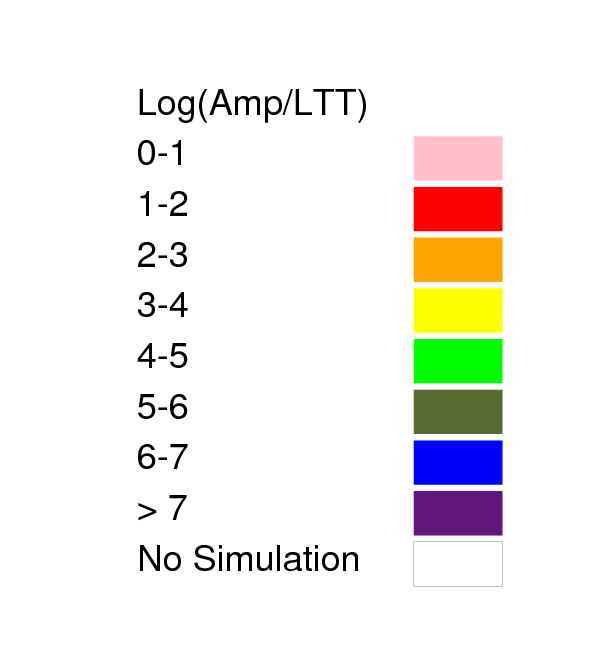}}\\
  \end{tabular}
%
  \put(-450,233){{\large $N = 874$}}
  \put(-450,221){{\large $M_o = M_{\oplus}$}}
  \put(-200,233){{\large $N = 50$}}
  \put(-200,221){{\large $M_o = M_{\oplus}$}}
  \put(-450,-29){{\large $N = 874$}}
  \put(-450,-41){{\large $M_o = 10 M_{\oplus}$}}
%
%
  \caption{Median log of the ratio of RMS amplitude to the contribution from Light Travel Time (LTT) for
5 different initial orbital configurations for 874 (left panels) and 50 (right panel) consecutive transits, and for an outer planet mass of one (upper panels) or ten (lower panel) Earth masses.  Note that LTT makes a contribution of at least a few percent in most areas of phase space explored here, and dominates the most hierarchical and lowest eccentricity secular regime for Super-Earth external perturbers.}
  \label{LTT}
\end{figure*}


\subsection{Extensions}

This study has attempted to thoroughly quantify the TTV properties in an observationally relevant region of phase space.  However, additional regions of interest exist, and future studies could help characterize alternate architectures.  Motivated by the recent discoveries of retrograde transiting planets, \cite{payetal2010} complements this work by considering the nonzero relative inclinations between two planets, and with a transiting planet that can be either internal or external to the additional perturber in the system.  Subsequent studies aim to characterize the prospects for constraining the frequency of multiple stellar systems among planet-hosting stars with TTVs \citep{montalto2010}, to detect Trojan perturbers (e.g. Haghighipour \& Capen 2010, in preparation), to break the degeneracy in the inverse problem by observing {\it either} each of {\it two} planets transiting in the same system \citep{raghol2010} or a transiting planet and a transiting moon \citep{kipping2010b}, and to combine Markov Chain Monte Carlo analyses of radial velocity data with TTVs for the recently scrutinized system HAT-P-13 (e.g. Payne \& Ford 2010, submitted).  Further, as already mentioned in Section 1, Transit Duration Variations (TDVs) may help constrain the degeneracies from TTVs.  \cite{kipping2010a} has recently presented new analytical formula for TDVs, and \cite{dvoetal2010} demonstrates a correlation between TDVs and the relative inclination of the planets in another system of recent interest, CoRoT-7.  Beyond these studies, other possibilities abound.  TTVs in systems with three or more planets will provide both an opportunity and a challenge for theorists to explain and interpret, and potentially habitable planets might best be identified through TTVs orbiting K or M-type stars.  Additionally, non-continuous transit observations, which are particularly relevant for ground-based observatories, may have different dependencies on signal amplitude and shape than those displayed here, and will likely be important to consider on a case-by-case basis.

\section{Conclusion}
Our goal was to illustrate and quantify the challenges involved in characterizing the mass and orbital parameters of an external perturber in a system with an observed transiting hot Jupiter through the use of Transit Timing Variations.  TTV signals may vary by orders of magnitude due to $\sim 10^{-3}$ AU shifts in semimajor axes, $\sim 0.005$ shifts in eccentricity, or $\sim 1^{\circ}$ shifts in orbital angles.  However, TTVs are a sensitive probe of hidden planets, and can suggest the existence of external perturbers due to signatures produced at high-order period commensurabilities (PCs; $p$:$q$, where $p \lesssim 20$ and $q \lesssim 3$).  More specific conclusions from this work are: 1) moderate (tens of seconds) values of signal amplitudes could indicate that the outer planet eccentricity could be as high as the Hill eccentricity, 2) high amplitude ($> 10^3$ s) signals don't necessarily imply close proximity to a PC, 3) high amplitude ($> 10^3$ s) signals are always indicative of a high value (at least 4/5ths of the Hill eccentricity) of the outer planet eccentricity, 4) near-PC amplitudes are often {\it higher} than in-PC amplitudes, a result largely independent of the number of observed transits, $N$, 5) signal amplitude is a non-monotonic function of $N$, 6) increasing the external mass generally increases the TTV signal, 7) increasing the internal (hot Jupiter) mass generally decreases the number of easily discernible PCs, and 8) the distinctive signal amplitudes for systems at PCs don't necessarily imply that those systems are actually captured in a mean motion resonance.  We propose using the shape data of a TTV curve through the autocorrelation function as a method to help characterize the external planets, and believe that the method could provide a promising avenue of future study.

\acknowledgments{We thank the referee for insightful observations and Eric Agol, Dan Fabrycky, Nader Haghighipour, Matt Holman, David Nesvorn{\'y}, and Jason Steffen for valuable and extensive discussions.  This material is based upon work supported by the National Science Foundation under Grant No. 0707203.

\clearpage

\appendix
\section{Appendix}

Here we derive the estimates for the libration widths drawn in Fig. \ref{fiducial} for selected PCs of up to 8th order and discuss the goodness of the approximation.  We take ``libration width'' to mean roughly the $a_o$ range for a given $e_o$ in which MMR locking between two planets is likely.  Because of the hierarchical nature of the three masses for the systems we consider, and the initially circular orbit of the more massive planet, we assume that the one argument given by Eq. (\ref{resangle}) of the disturbing function will dominate the gravitational potential.  This approximation is vital to obtaining a tractable analytic expression.

We can assess the goodness of the approximation of neglecting terms that include $e_i(t)$ by using \citeauthor{veras2007}'s (\citeyear{veras2007}) term-based integrator. We tested systems locked in the $3$:$1$ and $5$:$1$ MMRs by including different numbers of terms up to 4th-order.  For all systems, we set $M_{\star} = M_{\odot}$, $M_i = M_J$, $M_o = M_{\oplus}$, $a_i = 0.05$, and $a_o = a_i(p/q)^{2/3}$.  For the $3$:$1$ MMR, we set $e_o = 0.30$, $\lambda_i = 100^{\circ}, \lambda_o = 0^{\circ}, \varpi_i = 340^{\circ}$ and $\varpi_o = 180^{\circ}$ in order to produce resonant behavior.  For the $5$:$1$ MMR, we set $e_o = 0.45$, $\lambda_i = 5^{\circ}, \lambda_o = 170^{\circ}, \varpi_i = 3^{\circ}$ and $\varpi_o = 19^{\circ}$ in order to produce resonant behavior. Importantly, for the $3$:$1$ MMR, we set $e_i = 0.00$, but for the $5$:$1$ MMR, we set $e_i = 0.01$.  Initializing the hot Jupiter with a non-zero eccentricity might cause additional disturbing function terms to become significant, and is realistic given that an external terrestrial-mass perturber can typically force the hot Jupiter's eccentricity to values of $\approx 0.005$. 

We test these systems by including all arguments that, in their coefficients, include $e_i(t)$ (case I), and by neglecting all such arguments (case II).  For both cases and for both MMRs, the resonant angle librates, and does so about $180^{\circ}$.  In Case I, for the $3$:$1$ and $5$:$1$ MMRs, the libration amplitudes are $\approx 55^{\circ}$ and $\approx 65^{\circ}$; Case II changes these values by at most a few degrees.  The range of $e_o(t)$ over 3 years in Cases I and II is $0.3040-0.2639=0.0401$ and $0.3035-0.2586=0.0449$ for the $3$:$1$ MMR, and $0.4669-0.4346=0.0323$ and $0.4669-0.4346=0.0323$ for the $5$:$1$ MMR.  These results demonstrate that neglecting these additional arguments affects the motion in too small a manner to necessitate inclusion in this general, qualitative study.

With a single disturbing function argument, $\phi$, we can approximate a libration width according to the same prescription found in \cite{murder2000}.  The angle behaves like a pendulum so that:

\begin{equation}
\ddot{\phi} = \omega^2 \sin{\phi}
\end{equation}

\noindent{}One can find the value of $\omega$ through a combination of 1) the definition of $\phi$, 2) use of the disturbing functions, $R = R(\phi)$, and 3) use of Lagrange's Planetary Equations \citep{murder2000}.  We can define two disturbing functions for the systems we study here as:

\begin{eqnarray}
& &
R_1 = \frac{G M_i}{a_i} e_{o}^{p-q} \left( \alpha f_d + f_i \right) \cos{\phi}
\\
& &
R_2 = \frac{G M_o}{a_o} e_{o}^{p-q} \left( f_d + f_e \right) \cos{\phi}
\end{eqnarray}

\noindent{where} $G$ is the Gravitational Constant, $\alpha = a_i/a_o$,  and $f_d$, $f_{id}$ and $f_e$ are functions of $\alpha$ and $e_o$. By defining $\sigma_k = G \left( M_{\star} + M_{k} \right)$, for $k = i,o$, we find:

\begin{eqnarray}
& &
\ddot{\phi} \approx p \dot{n_{o}} - q \dot{n_{i}}
\\
& & = -\frac{3}{2} p \sigma_{o}^{1/2} a_{o}^{-5/2} \dot{a}_o - \frac{3}{2} q \sigma_{i}^{1/2} a_{i}^{-5/2} \dot{a}_i
\\
& & = \frac{-3 p}{a_{o}^2} \frac{\partial R_1}{\partial \lambda_o} - \frac{3 q}{a_{i}^2} \frac{\partial R_2}{\partial \lambda_i} 
\end{eqnarray}

\noindent{} so that:

\begin{equation}
\omega^2  = \frac{3 G e_{o}^{p-q}}{a_o a_i} 
\left( 
\frac{p^2 M_i}{a_o} \left[ \alpha f_d + f_{id}  \right] +
\frac{q^2 M_o}{a_i} \left[ f_d + f_e  \right]
\right)
\end{equation}

The energy of a pendulum is $E = (1/2)\dot{\phi}^2 + 2 \omega^2 \sin^2(\phi/2)$.  Hence, the maximum energy is $2 \omega^2$, and equating this value with $E$ yields a relation between $\dot{\phi}$ and $\phi$.  One can combine this relation with the Lagrange Planetary Equation for $\dot{a}$ in order to obtain:

\begin{equation}
da_o = \pm \frac{p G a_{o}^{1/2} M_i}{\omega a_i \sigma_{o}^{1/2}} e_{o}^{p-q}
\left[ \frac{a_o}{a_i} f_d + f_{id} \right]  \frac{\sin{\phi}}{\cos{(\phi/2)}} d\phi
\end{equation}

\noindent{which} can be integrated to finally obtain the (maximum) libration width $\delta_a$:

\begin{equation}
\delta_a = 
\frac{
\frac{2}{\sqrt{3}}
\frac{p a_o M_i}{a_{i}^{1/2} \left( M_{\star} + M_o  \right)^{1/2}  }
e_{o}^{(p-q)/2}
\left(\alpha f_d + f_{id} \right)
}
{\sqrt{
\frac{p^2 M_i}{a_{o}} \left( \alpha f_d + f_{id} \right) +
\frac{q^2 M_o}{a_{i}} \left( f_d + f_e \right)
}
}
\end{equation}

\noindent{in} the approximation $M_o = 0$,

\begin{equation}
\delta_a = a_o e_{o}^{(p-q)/2} 
\sqrt{\left( \frac{4}{3 \alpha} \right)
      \left( \frac{M_i}{M_{\star}} \right)
      \left( \alpha f_d + f_{id} \right) }
\label{finallib}
\end{equation}

For most MMRs, $f_{id} = f_e = 0$.  However, when $q=1$, one of the most important classes of MMRs for TTVs, both $f_{id}$ and $f_e$ are nonzero. Formulas for $f_d$, $f_{id}$ and $f_e$ are provided in \cite{murder2000} up to 4th order in $e_o$.  Stan Dermott, private communication, has provided us with terms up to 8th order.  Explicit formulas for $f_d$ in terms of $\alpha$ may be found in \cite{verarm2007}.  


The eccentricity at which Eq. (\ref{finallib}) holds is restricted by the Sundman criterion \citep{ferrazmello1994,sidnes1994}, a fundamental convergence criterion on the planar expansion of the disturbing function from \cite{ellmur2000}.  This criterion can be expressed as:

\begin{equation}
a_i D(e_i) < a_o d(e_o),
\end{equation}

\noindent{where}

\begin{eqnarray}
D(y) &=& \sqrt{1 + y^2} \cosh{z} + y + \sinh{z}
,
\\
d(y) &=& \sqrt{1 + y^2} \cosh{z} - y - \sinh{z}
\end{eqnarray}

\noindent{such} that $z$ is implicitly defined as $z = q \cosh{z}$.  This restriction prevents computation of the libration width for all eccentricities up to the Hill Stability limit.  Figure 5 of \cite{nesmor2008} shows a detailed view of the difference between the Hill Stability limit and the Sundman convergence limit for several different mass ratios.


\clearpage


\begin{thebibliography}{}
%



\bibitem[Adams \& Laughlin(2006)]{adalau2006} 
Adams, F.~C., \& Laughlin, G.\ 2006, \apj, 649, 1004 

\bibitem[Adams et al.(2010)]{adaetal2010}
Adams, E.~R., L{\'o}pez-Morales, M., Elliot, J.~L., Seager, S., 
\& Osip, D.~J.\ 2010, \apj, 714, 13 

\bibitem[Agol et al.(2005)]{agoetal2005} 
Agol, E., Steffen, J., Sari, R., \& Clarkson, W.\ 2005, \mnras, 359, 567 

\bibitem[Agol \& Steffen(2007)]{agoste2007} 
Agol, E., \& Steffen, J.~H.\ 2007, \mnras, 374, 941

\bibitem[Aigrain et al.(2008)]{aigetal2008} 
Aigrain, S., et al.\ 2008, \aap, 488, L43 

\bibitem[Alonso et al.(2008)]{aloetal2008} 
Alonso, R., et al.\ 2008, \aap, 482, L21

\bibitem[Bakos et al.(2009)]{baketal2009} 
Bakos, G.~{\'A}., et al.\ 2009, \apj, 707, 446 

\bibitem[Barge et al.(2008)]{baretal2008} 
Barge, P., et al.\ 2008, \aap, 482, L17 

\bibitem[Barnes \& Greenberg(2006a)]{bargre2006a} 
Barnes, R., \& Greenberg, R.\ 2006a, \apjl, 647, L163 

\bibitem[Barnes \& Greenberg(2006b)]{bargre2006b} 
Barnes, R., \& Greenberg, R.\ 2006b, \apjl, 652, L53 

\bibitem[Barnes \& Greenberg(2007)]{bargre2007} 
Barnes, R., \& Greenberg, R.\ 2007, \apjl, 665, L67 

\bibitem[Beaug{\'e} \& Michtchenko(2003)]{beamic2003} 
Beaug{\'e}, C., \& Michtchenko, T.~A.\ 2003, \mnras, 341, 760 

\bibitem[Beaug{\'e} et al.(2006)]{beaetal2006} 
Beaug{\'e}, C., Michtchenko, T.~A., \& Ferraz-Mello, 
S.\ 2006, \mnras, 365, 1160

\bibitem[Borkovits et al.(2003)]{boretal2003} 
Borkovits, T., {\'E}rdi, B., Forg{\'a}cs-Dajka, E., \& Kov{\'a}cs, T.\ 2003, \aap, 398, 1091 

\bibitem[Borucki et al.(2010)]{boretal2010} 
Borucki, W.~J., et al.\ 2010, Science, 327, 977

\bibitem[Butler et al.(2006)]{butetal2006} 
Butler, R.~P., et al.\ 2006, \apj, 646, 505 

\bibitem[Chambers(1999)]{chambers1999} 
Chambers, J.~E.\ 1999, \mnras, 304, 793

\bibitem[Chiang et al.(2001)]{chietal2001} 
Chiang, E.~I., Tabachnik, S., \& Tremaine, S.\ 2001, \aj, 122, 1607

\bibitem[Coughlin et al.(2008)]{couetal2008} Coughlin, J.~L., 
Stringfellow, G.~S., Becker, A.~C., L{\'o}pez-Morales, M., Mezzalira, F., 
\& Krajci, T.\ 2008, \apjl, 689, L149 

\bibitem[Crida et al.(2008)]{crietal2008} 
Crida, A., S{\'a}ndor, Z., \& Kley, W.\ 2008, \aap, 483, 325 

\bibitem[Csizmadia et al.(2010)]{csietal2010} 
Csizmadia, S., et al.\ 2010, \aap, 510, A94 

\bibitem[Deeg(2002)]{deeg2002} 
Deeg, H.~J.\ 2002, Earth-like Planets and Moons, 514, 237 

\bibitem[Deleuil et al.(2008)]{deletal2008} 
Deleuil, M., et al.\ 2008, \aap, 491, 889

\bibitem[D{\'{\i}}az et al.(2008)]{diazetal2008} D{\'{\i}}az, R.~F., 
Rojo, P., Melita, M., Hoyer, S., Minniti, D., Mauas, P.~J.~D., 
\& Ru{\'{\i}}z, M.~T.\ 2008, \apjl, 682, L49

\bibitem[Doyle \& Deeg(2004)]{doydee2004} 
Doyle, L.~R., \& Deeg, H.-J.\ 2004, Bioastronomy 2002: Life Among the Stars, 213, 80 

\bibitem[Dvorak et al.(2010)]{dvoetal2010} 
Dvorak, R., Schneider, 
J., \& Eybl, V.\ 2010, arXiv:1004.4129 

\bibitem[Ellis \& Murray(2000)]{ellmur2000} 
Ellis, K.~M., \& Murray, C.~D.\ 2000, Icarus, 147, 129 

\bibitem[Fabrycky \& Murray-Clay(2010)]{fabmur2010} 
Fabrycky, D.~C., \& Murray-Clay, R.~A.\ 2010, \apj, 710, 1408 

\bibitem[Ferraz-Mello(1994)]{ferrazmello1994} 
Ferraz-Mello, S.\ 1994, Celestial Mechanics and Dynamical Astronomy, 58, 37 

\bibitem[Ferraz-Mello et al.(2003)]{feretal2003} 
Ferraz-Mello, S., Beaug{\'e}, C., \& Michtchenko, T.~A.\ 2003, 
Celestial Mechanics and Dynamical Astronomy, 87, 99

\bibitem[Fischer et al.(2003)]{fisetal2003} 
Fischer, D.~A., et al.\ 2003, \apj, 586, 1394 

\bibitem[Ford(2005)]{ford2005} 
Ford, E.~B.\ 2005, \aj, 129, 1706 

\bibitem[Ford(2006)]{ford2006} 
Ford, E.~B.\ 2006, \apj, 642, 505 

\bibitem[Ford \& Holman(2007)]{forhol2007} 
Ford, E.~B., \& Holman, M.~J.\ 2007, \apjl, 664, L51 

\bibitem[Ford et al.(2005)]{foretal2005} 
Ford, E.~B., Lystad, V., \& Rasio, F.~A.\ 
2005, \nat, 434, 873 

\bibitem[Ford \& Rasio(2007)]{forras2007} 
Ford, E.~B., \& Rasio, F.~A.\ 2007, 
ArXiv Astrophysics e-prints, arXiv:astro-ph/0703163

\bibitem[Fridlund et al.(2010)]{frietal2010} 
Fridlund, M., et al.\ 2010, arXiv:1001.1426 

\bibitem[Gladman(1993)]{gladman1993} 
Gladman, B.\ 1993, Icarus, 106, 247 

\bibitem[Go{\'z}dziewski(2003)]{gozdziewski2003} 
Go{\'z}dziewski, K.\ 2003, \aap, 398, 1151 

\bibitem[Go{\'z}dziewski \& Maciejewski(2003)]{gozmac2003} 
Go{\'z}dziewski, K., \& Maciejewski, A.~J.\ 2003, \apjl, 586, L153 

\bibitem[Gregory(2007a)]{gregory2007a} 
Gregory, P.~C.\ 2007a, \mnras, 381, 1607 

\bibitem[Gregory(2007b)]{gregory2007b} 
Gregory, P.~C.\ 2007b, \mnras, 374, 1321 

\bibitem[Heyl \& Gladman(2007)]{heygla2007} 
Heyl, J.~S., \& Gladman, B.~J.\ 2007, \mnras, 377, 1511 

\bibitem[Holman \& Murray(2005)]{holmur2005} 
Holman, M.~J., \& Murray, N.~W.\ 2005, Science, 307, 1288 

\bibitem[Holman et al.(2010)]{holmanetal2010} 
Holman, M.~J., et al.\ 2010, Science, 330, 51 

\bibitem[Hrudkov{\'a} et al.(2008)]{hruetal2008} 
Hrudkov{\'a}, M., Skillen, I., Benn, C., Pollacco, D., Gibson, N., 
Joshi, Y., Harmanec, P., \& Tulloch, S.\ 2008, arXiv:0807.1000 

\bibitem[Ji et al.(2003)]{jietal2003} 
Ji, J., Liu, L., Kinoshita, H., Zhou, J., 
Nakai, H., \& Li, G.\ 2003, \apjl, 591, L57 

\bibitem[Jones et al.(2006)]{jonetal2006} 
Jones, H.~R.~A., Butler, R.~P., Tinney, C.~G., Marcy, G.~W., Carter, B.~D., Penny, A.~J., McCarthy, 
C., \& Bailey, J.\ 2006, \mnras, 369, 249 

\bibitem[Kipping(2009a)]{kipping2009a} Kipping, D.~M.\ 2009a, \mnras, 
392, 181 

\bibitem[Kipping(2009b)]{kipping2009b} Kipping, D.~M.\ 2009b, \mnras, 
396, 1797 

\bibitem[Kipping(2010a)]{kipping2010a} Kipping, D.~M.\ 2010, 
arXiv:1004.3819 

\bibitem[Kipping(2010b)]{kipping2010b} Kipping, D.~M.\ 2010, 
arXiv:1010.2492 

\bibitem[Kiseleva-Eggleton et al.(2002)]{kisetal2002} 
Kiseleva-Eggleton, L., Bois, E., Rambaux, N., 
\& Dvorak, R.\ 2002, \apjl, 578, L145 

\bibitem[Kley et al.(2005)]{kleetal2005} 
Kley, W., Lee, M.~H., Murray, N., \& Peale, S.~J.\ 2005, \aap, 437, 727 

\bibitem[Knutson et al.(2007)]{knuetal2007} 
Knutson, H.~A., Charbonneau, D., Noyes, R.~W., Brown, T.~M., 
\& Gilliland, R.~L.\ 2007, \apj, 655, 564 

\bibitem[Laughlin \& Chambers(2001)]{laucha2001} 
Laughlin, G., \& Chambers, J.~E.\ 2001, \apjl, 551, L109 

\bibitem[Lee(2004)]{lee2004} 
Lee, M.~H.\ 2004, \apj, 611, 517 

\bibitem[Lee \& Peale(2003)]{leepea2003} 
Lee, M.~H., \& Peale, S.~J.\ 2003, \apj, 592, 1201 

\bibitem[L{\'e}ger et al.(2009)]{legetal2009} 
L{\'e}ger, A., et al.\ 2009, \aap, 506, 287 

\bibitem[Libert \& Henrard(2007)]{libhen2007} 
Libert, A.-S., \& Henrard, J.\ 2007, \aap, 461, 759 

\bibitem[Libert \& Henrard(2008)]{libhen2008} 
Libert, A.-S., \& Henrard, J.\ 2008, Celestial Mechanics and Dynamical Astronomy, 100, 209 

\bibitem[Malhotra(2002)]{malhotra2002} 
Malhotra, R.\ 2002, \apjl, 575, L33 


\bibitem[Meschiari \& Laughlin(2010)]{marlau2010} 
Meschiari, S., \& Laughlin, G.\ 2010, arXiv:1005.5396

\bibitem[Mardling(2008)]{Mardling2008} 
Mardling, R.~A.\ 2008, 
Lecture Notes in Physics, Berlin Springer Verlag, 760, 59

\bibitem[Mardling(2010)]{Mardling2010}
Mardling, R. submitted to \mnras

\bibitem[Mayor \& Queloz(1995)]{mayque1995} 
Mayor, M., \& Queloz, D.\ 1995, \nat, 378, 355

\bibitem[Michtchenko et al.(2006a)]{micetal2006a} 
Michtchenko, T.~A., Beaug{\'e}, C., \& Ferraz-Mello, S.\ 2006, 
Celestial Mechanics and Dynamical Astronomy, 94, 411

\bibitem[Michtchenko et al.(2008a)]{micetal2008a} 
Michtchenko, T.~A., Beaug{\'e}, C., \& Ferraz-Mello, S.\ 2008a, \mnras, 387, 747 

\bibitem[Michtchenko et al.(2008b)]{micetal2008b} 
Michtchenko, T.~A., Beaug{\'e}, C., \& Ferraz-Mello, S.\ 2008b, \mnras, 391, 215 

\bibitem[Miller-Ricci et al.(2008a)]{miletal2008a} Miller-Ricci, E., 
et al.\ 2008a, \apj, 682, 586 

\bibitem[Miller-Ricci et al.(2008b)]{miletal2008b} Miller-Ricci, E., 
et al.\ 2008b, \apj, 682, 593 

\bibitem[Miralda-Escud{\'e}(2002)]{miraldaescude2002} 
Miralda-Escud{\'e}, J.\ 2002, \apj, 564, 1019 

\bibitem[Montalto(2010)]{montalto2010} 
Montalto, M.\ 2010, arXiv:1006.3026 

\bibitem[Murray \& Dermott(2000)]{murder2000} 
Murray, C.~D., \& Dermott, S.~F.\ 2000, Solar System Dynamics, by C.D.~Murray

\bibitem[Naef et al.(2001)]{naefetal2001} 
Naef, D., et al.\ 2001, \aap, 375, L27 

\bibitem[Nesvorn{\'y}(2009)]{nesvorny2009} 
Nesvorn{\'y}, D.\ 2009, \apj, 701, 1116 

\bibitem[Nesvorn{\'y} \& Beaug{\'e}(2010)]{nesbea2010}
Nesvorn{\'y}, D., \& Beaug{\'e}, C.\ 2010, \apjl, 709, L44 

\bibitem[Nesvorn{\'y} \& Morbidelli(2008)]{nesmor2008} 
Nesvorn{\'y}, D., \& Morbidelli, A.\ 2008, \apj, 688, 636

\bibitem[P{\'a}l \& Kocsis(2008)]{palkoc2008} 
P{\'a}l, A., \& Kocsis, B.\ 2008, \mnras, 389, 191 

\bibitem[Pan \& Sari(2004)]{pansar2004} 
Pan, M., \& Sari, R.\ 2004, \aj, 128, 1418

\bibitem[Payne et al.(2010)]{payetal2010} 
Payne, M.~J., Ford, E.~B., \& Veras, D.\ 2010, \apjl, 712, L86 

\bibitem[Queloz et al.(2009)]{queetal2009} 
Queloz, D., et al.\ 2009, \aap, 506, 303

\bibitem[Ragozzine \& Holman(2010)]{raghol2010} 
Ragozzine, D., \& Holman, M.~J.\ 2010, arXiv:1006.3727 

\bibitem[Rasio \& Ford(1996)]{rasfor1996} 
Rasio, F.~A., \& Ford, E.~B.\ 1996, Science, 274, 954

\bibitem[Rauer et al.(2009)]{rauetal2009} 
Rauer, H., et al.\ 2009, \aap, 506, 281 

\bibitem[Raymond et al.(2008)]{rayetal2008} Raymond, S.~N., Barnes, 
R., \& Gorelick, N.\ 2008, \apj, 689, 478 

\bibitem[Ribas et al.(2008)]{ribetal2008} Ribas, I., Font-Ribera, 
A., \& Beaulieu, J.-P.\ 2008, \apjl, 677, L59 

\bibitem[Rodr{\'{\i}}guez \& Gallardo(2005)]{rodgal2005} 
Rodr{\'{\i}}guez, A., \& Gallardo, T.\ 2005, \apj, 628, 1006

\bibitem[S{\'a}ndor et al.(2007)]{sanetal2007} 
S{\'a}ndor, Z., Kley, W., \& Klagyivik, P.\ 2007, \aap, 472, 981 

\bibitem[Sidlichovsky \& Nesvorny(1994)]{sidnes1994} 
Sidlichovsky, M., \& Nesvorny, D.\ 1994, \aap, 289, 972 

\bibitem[Simon et al.(2007)]{simetal2007} 
Simon, A., Szatm{\'a}ry, K., \& Szab{\'o}, G.~M.\ 2007, \aap, 470, 727 

\bibitem[Steffen \& Agol(2005)]{steago2005} 
Steffen, J.~H., \& Agol, E.\ 2005, \mnras, 364, L96 

\bibitem[Steffen et al.(2007)]{steetal2007} Steffen, J.~H., Gaudi, 
B.~S., Ford, E.~B., Agol, E., \& Holman, M.~J.\ 2007, arXiv:0704.0632 

\bibitem[Steffen et al.(2010)]{steetal2010} Steffen, J.~H., et al.\ 
2010, arXiv:1006.2763 

\bibitem[Stringfellow et al.(2009)]{stretal2009} 
Stringfellow, G.~S., Coughlin, J.~L., L{\'o}pez-Morales, M., 
Becker, A.~C., Krajci, T., Mezzalira, F., 
\& Agol, E.\ 2009, American Institute of Physics Conference Series, 1094, 481 

\bibitem[Veras(2007)]{veras2007} 
Veras, D.\ 2007, Celestial Mechanics and Dynamical Astronomy, 99, 197 

\bibitem[Veras \& Armitage(2004)]{verarm2004} 
Veras, D., \& Armitage, P.~J.\ 2004, Icarus, 172, 349 

\bibitem[Veras \& Armitage(2007)]{verarm2007} 
Veras, D., \& Armitage, P.~J.\ 2007, \apj, 661, 1311 

\bibitem[Veras \& Ford(2009)]{verfor2009} 
Veras, D., \& Ford, E.~B.\ 2009, IAU Symposium, 253, 486 

\bibitem[Voyatzis \& Hadjidemetriou(2006)]{voyhad2006} 
Voyatzis, G., \& Hadjidemetriou, J.~D.\ 2006, 
Celestial Mechanics and Dynamical Astronomy, 95, 259 

\bibitem[Weidenschilling \& Marzari(1996)]{weimar1996} 
Weidenschilling, S.~J., \& Marzari, F.\ 1996, \nat, 384, 619 

\bibitem[Zhou \& Sun(2003)]{zhosun2003} 
Zhou, J.-L., \& Sun, Y.-S.\ 2003, \apj, 598, 1290


\end{thebibliography}
\end{document}